\documentclass[twocolumn,trackchanges]{aastex631}

\usepackage{footnote}

\usepackage{amsmath}

\newcommand{\kms}{{\hbox {\,km\,s$^{-1}$}}}

\usepackage{amsmath}

\def\aco  {\ifmmode{{\rm CO}(J=1\to0)}\else{${\rm CO}(J=1\to0)$}\fi}
\def\bco  {\ifmmode{{\rm CO}(J=2\to1)}\else{${\rm CO}(J=2\to1)$}\fi}
\def\cco   {\ifmmode{{\rm CO}(J=3\to2)}\else{${\rm CO}(J=3\to2)$}\fi}
\def\dco  {\ifmmode{{\rm CO}(J=4\to3)}\else{${\rm CO}(J=4\to3)$}\fi}
\def\eco  {\ifmmode{{\rm CO}(J=5\to4)}\else{${\rm CO}(J=5\to4)$}\fi}

\received{2025 March 5}
\revised{2025 June 26}
\accepted{2025 July 18}
\published{2025 September 16}

\shorttitle{HerS-3: An Einstein Cross with a Central Image}
\shortauthors{Cox et al.}
\graphicspath{{./}{figures/}}

\begin{document}

\title{HerS-3: An Exceptional Einstein Cross Reveals a Massive Dark Matter Halo}

\author[0000-0003-2027-8221]{P. Cox}
\affiliation{Sorbonne Universit{\'e}, UPMC Paris 6 \& CNRS, UMR 7095, Institut d'Astrophysique de Paris, 98b bd. Arago, 75014 Paris, France} \email{pierre.cox@iap.fr}

\author[0000-0001-7387-0558]{K.~M. Butler}
\affiliation{Departement of Space, Earth \& Environement, Chalmers University of Technology, Chalmersplatsen 4, Gothenburg SE-412 96, Sweden}

\author[0000-0001-6812-2467]{C.~R. Keeton}
\affiliation{Department of Physics and Astronomy, Rutgers University -- New Brunswick, 136 Frelinghuysen Road, Piscataway, NJ 08854-8019, USA}

\author[0009-0003-5696-9355]{L. Eid}
\affiliation{Department of Physics and Astronomy, Rutgers University -- New Brunswick, 136 Frelinghuysen Road, Piscataway, NJ 08854-8019, USA}

\author[0009-0009-9483-8763]{E. Borsato}
\affiliation{Dipartimento di Fisica e Astronomia ‘G. Galilei’, Università di Padova, vicolo dell’Osservatorio 3, I-35122 Padova, Italy}

\author[0000-0002-5268-2221]{T.~J.~L.~C. Bakx}
\affiliation{Departement of Space, Earth \& Environement, Chalmers University of Technology, Chalmersplatsen 4, Gothenburg SE-412 96, Sweden}

\author[0000-0002-7176-4046]{R. Neri}
\affiliation{Institut de Radioastronomie Millim\'etrique (IRAM), 300 rue de la Piscine, 38406 Saint-Martin-d'H{\`e}res, France}

\author[0000-0002-0675-0078]{B.~M. Jones}
\affiliation{I. Physikalisches Institut, Universit\"at zu K\"oln, Z\"ulpicher Strasse 77, D-50937 K\"oln, Germany}

\author[0000-0002-3094-1077]{P. Prajapati}
\affiliation{I. Physikalisches Institut, Universit\"at zu K\"oln, Z\"ulpicher Strasse 77, D-50937 K\"oln, Germany}
\affiliation{Max-Planck-Institut f{\"u}r Radioastronomie, Auf dem H{\"u}gel 69, 53121 Bonn, Germany}

\author[0000-0002-7892-396X]{A.~J. Baker}
\affiliation{Department of Physics and Astronomy, Rutgers University -- New Brunswick, 136 Frelinghuysen Road, Piscataway, NJ 08854-8019, USA}

\affiliation{Department of Physics and Astronomy, University of the Western Cape, Robert Sobukwe Road, Bellville 7535, Cape Town, South Africa}

\author[0000-0002-0320-1532]{S. Berta}
\affiliation{Institut de Radioastronomie Millim\'etrique (IRAM), 300 rue de la Piscine, 38406 Saint-Martin-d'H{\`e}res, France}

\author[0000-0002-3892-0190]{A. Cooray}
\affiliation{University of California Irvine, Department of Physics \& Astronomy, FRH 2174, Irvine, CA 92697, USA}

\author[0000-0003-3460-5633]{E.~M. Corsini}
\affiliation{Dipartimento di Fisica e Astronomia ‘G. Galilei’, Università di Padova, vicolo dell’Osservatorio 3, I-35122 Padova, Italy}
\affiliation{INAF – Osservatorio Astronomico di Padova, vicolo dell’Osservatorio 5, I-35122 Padova, Italy}

\author[0000-0003-3948-7621]{L. Marchetti}
\affiliation{IDIA \& Departement of Astronomy, University of Cape Town, 7701 Rondebosch, Cape Town, South Africa}
\affiliation{Istituto Nazionale di Astrofisica, Istituto di Radioastronomia - Italian ARC, Via Piero Gobetti 101, 40129 Bologna, Italy}

\author[0000-0002-4721-3922]{A. Omont}
\affiliation{Sorbonne Universit{\'e}, UPMC Paris 6 \& CNRS, UMR 7095, Institut d'Astrophysique de Paris, 98b bd. Arago, 75014 Paris, France}

\author[0000-0003-3201-0185]{A. Beelen}
\affiliation{Aix-Marseille Universit\'{e}, CNRS \& CNES, LAM, 38, rue Frédéric Joliot-Curie, 13388 Marseille, France}

\author[0000-0002-5540-6935]{R. Gavazzi}
\affiliation{Aix-Marseille Universit\'{e}, CNRS \& CNES, LAM, 38, rue Frédéric Joliot-Curie, 13388 Marseille, France}

\author[0009-0007-2281-4944]{D. Ismail}
\affiliation{Université de Strasbourg, CNRS, Observatoire astronomique de Strasbourg, UMR 7550, 67000 Strasbourg, France}

\author[0000-0001-5118-1313]{R.~J. Ivison}
\affiliation{Institute for Astronomy, University of Edinburgh, Royal Observatory, Blackford Hill, Edinburgh EH9 3HJ, UK}
\affiliation{School of Cosmic Physics, Dublin Institute for Advanced Studies, 31 Fitzwilliam Place, Dublin D02 XF86, Ireland}

\author[0000-0001-6971-4851]{M. Krips}
\affiliation{Institut de Radioastronomie Millim\'etrique (IRAM), 300 rue de la Piscine, 38406 Saint-Martin-d'H{\`e}res, France}

\author[0000-0003-1939-5885]{M.D. Lehnert}
\affiliation{Universit\'{e} Lyon 1, ENS de Lyon, Centre de Recherche Astrophysique de Lyon (UMR5574), 69230 Saint-Genis-Laval, France}
\affiliation{Sorbonne Universit{\'e}, UPMC Paris 6 \& CNRS, UMR 7095, Institut d'Astrophysique de Paris, 98b bd. Arago, 75014 Paris, France}

\author[0000-0002-2985-7994]{H. Messias}
\affiliation{European Southern Observatory, Alonso de Córdova 3107, Casilla 19001, Vitacura, Santiago, Chile}
\affiliation{Joint ALMA Observatory, Alonso de Córdova, 3107, Vitacura, Santiago 763-0355, Chile}

\author[0000-0001-9585-1462]{D. Riechers}
\affiliation{I. Physikalisches Institut, Universit\"at zu K\"oln, Z\"ulpicher Strasse 77, D-50937 K\"oln, Germany}

\author[0000-0003-3745-4228]{C. Vlahakis}
\affiliation{National Radio Astronomy Observatory, 520 Edgemont Rd., Charlottesville, VA 22901, USA}

\author[0000-0003-4678-3939]{A. Wei\ss}
\affiliation{Max-Planck-Institut f{\"u}r Radioastronomie, Auf dem H{\"u}gel 69, 53121 Bonn, Germany}

\author[0000-0001-5434-5942]{P. van der Werf}
\affiliation{Leiden Observatory, Leiden University, PO Box 9513, 2300 RA Leiden, The Netherlands}

\author[0000-0002-8117-9991]{C. Yang}
\affiliation{Departement of Space, Earth \& Environement, Chalmers University of Technology, Chalmersplatsen 4, Gothenburg SE-412 96, Sweden}





\begin{abstract}
We present a study of HerS-3, a dusty star-forming galaxy at $z_{\rm spec} = 3.0607$, which is gravitationally amplified into an Einstein cross with a fifth image of the background galaxy seen at the center of the cross. Detailed 1-mm spectroscopy and imaging with NOEMA and ALMA resolve the individual images and show that each of the five images display a series of molecular lines that have similar central velocities, unambiguously confirming that they have identical redshifts. The {\it HST} F110W image reveals a foreground lensing group of four galaxies with a photometric redshift $z_{\rm phot} \sim 1.0$. Lens models that only include the four visible galaxies are unable to reproduce the properties of HerS-3. By adding a fifth massive component, lying south-east of the brightest galaxy of the group, the source reconstruction is able to match the peak emission, shape and orientation for each of the five images. The fact that no galaxy is detected near that position indicates the presence of a massive dark matter halo in the lensing galaxy group. In the source plane, HerS-3 appears as an infrared luminous starburst galaxy seen nearly edge-on. The serendipitous discovery of this exceptional Einstein cross offers a potential laboratory for exploring at small spatial scales a nuclear starburst at the peak of cosmic evolution and studying the properties of a massive dark matter halo associated with the lensing galaxy group.

\end{abstract}



\section{Introduction} \label{sec:intro}
Strong gravitational lensing occurs when light from a distant galaxy passes by a massive galaxy (or a group of galaxies) that distorts spacetime and causes the path of light of the background source to bend, making it appear brighter and producing multiple images, arcs, or a ring. The resulting amplification enables studies of galaxies in the early universe at spatial scales below 100~pc with current facilities. Such systems are powerful astrophysical laboratories for constraining the properties of dark matter in galaxies, groups, and clusters \citep[e.g.,][]{Hezaveh2016, Gilman2019, Sengul2022, Inoue2023, Nierenberg2024, Natarajan2024, Sheu2024}. Multiply imaged sources with time variability related to quasars or supernova explosions can also be used to monitor time delays and measure the Hubble constant \citep[e.g.,][]{Refsdal1964, Kochanek2004, Chen2019, Schmidt2023}.

For a nearly perfect alignment and an elliptical lens mass distribution, the background source will appear as a quadruply imaged system. The archetype of such systems is the Einstein cross, where four distinct images of the background source form a cross-like pattern with a high degree of symmetry. It remains a challenge finding and confirming such rare sources, and, as of today, the majority of cases have been discovered using optical or near-infrared imaging. 

The first calculations of gravitational lensing for stars (as both lens and source) were published by \cite{Chwolson1924} and, later, by \cite{Einstein1936}, who noted that ``there is no great chance of observing this phenomenon". This led \cite{Zwicky1937} to point out that extragalactic {\em nebulae} (i.e., galaxies) ``offer a much better chance than stars for the observation of gravitational lens effects". Forty years later, the first known lens was discovered in the optical by \cite{Walsh1979}, who showed that the quasar QSO-0956+561, at $z$=1.405, was lensed in two images separated by $6''$. A second discovery of a gravitationally amplified source was published soon thereafter, the quasar PG 1115+080, a quadruple lensed system at $z$=1.722 \citep{Weymann1980, Hege1981}. The third detection was made a few years later by \cite{Huchra1985} who reported a quasar lensed in a cross pattern, QSO-2237+0305 at $z$=1.695, known as the Einstein Cross \citep[see also][]{Yee1988, Adam1989}.  Many other examples of quasars lensed by foreground galaxies were found since then, including further cases of Einstein crosses. \cite{Lucey2018} reported two diamond shaped systems of quadruply imaged quasars; \cite{Witsotzki2002} discovered a lensed $z$=1.69 quasar (HE~0435$-$1223) with four images in a cross-shaped arrangement around a massive elliptical galaxy at $z$=0.45 \citep[see also][]{Wong2017}; one Einstein cross was identified in a sample of six quadruple-image lensed quasars by \cite{Wong2020}; and \cite{Stern2021} presented a dozen quadruply imaged systems discovered in the {\it Gaia} Data Release 2, including three Einstein crosses.  

Einstein crosses produced by galaxy-galaxy strong lensing are equally rare with only a dozen confirmed cases among a few hundred known lensed galaxies. \cite{Ratnatunga1995} reported the serendipitous discovery of two cases of such lenses using the Hubble Space Telescope ({\it HST}). \cite{Bolton2006} identified a cruciform configuration of a Ly$\alpha$ emitter galaxy at $z$=2.701 lensed by a luminous red galaxy at z$\sim$0.33. \cite{Pettini2010}  presented the case of CSWA~20, a blue star-forming galaxy at $z$=1.433 lensed by a luminous red galaxy at $z$=0.741 into a four-imaged configuration reminiscent of an Einstein cross, with a large separation ($\sim 6''$) between the images. \cite{Bettoni2019} demonstrated the lens nature of a cross-like system by deep spectroscopic observations, showing that it consisted of a Lyman-break galaxy at $z$=3.03 lensed by a galaxy at $z$=0.556. \cite{Napolitano2020} discovered two Einstein crosses in the footprint of the Kilo-Degree Survey, and \cite{Cikota2023} reported on a strong lensing system, DESI~253.2534+26.8843, a starburst galaxy at $z$=2.597 lensed into an Einstein cross by an elliptical galaxy at $z$=0.636.

Strongly lensed systems have also been discovered using sub-millimeter facilities, mostly through follow-up observations of infrared luminous dust-enshrouded starburst galaxies in the early Universe selected from the {\it Herschel}, Planck, and South Pole Telescope (SPT) surveys \citep{Eales2010, Oliver2012, Planck2015, Vieira2010}. The majority of these gravitational lensed systems, now in excess of 300 cases, appear as complete or partial Einstein rings or extended arcs, with no case known to date of an Einstein cross. One famous example of a lensed sub-millimeter galaxy is SDP.81 at $z$=3.042, magnified by a galaxy at $z$=0.3 \citep{Negrello2010, Omont2013}, which has been studied in exquisite sensitivity and angular resolution (23~mas) with the Atacama Large Millimeter/submillimeter Array (ALMA) to reveal the full details of a nearly complete Einstein ring \citep{ALMA-Partnership2015}. These high quality ALMA data have also provided evidence of dark matter substructure in the foreground lensing galaxy \citep{Hezaveh2016}. Another example is the Cloverleaf, a quadruply imaged starburst at $z$=2.55, harbouring a broad absorption line quasar, lensed by a $z \sim$1.7 galaxy \citep{Barvainis1994, Kneib1998, Chartas2004}, in which {\it HST} and ALMA images revealed a partial Einstein ring linking the four image components \citep{Chantry2007, Zhang2023}. Other quadruple lensed sources, which were found in the sub-millimeter, all show lensing arcs connecting the images, e.g, HLSW-01 at $z=2.95$ \citep{Conley2011}, G09v.1.97 at $z=3.63$ \citep{Yang2019}, or J0314$-$4452 at $z=2.93$ and J0150$-$5924 at $z=2.78$ \citep{Riechers2025}.

Detailed follow-up observations of high-$z$ galaxies detected in far-infrared and sub-millimeter continuum surveys are only possible when robust spectroscopic redshifts are available. Using the IRAM NOrthern Extended Millimeter Array (NOEMA), the $z$-GAL spectroscopic redshift survey of bright {\it Herschel}-selected dusty star-forming galaxies (DSFGs) has recently delivered precise redshifts for 165 individual galaxies, representing the largest flux-limited sample of DSFGs with unambiguous redshifts obtained to date \citep{Cox2023, Neri2020}. The $z$-GAL survey provides a foundation for exploring in detail the properties of these high-$z$ galaxies. 

In a recent follow-up study to trace fueling and feedback in $z$-GAL selected galaxies (Butler et al., in prep.), one gravitationally amplified galaxy stood out for its unusual lensed morphology. The source, HerS-3 at $z$=3.0607, was revealed to be an exceptional Einstein cross that, uniquely compared to all previous examples, shows a central fifth image.  

In this paper, we present a multi-wavelength study of HerS-3 and its surroundings, including sub-millimeter, radio and optical/near-infrared observations, focusing on the origin of the remarkable lensed image of this galaxy. In Section~\ref{sec:obs}, we describe the series of observations that are used in this study. In Section~\ref{sec:results}, we report the main results from the multi-wavelength observations for the molecular emission lines as well as the optical/near-infrared, sub-millimeter, and radio continuum emission. Section~\ref{sec:Lensing-Model} presents the gravitational lensing model and the results of the analysis, including the implied mass distribution of the foreground lensing system. Section~\ref{sec:Discussion} outlines the implications of the results of this study, addressing the main properties of HerS-3 in the source plane informed by the lensing model (Sect.~\ref{sec:Source-Properties}), discussing the nature of the additional mass component needed to reproduce the characteristics of the five images of the Einstein cross, and presenting them in a broader context by highlighting the potential use of this cosmic laboratory (Sect.~\ref{sec:dark-matter-halo} \& \ref{sec:cosmological-probe}). Finally, Section~\ref{sec:Conclusions} summarizes the main findings of this paper and outlines future observational prospects. 

Throughout this paper, we adopt a spatially flat $\Lambda$CDM cosmology with $H_{0}=67.4\,{\rm km\,s^{-1}\,Mpc^{-1}}$ and $\Omega_\mathrm{M}=0.315$ \citep{Planck2020}. With these parameters, $1\farcs0$ corresponds to $\rm \sim 7.8 \, kpc$ at $z=3.06$ in the source plane and the luminosity distance to the source is $D_{\rm L} = 2.7 \times 10^4 \, \rm Mpc$; in the lens plane, at $z\sim1.0$ (see Sect.~\ref{sec:Foreground-Group-Galaxies}), $1\farcs0$ corresponds to $\rm \sim 8.2 \, kpc$.

\section{Observations} \label{sec:obs}
\subsection{NOEMA} \label{sec:obs-NOEMA}
We used NOEMA to target high-frequency molecular lines in HerS-3, redshifted into the 1-mm band. The observations were carried out under project W21DF (P.I.: P. Cox). The project was completed on 2022 January 15 with twelve antennas using the intermediate C-configuration.  The source HerS-3 was observed for a total observing time of 1.13~hours, which resulted in a coherent set of high-sensitivity observations in Band 3 covering the two strongest ground-state OH$^+$ lines, corresponding to the redshifted transitions of $\rm OH^+(1_1$--$0_1)$ and $\rm OH^+(1_2$--$0_1)$ ($\rm \nu_{rest} = 1033.118$~GHz and 971.803~GHz, respectively), together with the $\rm ^{12}CO(9$--$8)$ and $\rm H_2O(2_{02}$--$1_{11})$ emission lines ($\rm \nu_{rest} = 1036.912$~GHz and 987.927~GHz, respectively), and the underlying dust continuum emission.

Observing conditions were excellent with an atmospheric phase stability rms of 20$\rm ^o$ on the longest baseline (400~m) and 1.5~mm of precipitable water vapor. The correlator was operated in the low resolution mode to provide spectral channels with a nominal resolution of 2~MHz. The NOEMA antennas were equipped with 2SB receivers that cover a spectral window of 7.744~GHz in each sideband and polarization. We covered the frequency range from 237.4 to 245.2~GHz in the lower sideband (LSB) and 252.8 to 260.5~GHz in the upper sideband (USB). 

The flux calibrators used were MWC349 and LkH$\alpha$101, and the phase calibrator was 0106+013. The data were calibrated, averaged in polarization, mapped, and analyzed in the GILDAS software package. The absolute flux calibration is accurate to within 10\%. The sensitivities reached in 40~MHz ($\sim 50 \, \rm km \, s^{-1}$) channels are 0.90 mJy\,beam$^{-1}$ for the USB and 0.88 mJy\,beam$^{-1}$ for the LSB, respectively.  

A continuum uv-table of HerS-3 was produced by averaging the visibilities from the line-free sections of the USB and LSB sidebands. The USB continuum visibilities and their weights were scaled based on an assumed spectral index of 2.28 and aligned in the uv-plane to the reference frequency of 239.301 GHz. The continuum flux density of HerS-3 was sufficiently strong to allow phase self-calibration down to the thermal noise level of the visibilities. A continuum dirty map was constructed using calibrated visibilities, applying a Briggs robust weighting parameter of 0.1. The map was cleaned down to the thermal noise level with the CLARK algorithm, using a mask to exclude regions without continuum emission. This process achieved high image fidelity and an imaging dynamic range, defined as the ratio of peak flux density to rms noise in the map, which exceeded 30 in the dust continuum emission. The resulting map has an angular resolution of $0\farcs 92 \times 0\farcs 52$ (PA $\rm 23^o$), and a 1$\sigma$ sensitivity of $69\,\mu$Jy\,beam$^{-1}$.

For our spectral analysis of the data, we created separate uv tables of the USB and LSB. We imaged both sidebands using the Hogbom algorithm and a natural weighting scheme to maximize the SNR. A mask was applied to exclude regions outside the continuum and line emission regions before cleaning down to the thermal noise of each side band. The resulting data cubes are binned in frequency by a factor of 30, resulting in channel widths of 60~MHz or 70.7~\kms\ at the red-shifted rest frequency of the $\rm OH^+(1_1$--$0_1)$ line.

\begin{figure*}[ht!]
\centering
\includegraphics[width=\textwidth]{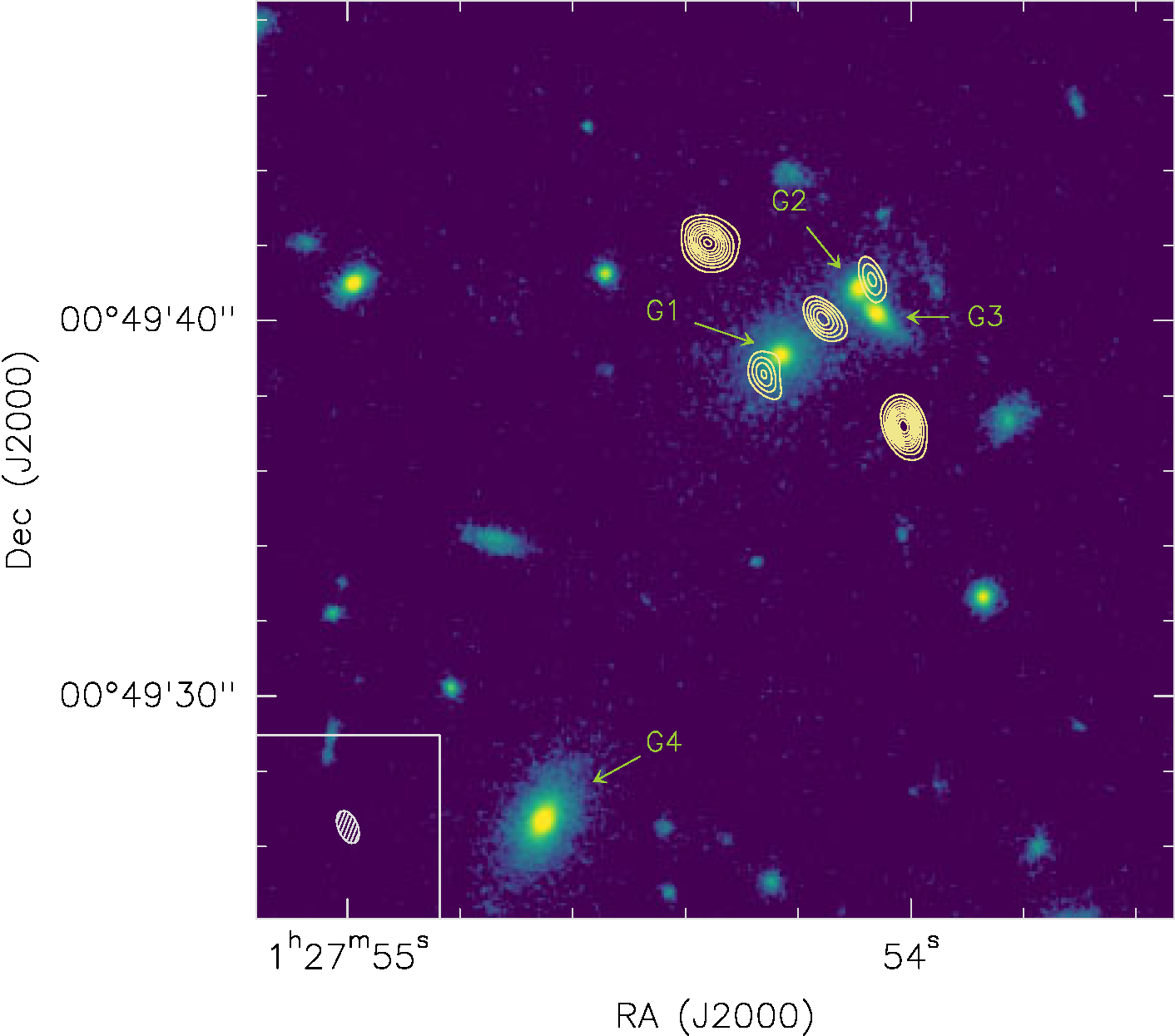}
\caption{The 1.2~mm dust continuum emission of HerS-3 observed with NOEMA and obtained with a robust weighting of 0.1 (shown in yellow contours) reveals that the background starburst galaxy at $z$=3.0607 is lensed into an Einstein cross with a fifth image in the middle of the cross. The {\it HST} \textit{F110W} image shows the foreground group of four lensing galaxies (labeled G1, G2, G3 and G4) at $z \sim 1.0$ - see Sect.~\ref{sec:Foreground-Group-Galaxies}. The $0\farcs92 \times 0\farcs52$ (PA $\rm 23^\circ$) NOEMA beam is shown in the lower left corner. The continuum contours are plotted starting at 5$\sigma$ in steps of 3$\sigma$ (the 1$\sigma$ noise level is $69\,\mu$Jy~beam$^{-1}$). The NOEMA continuum map has been corrected for primary-beam attenuation.
}
\label{fig:NOEMA-HST}. 
\end{figure*}

\subsection{ALMA} \label{sec:ALMA-cont}
The source HerS-3 was part of the ALMA Cycle 9 project ADS/JAO.ALMA~2022.1.00145.S (P.I.: T. Bakx) to obtain short Band-7 continuum observations of {\it Herschel}-selected high-$\it z$ galaxies with robust spectroscopic redshifts. The source was observed at 291.8~GHz for a total on-source time of $\sim$3.0~min with 4x128 dual polarization channels covering a bandwidth of 7.875~GHz with a 4~GHz sideband separation. The observations were done on 2023 June 1. The bandpass and phase calibrators used were J0006$-$0623 and J0125$-$0005, respectively.  Observing conditions were excellent with a median precipitable water vapor at zenith of $\rm 0.67\pm0.03 \, mm$, a phase stability of 9.1$\rm ^o$ rms and a system temperature $\rm T_{sys}=91~K$ in configuration C43-6 with a baseline length between 27~m and 3637~m.  The observations were carried out with 45 antennas and a maximum baseline of 1~km, yielding an angular resolution of $0\farcs12 \times 0\farcs10$ using a $\rm robust = 2.0$ weighting and a continuum sensitivity of $\rm 63~\mu Jy \, beam^{-1}$ (1$\sigma$). The ALMA data were reduced and imaged using the Common Astronomy Software Application\footnote{\url{http://casa.nrao.edu}} (CASA v6.5.4.9) \citep{McMullin2007, CASA-Team2022}. 

The continuum flux densities of each of the five images were extracted from the primary-beam corrected fields, using apertures that fit all of the continuum emission and extend two beams beyond the $2 \sigma$ contours, in an effort to recover all the flux. The errors on these extracted flux densities were then calculated from the per-beam error, multiplied by the square root of the number of beams within the aperture. Finally, an additional 10\% error was added in quadrature to account for calibration errors, as recommended by the ALMA Technical Handbook \citep{Cortes2024}. Although the images are extended, the peak flux positions are used as the central positions listed in Table~\ref{tab:cont-individual-images}.

\subsection{VLA} \label{sec:VLA}
We observed the radio continuum emission of HerS-3 using the NSF’s Karl G. Jansky Very Large Array (VLA) as part of a 6~GHz continuum survey of all the $z$-GAL sources (program I.D.:VLA/22A-211 and VLA/23A-030 - P.I.: T. Bakx). A complete description of the results of this survey will be presented in Bakx et al. (in prep.). The observations were done in the A-configuration in order to achieve an angular resolution of $\sim 0\farcs4$. We used a 4~GHz bandwidth and 3-bit samplers to maximize bandwidth and continuum sensitivity, using the standard 1~MHz ($\rm \sim 50 \, km \, s^{-1}$) frequency resolution (dual polarization) in broadband mode. The data were acquired in November 2023 under stable atmospheric conditions and HerS-3 was observed for a total time of 40~min (27 min on-source integration). 3C48 was used for bandpass and flux calibration, and J0115-0127 as the gain calibrator. 

The data were reduced, calibrated and imaged using CASA v6.5.4.9 \citep{McMullin2007, CASA-Team2022}. The accuracy of the flux density calibration is within 10–15\%, when comparing the measured fluxes for the calibrators from our observations to the calibrator models. The continuum map was created and cleaned using the TCLEAN function of CASA with natural weighting.

\begin{figure*}[ht!]
\centering
\includegraphics[width=\textwidth]{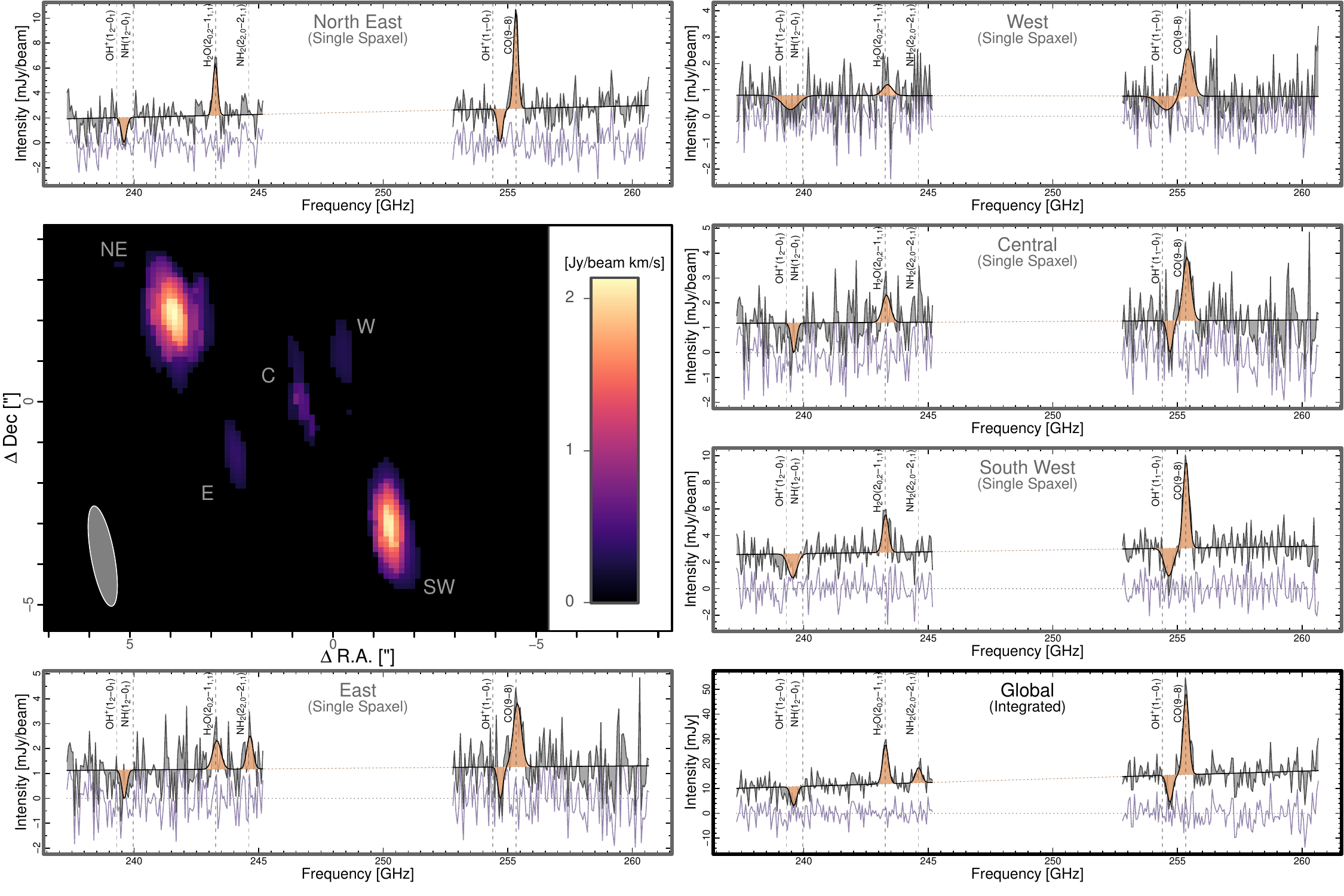}
\caption{The spectra for each of the five images of the Einstein cross of HerS-3 (NE: North East; E: East; C: Central; W: West; and SW: South West) observed with NOEMA in the frequency ranges between 237.4 to 245.2~GHz (LSB) and 252.8 to 260.5~GHz (USB).  The global integrated spectrum is displayed in the lower right panel. The $\rm ^{12}CO(9$--$8)$ moment-0 map is shown in the center left panel (with the $2\farcs3 \times 0\farcs9$ (PA $\rm 23^\circ$) beam in the lower left corner and axes centered on the NOEMA phase center). The spectra of the five Einstein cross images are displayed together with the continuum, including the fits to the continuum and the emission and absorption molecular lines, and the residuals (magenta). The spectra and continuum displayed are for individual spaxels of each of these regions. The global spectrum is spatially integrated over the $>3\sigma $ continuum pixels. The velocity channels have been binned to 30~$\rm km \, s^{-1}$. In each of the spectral panels, the vertical lines indicate the red-shifted frequencies for $z$=3.0607 of the molecular lines covered in these observations: $\rm ^{12}CO(9$--$8)$, $\rm OH^+(1_1$--$0_1)$, $\rm OH^+(1_2$--$0_1)$, $\rm H_2O(2_{02}$--$1_{11})$, $\rm NH_2(2_{20}$--$2_{11})$, and $\rm NH(1_2$--$0_1)$.}
\label{fig:Spectral-Fits}. 
\end{figure*}

\begin{figure*}[ht!]
\centering 
\includegraphics[width=0.95\textwidth]{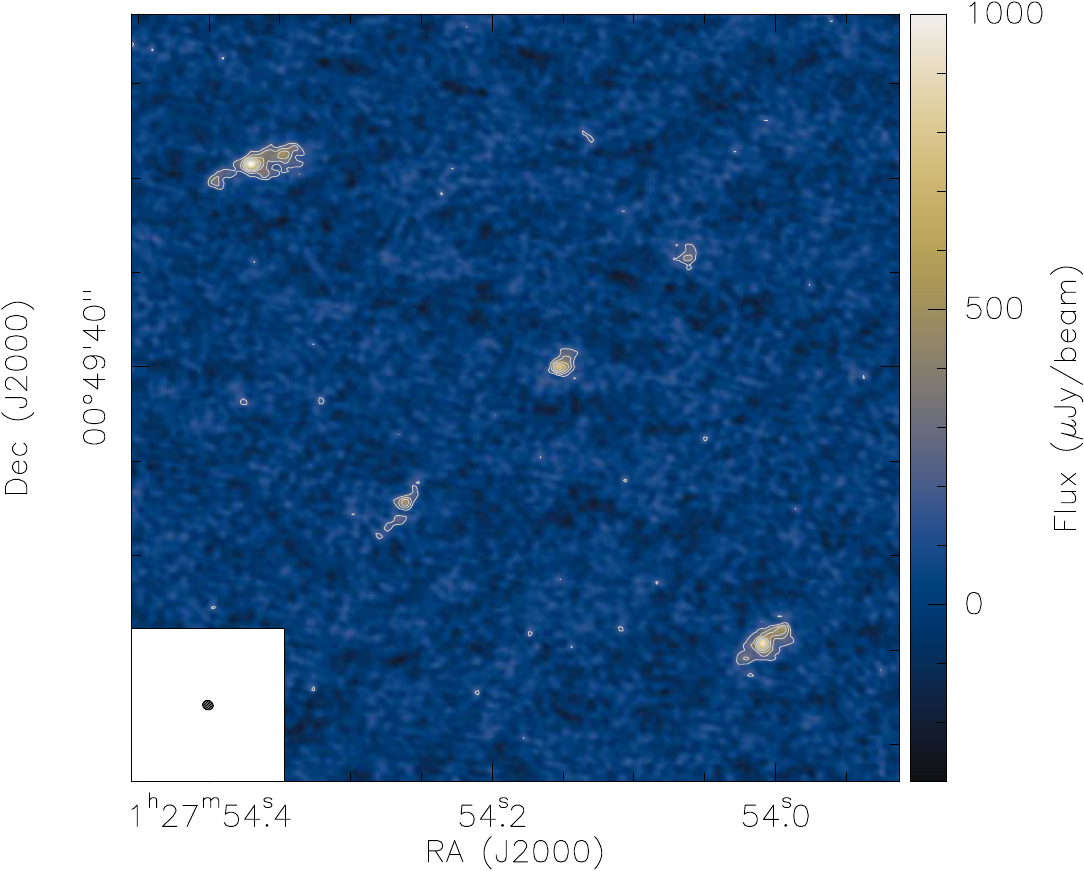}%
\caption{The 291.8~GHz dust continuum emission of HerS-3 observed with ALMA in Band 7 provides a detailed view of the extent and orientation of each of the five images of the Einstein cross. The continuum contours are plotted starting at 3$\sigma$ in steps of 2$\sigma$ levels (with $\rm 1\sigma = 63 \, \mu \, Jy~beam^{-1}$). The $0\farcs12 \times 0\farcs10$ (PA $74^\circ$) beam is shown in the lower-left corner of the panel.}
\label{fig:ALMA-continuum}. 
\end{figure*}

\subsection{{\it HST} imaging} \label{sec:HST}
The {\it HST} imaging of HerS-3 is part of a snapshot observations project, ID~15242 (P.I.: L. Marchetti), which was carried out from October 2017 to June 2018. The observations were performed using the wide-YJ filter \textit{F110W} of the Wide Field Camera 3 (WFC3) IR channel. HerS-3 was observed with a total exposure time of 711.74~s. The {\it HST} data were obtained from the Mikulski Archive for Space Telescopes (MAST) and the specific observations here analyzed can be accessed via \dataset[doi:10.17909/vg9e-2714]~\footnote{\url{https://doi.org/10.17909/vg9e-2714}}. However, these {\it HST} data were re-reduced, as explained in \cite{Borsato2024}, adopting, in particular, a different drizzling strategy than the standard one used for the MAST data products by setting the output pixel scale and linear drop size equal to $\rm 0\farcs064 \, pixel^{-1}$ and 0.8, respectively. Details on the observations, data reduction, and computation of the background, noise map, source mask, and PSF are provided in \cite{Borsato2024}. The final \textit{HST} pixel scale is $0\farcs064$. Importantly, astrometric corrections were performed by matching the {\it Gaia} early Data Release 3 sources \citep{Gaia2021} with their nearest {\it HST} counterparts. We found 3 \textit{Gaia} sources for which we computed the mean absolute offsets $\langle d_N=0\farcs36 \rangle$  and $\langle d_E=0\farcs17 \rangle$ along the North and East directions. Computing the rms of the residual offsets after correcting for the mean shifts, we obtain a value of $\rm rms = 0\farcs022$. Since errors of both the \textit{Gaia} DR3 astrometry and the relative \textit{HST} sources are much smaller than the residual rms, we adopted the latter as the error on the absolute astrometry of the \textit{HST} data.

\subsection{Subaru and SDSS imaging} \label{sec:optical}
In order to recover the optical rest-frame emission of the lensing galaxies, we used stacked SDSS and Subaru/HSC observations. We downloaded Subaru/HSC data from the Subaru HSC PDR3 (public data release 3) page. We retrieved $\sim 10\times10$ arcmin$^2$ cutouts around the centroid of the SPIRE detection of HerS-3 in the \textit{g}, \textit{r}, \textit{i}, \textit{z}, and \textit{Y} bands. We also retrieved similar cutouts of the noise maps and, through the HSC PSF picker \citep{Bosch2018}, cutouts of PSF models near the position of the HerS-3 system. The Subaru/HSC pixel scale is $0\farcs168$. To add some constraints in the blue, we used stacked SDSS data. We downloaded the deep stacks computed by \citet{Jiang2014}\footnote{Available in \url{http://das.sdss.org/ge/sample/stripe82/}} in the \textit{u} band. Each stacked image is a $\sim 14\times 18$ arcmin$^2$ band covering the SDSS Stripe 82. The details on the stacking procedure are available in \citet{Jiang2014}. To construct a noise map and the PSF for the SDSS data, we followed the same procedure applied in \citet{Borsato2024}. The SDSS pixel scale is $0\farcs396$.

\vspace{4cm}
   \begin{table*}[ht!]
        \caption{Coordinates and continuum flux densities of the individual images of the HerS-3 Einstein cross} 
        \label{tab:cont-individual-images}          
        \begin{center}
        \begin{tabular}{lcccccc} 
            \hline\hline       
        Image    & \multicolumn{2}{c}{Coordinates} & \multicolumn{2}{c}{NOEMA} & ALMA & VLA \\
                 & R.A.        & Dec.       & 241 GHz        & 258~GHz        &   292~GHz & 6~GHz\\
                 &  \multicolumn{2}{c}{J2000}  & \multicolumn{4}{c}{(mJy)}  \\
            \hline
            SW   & 01:27:54.01 & 00:49:37.1 & 4.05$\pm$0.16  & 5.81$\pm$0.18  & 5.70$\pm$0.39 & 0.075$\pm$0.015  \\
            NE   & 01:27:54.37 & 00:49:42.1 & 4.27$\pm$0.17  & 6.23$\pm$0.19  & 8.48$\pm$0.59 & 0.046$\pm$0.013 \\
            E    & 01:27:54.26 & 00:49:38.5 & 2.40$\pm$0.08  & 2.37$\pm$0.09  & 2.17$\pm$0.31 & 0.018$\pm$0.007 \\
            C    & 01:27:54.15 & 00:49:40.0 & 1.82$\pm$0.08  & 2.11$\pm$0.09  & 3.07$\pm$0.33 & 0.048$\pm$0.014  \\
            W    & 01:27:54.06 & 00:49:41.1 & 1.40$\pm$0.08  & 0.80$\pm$0.06  & 1.82$\pm$0.33 &   $<$0.0012        \\
        \hline      
        \end{tabular} 
        \end{center}
        \vspace{0.5cm}
        \tablenotetext{}{{\bf Notes.} The coordinates correspond to the center of each image of the HerS-3 Einstein cross, as derived from the ALMA 292~GHz dust continuum. The flux densities listed are the measured values with the $1 \sigma$ uncertainty. In case of a non-detection, the flux density is given as a $\rm 3\sigma$ upper limit.}
    \end{table*}

\begin{table*}[ht!]
    \caption{Photometric properties of the lensing galaxies G1, G2, G3, and G4.}
        \label{tab:SED_fit_results}
    \begin{center}
    \begin{tabular}{c c c c c c c c c c}
    \hline\hline
       Lens & $\rm \Delta R.A.$ & $\rm \Delta Dec.$ & $q$ & PA & $z_{\rm phot}$ & $A_V$ & $\log_{10}(M_*)$         & $\log_{10}(\rm SFR)$      & $Z$         \\
        & [arcsec] & [arcsec]  &   & [deg] &  &  [mag]  & [$\rm M_{\odot}$] & [$\rm M_{\odot} \, yr^{-1}$] & [$Z_{\odot}$] \\
    \hline
       G1 & $-0.004$ & $+0.032$ & $0.830_{-0.005}^{+0.005}$ & $38.74_{-0.77}^{+0.74}$ &  $1.03_{-0.04}^{+0.03}$ & $0.43_{-0.05}^{+0.05}$ &  $11.09_{-0.05}^{+0.09}$ & $0.17_{-0.12}^{+0.12}$ & $0.67_{-0.11}^{+0.10}$ \\
       G2 & $2.08$ & $1.83$ & $0.611_{-0.007}^{+0.006}$ & $51.51_{-0.52}^{+0.49}$ & $0.99_{-0.04}^{+0.05}$ & $0.44_{-0.05}^{+0.06}$ &  $10.72_{-0.07}^{+0.11}$ & $-0.51_{-0.41}^{+0.25}$ & $0.70_{-0.12}^{+0.12}$ \\
       G3 & $2.53$ & $1.15$ & $0.431_{-0.004}^{+0.003}$ & $132.57_{-0.19}^{+0.19}$ & $0.97_{-0.03}^{+0.04}$ & $0.43_{-0.06}^{+0.06}$ &  $10.78_{-0.07}^{+0.12}$ & $-0.68_{-2.00}^{+0.49}$ & $0.69_{-0.12}^{+0.12}$ \\
       G4 & $-6.33$ & $-12.34$ & $0.535_{-0.003}^{+0.003}$ & $62.84_{-0.25}^{+0.27}$ & $1.06_{-0.01}^{+0.01}$ & $0.25_{-0.05}^{+0.05}$ & $11.10_{-0.02}^{+0.02}$ & $-1.76_{-1.13}^{+0.61}$ & $0.79_{-0.11}^{+0.11}$ \\
\hline
    \end{tabular}
    \end{center}
      \vspace{-0.2cm}
    \tablenotetext{}{{\bf Notes.}  $\rm \Delta R.A.$ and $\rm \Delta Dec.$ are the relative positions of G1, G2, G3 and G4 to the maximum likelihood position of G1: $\rm R.A._{G1} = 01h27m54.23s$ and $\rm Dec._{G1} = 00^o49'39\farcs09$ (J2000). For the positions and shape information (axis ratio $q$ and position angle PA), we adopt the values measured from the \textit{HST} data with the highest spatial resolution, signal-to-noise ratio, and smallest pixel scales available. In the case of G1, we refer to the shape information of the resolved component used for the surface brightness modeling (Appendix~\ref{sec:zphot}). Errors on the positions are less than 1 mas. Details about the photometric redshift ($z_{\rm phot}$), dust extinction ($A_V$), stellar mass ($M_*$), star formation rate (SFR), and metallicity ($Z$) estimates are given in Appendix~\ref{sec:zphot}. }
\end{table*}

\section{Results} \label{sec:results}
The combination of high-quality multi-wavelength observations in the sub-millimeter, radio, and near-infrared allow us to study the configurations and properties of both the lensed background galaxy, HerS-3, and the foreground lensing galaxy group. In this section, we will describe the field around HerS-3 and the properties of the foreground galaxies (Sect.~\ref{sec:morphology-z}), then present the detailed characteristics of the Einstein cross (Sect.~\ref{sec:Einstein-Cross}), analyze the spectral energy distribution of HerS-3 (Sect.~\ref{sec:SED}), and outline the properties of the molecular lines in Sect.~\ref{sec:Molecular-Lines}.

\subsection{Overall View of the HerS-3 Field} \label{sec:morphology-z}
\subsubsection{The Einstein Cross} \label{sec:Overall-Einstein-Cross}
In the $z$-GAL survey, HerS-3 was observed at 3 and 2-mm at moderate angular resolution ($\sim$5$''$). The source was found to be extended over $\sim$10$''$ in the north-east/south-west direction, with two distinct peaks detected in the $^{12}$CO(3-2) and (5-4) emission lines, yielding a redshift of $z_{\rm spec}$=3.0607 for both components \citep{Cox2023}. The new NOEMA observations, done at 1-mm with a beam of $0\farcs92\times 0\farcs52$, reveal that HerS-3 is an Einstein cross that includes a bright central fifth image (Fig.~\ref{fig:NOEMA-HST}; the coordinates of each of the continuum images are listed in Table~\ref{tab:cont-individual-images}).

The north-east and south-west images, which are the brightest, are separated by $7\farcs5$. This is an unusually wide separation as compared to other Einstein crosses or quadruple imaged quasars, where the distance between the images is typically of a few arcsec -- notable exceptions are the lensed quadrupole quasars J1004+412 and J165105$-$041725 that have maximum separations between the components of $14\farcs6$ and 10$''$, respectively \citep{Inada2003, Stern2021}. The three images along the south-east/north-west direction have a smaller separation and extend over $\sim 4^{\prime\prime}$. It is also noteworthy that the alignment of the north-east and south-west images with the image in the middle is slightly bent. 

All five images, which have various molecular emission lines detected, are confirmed to be at the same redshift ($z_{\rm spec}$=3.0607) and relative velocities, demonstrating that they originate from the same physical source (Fig.~\ref{fig:Spectral-Fits} and Sect.~\ref{sec:Molecular-Lines}). This is the first time that a gravitationally amplified high-$z$ galaxy is seen in an Einstein cross system with an image of the background galaxy at the center of the cross and that an Einstein cross is detected at sub-millimeter wavelengths.

\subsubsection{The Foreground Group of Galaxies} \label{sec:Foreground-Group-Galaxies}
The {\it HST} F110W image (at 1.15~$\rm \mu m$) shows that the $25''\times25''$ field around HerS-3 contains dozens of galaxies with various levels of brightness (Appendix~\ref{sec:zphot} and Fig.~\ref{fig:HST-all-sources}). The near-infrared emission is dominated by the three close-by and brightest galaxies in the field, which are aligned along the minor axis of the Einstein cross, with one galaxy to the south-east (G1) and the two others to the north-west (G2 and G3) of the central image (Fig.~\ref{fig:NOEMA-HST}). The galaxies G2 and G3, which are heavily blended with their diffuse emission overlapping, are each separated from G1 by $2\farcs7$. All three galaxies show faint, extended diffuse emission in the {\it HST} image. Based on the photometry available from the {\it HST}, SDSS and Subaru/HSC data, the photometric redshifts of the galaxies G1, G2 and G3 are estimated to be in the range $0.97 < z_{\rm phot} < 1.03$, indicating that they are part of a galaxy group. The galaxy G4, which is located $13\farcs5$ south-east of G1, also belongs to this group with a best fitting photometric redshift of $z_{\rm phot} = 1.06\pm0.01$ (Table~\ref{tab:SED_fit_results} and Fig.~\ref{fig:Photo-z} in Appendix~\ref{sec:zphot}).

Through the SPectra Analysis $\&$ Retrievable Catalog Lab (SPARCL\footnote{The SPARCL client can be found at \url{https://sparclclient.readthedocs.io/en/latest/}}) service, we retrieved an optical spectrum (from 4000 to 10000~\AA) for the three galaxies G1, G2 and G3. Processing this spectrum using the standard BOSS-DR16 pipeline \citep{Stoughton2002, Ahumada2020}, we retrieved a spectroscopic redshift of $z_{\rm spec} = 1.0775$. Unfortunately, given the close vicinity of the three galaxies (particularly G2 and G3), we could not associate this redshift to any specific group member. Moreover, the spectrum displays a low median SNR computed for the unmasked pixels of 0.41 and was flagged as having a high fraction of points more than $5~\sigma$ away from the best-fitting model. For these reasons, this spectroscopic redshift is only indicative, although we note that the value remains consistent with the results obtained from the spectral energy distribution analysis (Fig.~\ref{fig:Photo-z} in Appendix~\ref{sec:zphot}).  

Based on these findings, HerS-3 appears to be lensed by a group of at least four galaxies, G1, G2, G3, and G4 at $z_{\rm phot} \sim 1.0$. By fitting the spectral energy distributions and modeling the surface brightness of each of these four lensing galaxies (Appendix~\ref{sec:zphot}), we derived their photometric properties that are listed in Table~\ref{tab:SED_fit_results}. The galaxies G1, G2, G3, and G4 have low star formation rates, $0.02 \lesssim \rm SFR\, / {\rm (M_{\odot}\, yr^{-1})} \lesssim 1.5$, and stellar masses of $\rm \sim 10^{11} M_{\sun}$, showing that they are massive quenched systems. 

Investigating the field within a $\sim 60''$ ($\sim500$~kpc) radius from G1 to search for potential additional members belonging to the lensing group, we found a total of 10 galaxies with photometric redshifts in the range $0.9<z<1.1$, which are consistent with those of the main lensing galaxies G1, G2, G3, and G4 (Appendix~\ref{sec:zphot} and Fig.~\ref{fig:members}). These other candidate members of the galaxy group have either low stellar masses or are located at far distances from the main mass distribution of the group. Therefore, these additional galaxies belonging to the group are unlikely to contribute significantly to the lensing of the background source, HerS-3.

\subsection{Detailed View of the Einstein Cross} \label{sec:Einstein-Cross}
The $0\farcs92\times 0\farcs52$ beam of the NOEMA data gives first indications that each of the images of the Einstein cross are extended in the east-west direction, in particular the north-east and south-west components (Fig.~\ref{fig:NOEMA-HST}). There is also evidence of changes in the orientation from one image to the other, again most clearly seen in the brightest north-east and south-west components. 

\begin{figure}[ht!]
\centering
\includegraphics[width=0.47\textwidth]{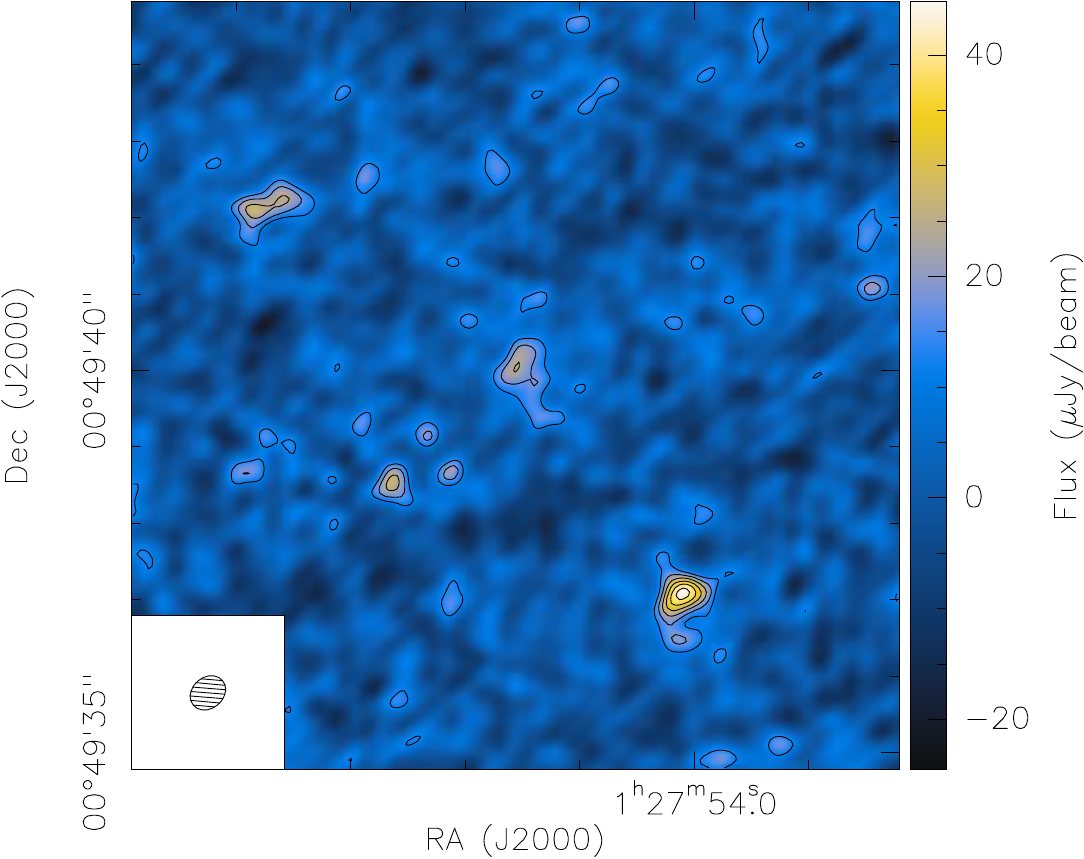}
\caption{The 6~GHz radio continuum emission of HerS-3 observed with the VLA. The continuum contours are plotted starting at 2$\sigma$ in steps of 1$\sigma = 6 \, \rm \mu Jy \, beam^{-1}$. The peak 6~GHz continuum flux is 45~$\rm \mu Jy/beam$. The $0\farcs50 \times 0\farcs41$ (PA $130^\circ$) beam is shown in the lower left corner.}
\label{fig:VLA-continuum}. 
\end{figure}

These characteristics are revealed in greater detail in the higher angular resolution ($0\farcs12 \times 0\farcs10$) ALMA continuum Band~7 (1~mm) image, which provides crucial information on the spatial extent and orientation of the individual images of the Einstein cross (Fig.~\ref{fig:ALMA-continuum}). The centroids of all five images are well aligned with those of the 1~mm NOEMA image. The north-east and south-west images are elongated with extensions of $1\farcs4$ and $1\farcs2$ and widths of $0\farcs3$, both showing a central emission peak, and orientations of PA$\sim 108^\circ$ and $\sim 126^\circ$, respectively. Of the three east-west oriented images, the central one is the strongest with a clear peak and emission extending over $\sim 0\farcs6$ with a PA of $\sim 138^\circ$; the east component is weaker, extending over $\sim 0\farcs8$ with a PA of $\sim 146^\circ$; the west image is the weakest of all five components with an extension comparable to the central image. 

The VLA 6~GHz continuum shows four of the five images (the weaker west component has only an upper limit), with characteristics that are comparable to those seen in the ALMA continuum image (Fig.~\ref{fig:VLA-continuum}). 

Table~\ref{tab:cont-individual-images} lists the centroids (derived from the ALMA image) and the continuum flux densities of each of the images of the HerS-3 Einstein cross as measured by NOEMA, ALMA, and the VLA.     

\begin{table}
\caption{Far-infrared, sub-millimeter and radio continuum flux densities of HerS-3.} 
\begin{center}
\begin{tabular}{cccc}
\hline\hline
$\lambda$ [mm]   & $\nu$ [GHz] &  $S_{\nu}$ $[$mJy$]$ &  Ref.     \\
\hline            
\multicolumn{4}{c}{\it Herschel}   \\
0.25  &              & 253$\pm$6          &     (1) \\ 
0.35  &              & 250$\pm$6          &      "  \\ 
0.50  &              & 191$\pm$7          &      "  \\ 
\multicolumn{4}{c}{ALMA}           \\
1.02  &   292       &  21.2$\pm$1.4   &     (2) \\ 
\multicolumn{4}{c}{NOEMA}             \\
1.16  &   258       &  18.6$\pm$0.1    &      (2) \\ 
1.24  &   241       &  14.2$\pm$0.1    &      (2) \\ 
1.94  &   154       &    3.1$\pm$0.5    &      (3) \\ 
2.15  &   139       &    2.2$\pm$0.4    &       "  \\ 
2.91  &   103       &    0.30$\pm$0.07 &       "  \\ 
3.15  &    95       &    $<$0.39    &       "  \\
3.44  &    87       &    0.43$\pm$0.08    &       "  \\
3.79  &    79       &    $<$0.21 &       "  \\
\multicolumn{4}{c}{VLA}          \\ 
11    &    28       &    0.10$\pm$0.03   &  (4) \\
50    &     6       &    0.19$\pm$0.02   &  (2)  \\ 
\hline
\end{tabular}
\end{center}
\vspace{-0.2cm}
\tablenotetext{}{{\bf Notes.} (1) \cite{Nayyeri2016} (2) This paper (3) \cite{Ismail2023} (4) Prajapati et al. (in prep.). The wavelengths and frequencies are the observed ones. The flux densities listed are the measured values with the $1 \sigma$ uncertainty. In case of a non-detection, the flux density is given as a $\rm 3\sigma$ upper limit.}
\label{tab:continuum}
\end{table}

\begin{figure}[ht!]
\centering
{\includegraphics[height=0.38\textwidth]{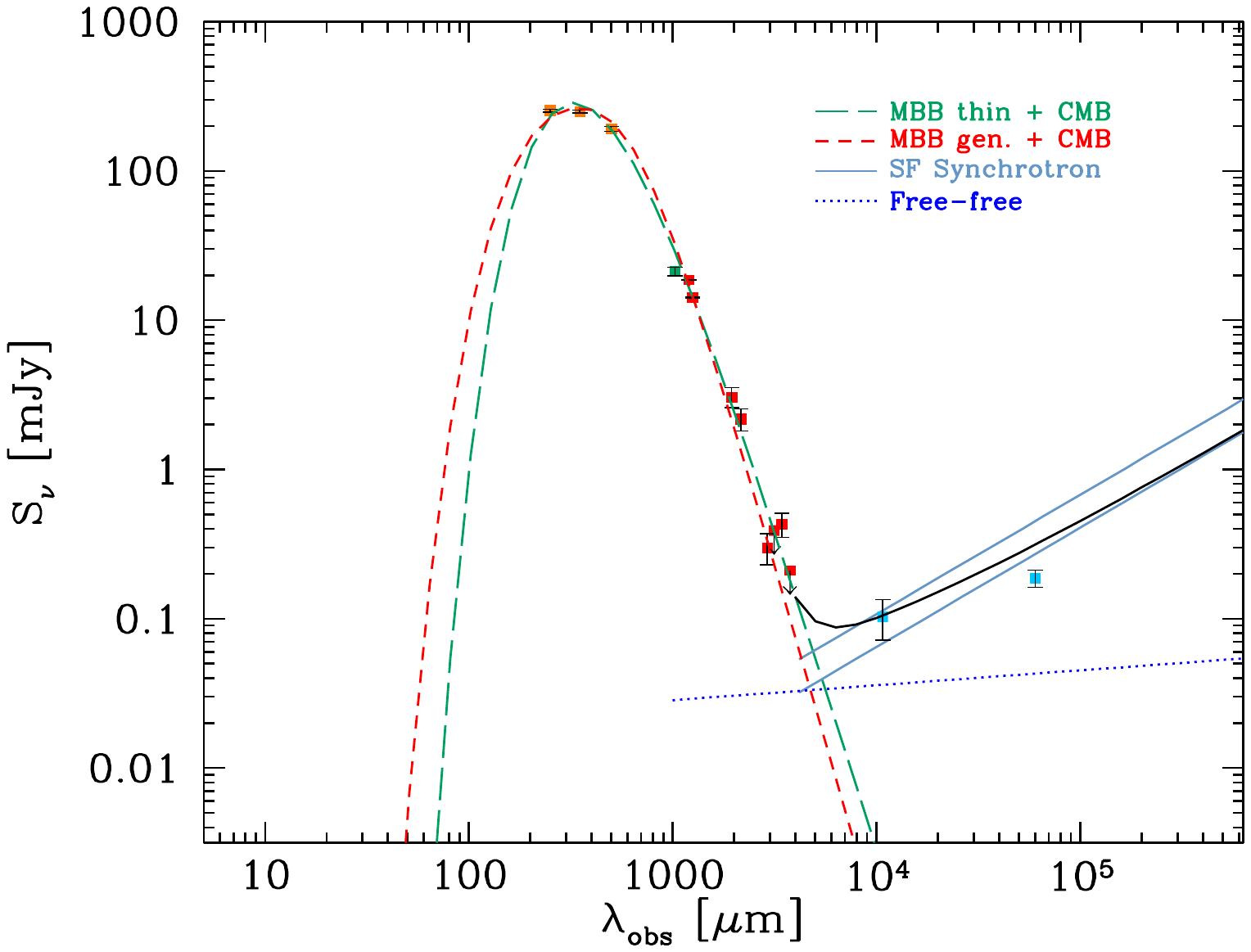}}
\caption{Spectral energy distribution of HerS-3 from infrared to radio wavelengths, showing the contributions of the starburst and the free-free and star-formation synchrotron emissions. The data include continuum flux densities from {\it Herschel} SPIRE (orange dots), ALMA (green dot), NOEMA (red dots), and VLA (pale blue dots) - the error bars correspond to the $1\sigma$ uncertainty (Table~\ref{tab:continuum}). The fits are based on the modified black body (MBB) thin and general models (green and red dashed lines, respectively), including corrections for the effects of the CMB (see text for details). The blue lines represent the synchrotron emission due to star formation, with a fixed spectral index $\alpha_\textrm{synchr} = -0.8$, normalized on the basis of the two radio/far-infrared correlations derived by \cite{Delhaize2017} (lower line), and \cite{Magnelli2015} (upper line). The dotted blue line shows the free-free emission, with a spectral index $\alpha_\textrm{ff}=-0.1$, normalized to fit the 28~GHz data after subtracting the \cite{Delhaize2017} synchrotron emission. The solid black line is the sum of the dust, the synchroton \citep{Delhaize2017}, and the free-free components in the radio regime.  
}
\label{fig:SED}. 
\end{figure}

\begin{figure*}[ht!]
\centering
\includegraphics[width=0.97\textwidth]{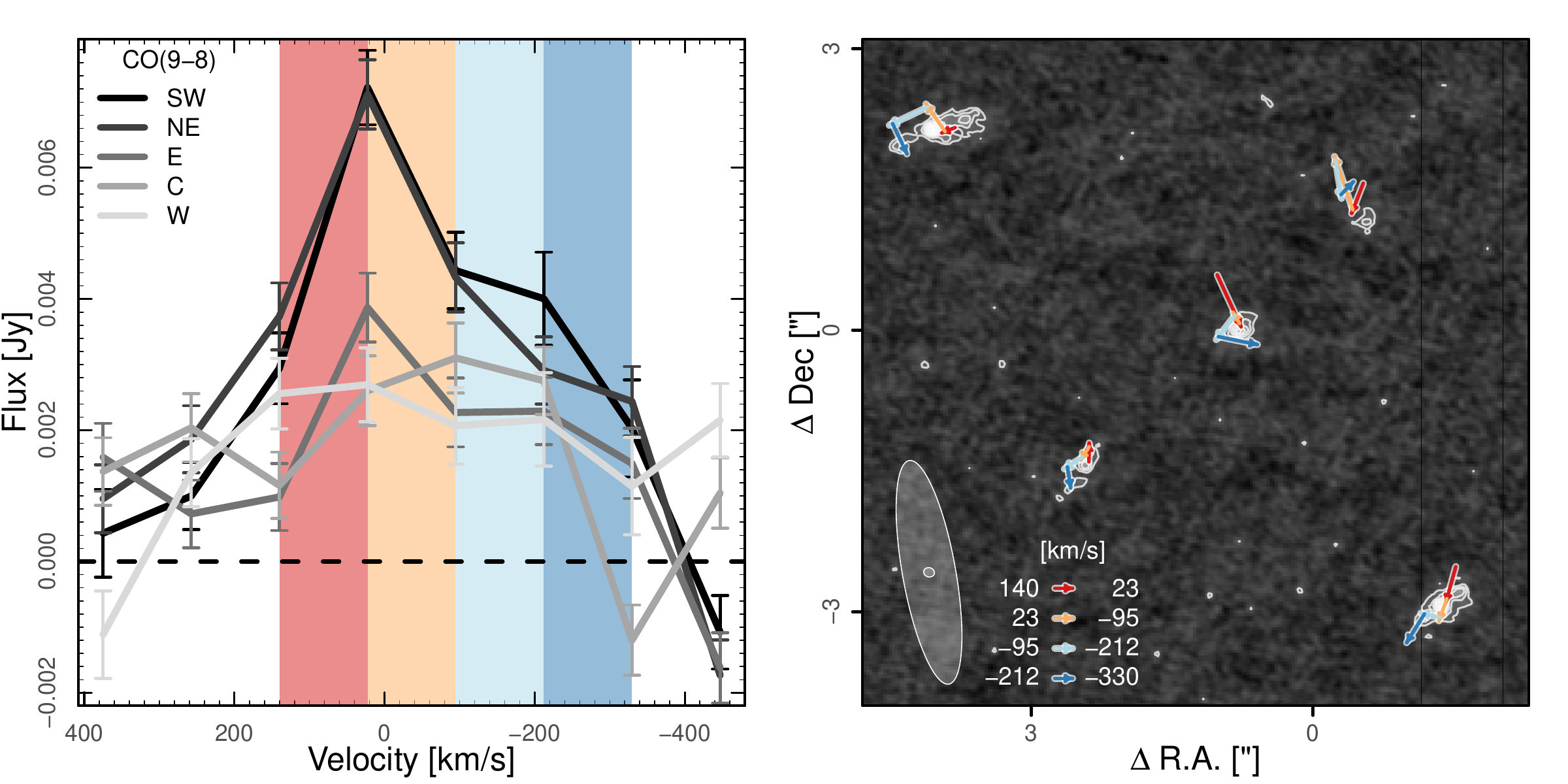}
\caption{Two-dimensional UV-plane fitting results of the $\rm ^{12}CO(9$--$8)$ emission line in bins of 118 \kms\, for the five HerS-3 images. \textit{Left:} $\rm ^{12}CO(9$--$8)$ emission line flux as a function of velocity offset from the systemic velocity of HerS-3. Each of the five images are highlighted by a different shade of grey, revealing their $\rm ^{12}CO(9$--$8)$ spectra. \textit{Right:} Movement between velocity bins of the $\rm ^{12}CO(9$--$8)$ emission flux centers of each image overlaid on the ALMA 291.8~GHz dust continuum emission map where the contours are plotted starting at $3 \sigma$ in steps of $2 \sigma$ (with $\rm 1 \sigma = 63 \, \mu Jy \, beam^{-1}$). The colored arrows indicate the direction of movement from bin to bin within 140 and $-$330\,\kms, corresponding to the velocity ranges shaded in the same colors in the left panel. The NOEMA and ALMA beams are shown together in the lower left corner and the axes centered on the NOEMA phase center as in Fig. \ref{fig:Spectral-Fits}.}
\label{fig:Velocity-Position}. 
\end{figure*}

\subsection{Continuum Spectral Energy Distribution} \label{sec:SED}
Together with the {\it Herschel} and $z$-GAL continuum flux densities \citep{Ismail2023}, the new NOEMA and ALMA data in the range between 240 and 290~GHz enable for a continuous sampling of the far-infrared and sub-millimeter spectral energy distribution (SED) of HerS-3. In addition, the VLA continuum data at 6~GHz and 28~GHz (Prajapati et al. in prep.) provides crucial information on the origin of the radio emission in HerS-3. Figure~\ref{fig:SED} shows all the continuum data available for HerS-3 from infrared to radio wavelengths, and Table~\ref{tab:continuum} lists the continuum flux densities.  

We fit the continuum data using two different SED dust models, namely: a single-temperature modified black-body (MBB) in the optically-thin approximation and a MBB model in its general form \cite[][and references therein]{Berta2021, Ismail2023}. The effect of the cosmic microwave background (CMB), which produces a steepening of the millimetric slope, has been taken into account following the prescriptions of \cite{daCunha2013}. We adopt the $\kappa_\nu$ re-normalization by \citet{draine2014}. The best fit solutions are shown in Fig.~\ref{fig:SED}. The radio continuum data point is compared, a posteriori, to a power-law spectrum $S_\nu \propto \nu^{-0.8}$, which represents the synchrotron emission normalized on the basis of the radio/far-infrared correlation \citep{Magnelli2015, Delhaize2017}.  The continuum at 6~GHz lies slightly below the expected star-formation synchrotron power law based on the \cite{Delhaize2017} relation. There is therefore no indication of a radio loud active galactic nucleus in HerS-3. The 28 GHz continuum flux density, on the other hand, is in excess with respect to the \cite{Delhaize2017} synchrotron emission. To account for this, we add a free-free component in the form $S_\nu \propto \nu^{-0.1}$.

When adopting an optically-thin (or general form, in parenthesis) MBB to model the dust continuum of HerS-3, we derive values for the following properties, where $\mu$ is the gravitational lens magnification: the spectral index is $\beta=2.31\pm0.08$ ($2.67\pm0.16$); the dust temperature $T_{\rm dust}=34.1 \pm 1.1 \, \rm K$ ($60.0 \pm 0.08 \, \rm K$); the dust mass $\mu M_{\rm dust}=7.43 \pm 0.28 \times 10^9 \, \rm M_{\sun}$ ($5.08 \pm 0.96 \times 10^9 \, \rm M_{\sun}$); and the far-infrared luminosity (from 50 to 1000~$\rm \mu m$) $\mu L_{\rm FIR}=5.28 \pm 0.09 \times 10^{13} \, L_{\sun}$ ($5.31 \pm 0.11 \times 10^{13} \, \rm L_{\sun}$), yielding an estimate of the total infrared luminosity (from 8 to 1000~$\rm \mu m$) of $\mu L_{\rm IR}=7.54 \pm 0.12 \times 10^{13} \, L_{\sun}$ ($7.58 \pm 0.16 \times 10^{13} \, \rm L_{\sun}$) by dividing $\mu L_{\rm FIR}$ by 0.70 \citep{Berta2023}. These dust properties are consistent with the values that were derived by \cite{Ismail2023}.

\subsection{Molecular emission and absorption lines} \label{sec:Molecular-Lines}
All five images of the Einstein cross of HerS-3 reveal a similarly rich spectrum, displaying the emission lines of $\rm ^{12}CO(9$--$8)$ and $\rm H_2O(2_{02}$--$1_{11})$, and the $\rm OH^+(1_1$--$0_1)$ and $\rm OH^+(1_2$--$0_1)$ lines that are both seen in absorption. In addition, there is a tentative detection of $\rm NH_2(2_{20}$--$2_{11})$ in the global spectrum of HerS-3 (Fig.~\ref{fig:Spectral-Fits}).

We performed spectral fitting of the spectra in each individual image and the stacked global HerS-3 spectrum. Using the \texttt{curve\_fit} procedure, we simultaneously fitted each of the molecular emission and absorption lines with a single Gaussian component and the underlying continuum with a linear curve. To reduce the number of free parameters and improve our sensitivity to the line parameters, we anchored the central velocity and line width of the two $\rm OH^+$ transitions to one another. The same was done for the molecular emission lines, as they are expected to trace the same molecular gas reservoir. The best-fit spectra are plotted in Fig.~\ref{fig:Spectral-Fits}, and the best-fit parameters for the global spectrum are presented in Table~\ref{tab:derived-global}.

The $\rm ^{12}CO(9$--$8)$ spectra observed in each of the five images do not show any significant variation in their profile or central velocity, unambiguously demonstrating that the five images have the same redshift ($z$=3.0607). The global spectra of the $\rm ^{12}CO(9$--$8)$ and $\rm H_2O(2_{02}$--$1_{11})$ emission lines have a full-width half maximum (FWHM) of $320\pm28 \, \rm km s^{-1}$, and are, within the errors, at the expected redshifted frequencies (Table~\ref{tab:derived-global}).

\begin{table}      
\caption{Molecular absorption and emission line properties of HerS-3, obtained by the line fitting of the global spectrum.}
        \label{tab:derived-global}              
        \begin{center}
        \vspace{-0.4cm}
        \begin{tabular}{lcccc} 
            \hline\hline       
            & $\rm S$&$\rm V$& FWHM & \\
              & $\rm [Jy\kms]$ & $\rm [\kms]$ & $\rm [\kms]$&  \\
            \hline
            \multicolumn{5}{c}{Absorption Lines}\\
            \hline
                $\rm OH^+(1_1$-$\rm 0_1)$ & $-3.78\pm0.81$  & $-350\pm32$  & $321\pm78$ & \\
                $\rm OH^+(1_2$-$\rm 0_1)$ & $-2.87\pm0.72$  & $\prime\prime$  & $\prime\prime$ & (1) \\
            \hline
            \multicolumn{5}{c}{Emission Lines}\\
            \hline
                $\rm ^{12}CO(9$--$\rm 8)$          & $11.3\pm0.95$  & $-9.3\pm12$  & $320\pm28$ & \\
                $\rm H_2O(2_{02}$--$\rm 1_{11})$ & $5.7\pm0.87$  & $\prime\prime$  & $\prime\prime$ & (2) \\
                $\rm NH_2(2_{20}$--$\rm 2_{11})$ & $2.3\pm0.84$  & $\prime\prime$  & $\prime\prime$ & (2) \\
            \hline      
        \end{tabular}
        \end{center}
        \vspace{-0.3cm}
            \tablenotetext{}{{\bf Notes.} The values listed are the primary-beam corrected line flux (S), the velocity shift (V) with respect to the redshift of HerS-3 ($z$=3.0607) and the full-width half maximum (FWHM) as derived from the line fitting of the global spectrum (Fig.~\ref{fig:Spectral-Fits}). (1) The absorption depth, central velocity and line width of the two $\rm OH^+$ transitions are anchored to one another. (2) The same procedure is applied to the central velocity and line width of the emission lines but their fluxes are left as free parameters. The uncertainties are taken from those given by the fitting procedure in \texttt{curve\_fit} and the $\rm NH_2$ emission is considered to be a tentative detection.}  
    \end{table}

The peaks of both OH$^+$ absorption lines are blue-shifted by $\rm \sim 350 \, km s^{-1}$ with respect to the systemic velocity of the main molecular gas reservoir as traced by the CO emission lines (Table~\ref{tab:derived-global}). Detected in absorption against the dust continuum, the gas seen in OH$^+$ is therefore interpreted as tracing material that is outflowing from the starburst along the line-of-sight.

The NOEMA data provides a first view of the molecular gas morphology and kinematics of HerS-3 as traced by the $\rm ^{12}CO(9$--$8)$ emission line, which is the strongest line detected in the 1~mm spectra. The morphology of the molecular gas is shown in Figure~\ref{fig:Spectral-Fits}, which displays the velocity-integrated moment-0 map of the $\rm ^{12}CO(9$--$8)$ emission line. To investigate the velocity-position information of the molecular gas, we first binned the continuum subtracted $uv$ table presented in Sec.~\ref{sec:obs-NOEMA} into channels of $\rm \Delta(v)=$118~\kms. We then used the 2D $uv$-plane source fitting procedure \texttt{UV\_FIT} to simultaneously fit 2D circular Gaussians to all five images in each velocity bin. This provides position and flux information of the $\rm ^{12}CO(9$--$8)$ emission for each of the images in every velocity bin. The $\rm ^{12}CO(9$--$8)$ emission spectra for all five images, extracted from the 2D Gaussian fitting, are shown in the left panel of Fig.~\ref{fig:Velocity-Position}, which highlights their similarity in spectral shape. In the right panel of Fig.~\ref{fig:Velocity-Position}, we show the shift in the central position of the $\rm ^{12}CO(9$--$8)$ emission between the four highest S/N velocity bins for each image. The arrows indicate the direction of the spatial shifts from one bin to the next, starting from 140 to $-330~ \rm km \, s^{-1}$ i.e., from red- to blue-shifted emission. 

In the north-eastern and south-western images, where the CO(9-8) emission is marginally resolved, the central position displays a general shift towards the south-east from red- to blue-shifted velocities for both images. The high-resolution ALMA dust continuum map is shown underneath the arrows indicating the spatial drift in Fig.~\ref{fig:Velocity-Position}. With this overlay, we see that the south-easterly shift in CO(9-8) emission from red to blue velocities follows the major axis of the galaxy's dust continuum emission. This may indicate rotation in the galaxy's disk with the north-west side of the galaxy rotating away from the observer and the south-east side moving towards the observer. High signal-to-noise spectral line data at higher spatial resolution is needed to confirm this result and further investigate the kinematics of HerS-3.

\begin{figure*}[ht!]
    \centering
    \includegraphics[width=0.6\linewidth]{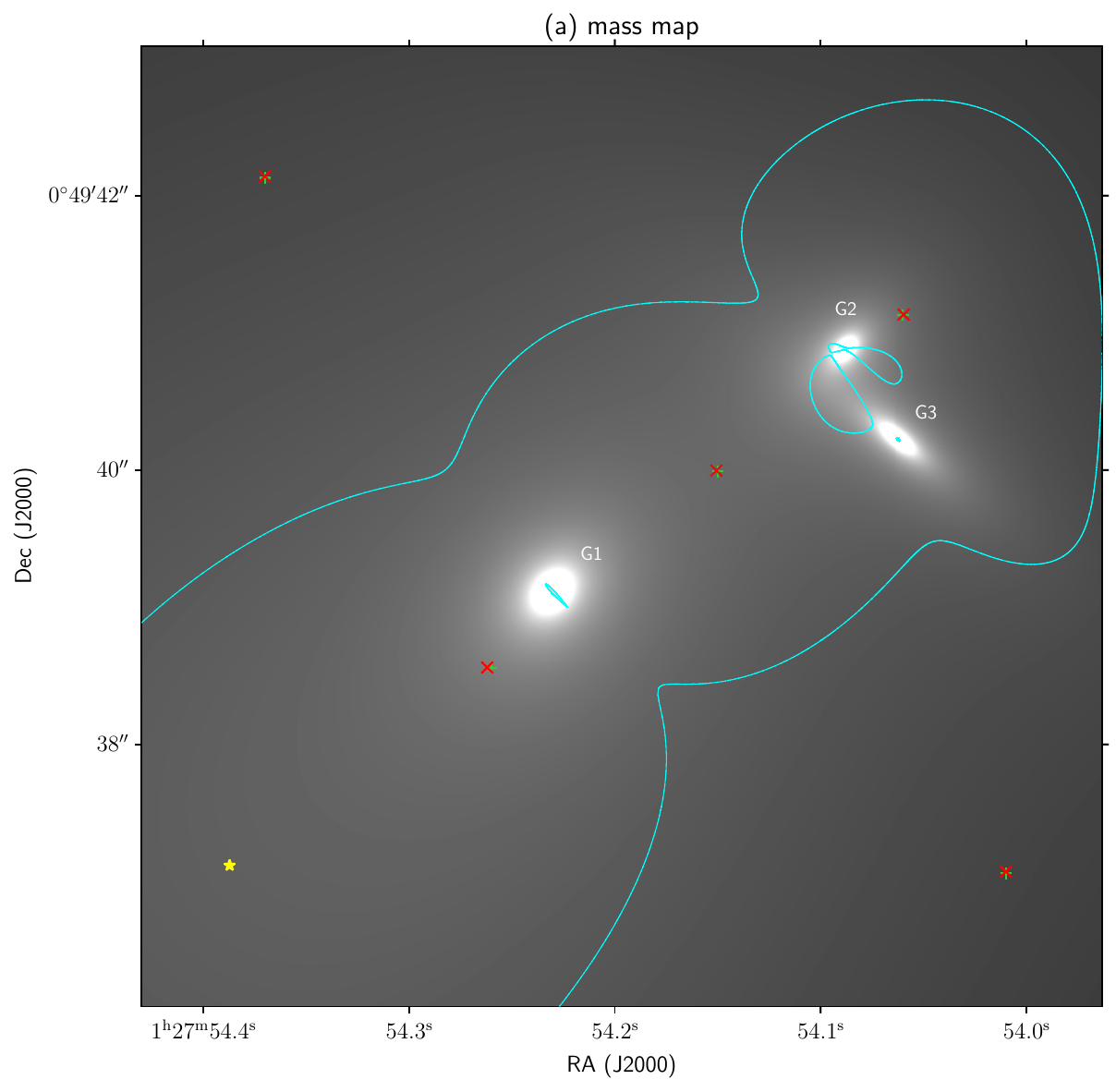}
    \vspace{0.5cm}
    \includegraphics[width=\linewidth]{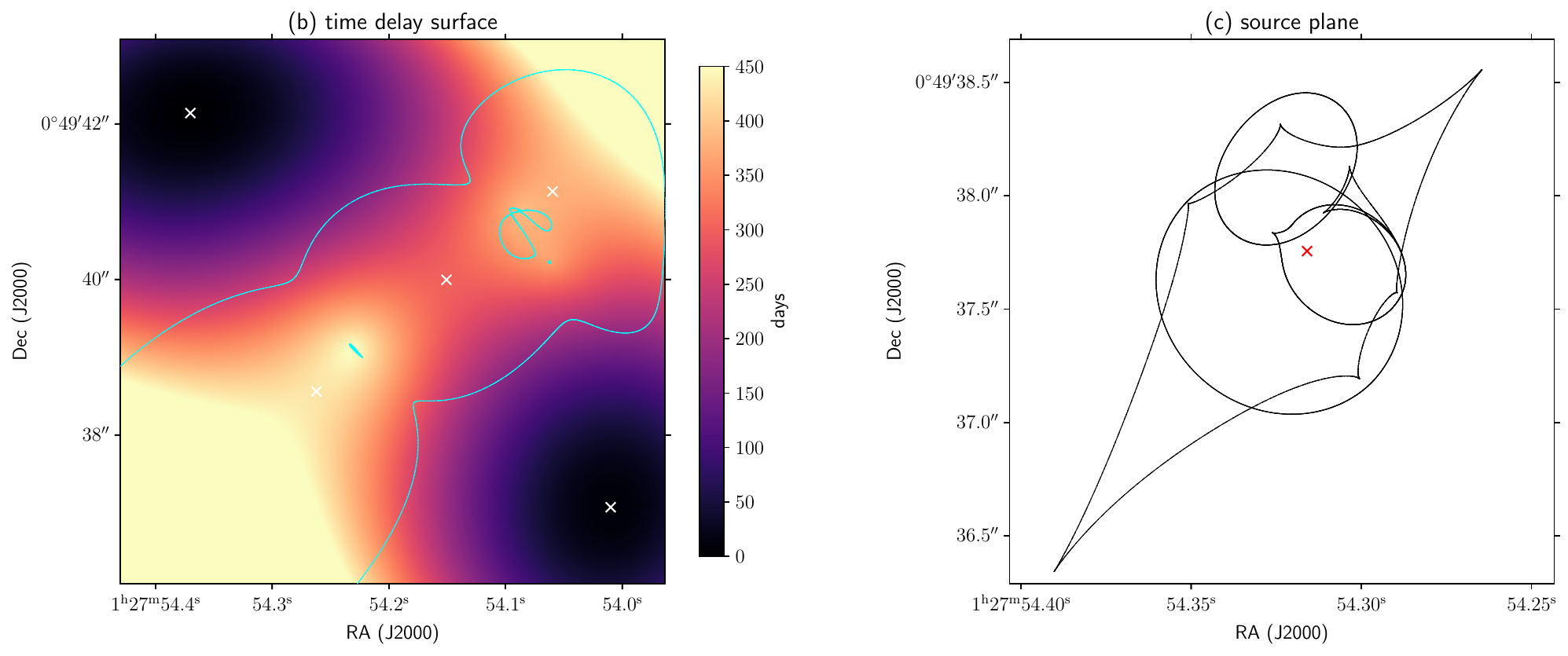}
    \vspace{0.5cm}
    \caption{
    The best single-plane lens model with a cored halo as the additional mass component. All mass components are assumed to be at the same redshift ($z_{\rm phot} \approx 1$). 
    (a) The gray scale shows the surface density map; the galaxies G1, G2, and G3 are labeled (G4 is out of the frame to left), and the center of the additional mass component is marked with a yellow $\star$. The green $+$ symbols show the observed positions of the five images of the Einstein cross, while the red $\times$ symbols show the positions of the images predicted by the lens model; the observed and predicted positions are in excellent agreement. The lensing critical curves are shown in cyan. 
    (b) The colormap shows the lensing time delay surface, assuming a lens redshift $z_l = 1$ along with the measured source redshift $z = 3.0607$. The lensed images appear at stationary points of this surface; NE and SW lie at local minima, while E, C, and W lie at saddle-points. The lensing critical curves are shown again for completeness.
    (c) The lensing caustics in the source plane; note that the spatial scale differs from the other panels. The inferred source position is indicated with the red $\times$ symbol.
    }
    \label{fig:model-core}
\end{figure*}

\begin{figure*}[ht!]
    \centering
    \includegraphics[width=1.0\linewidth]{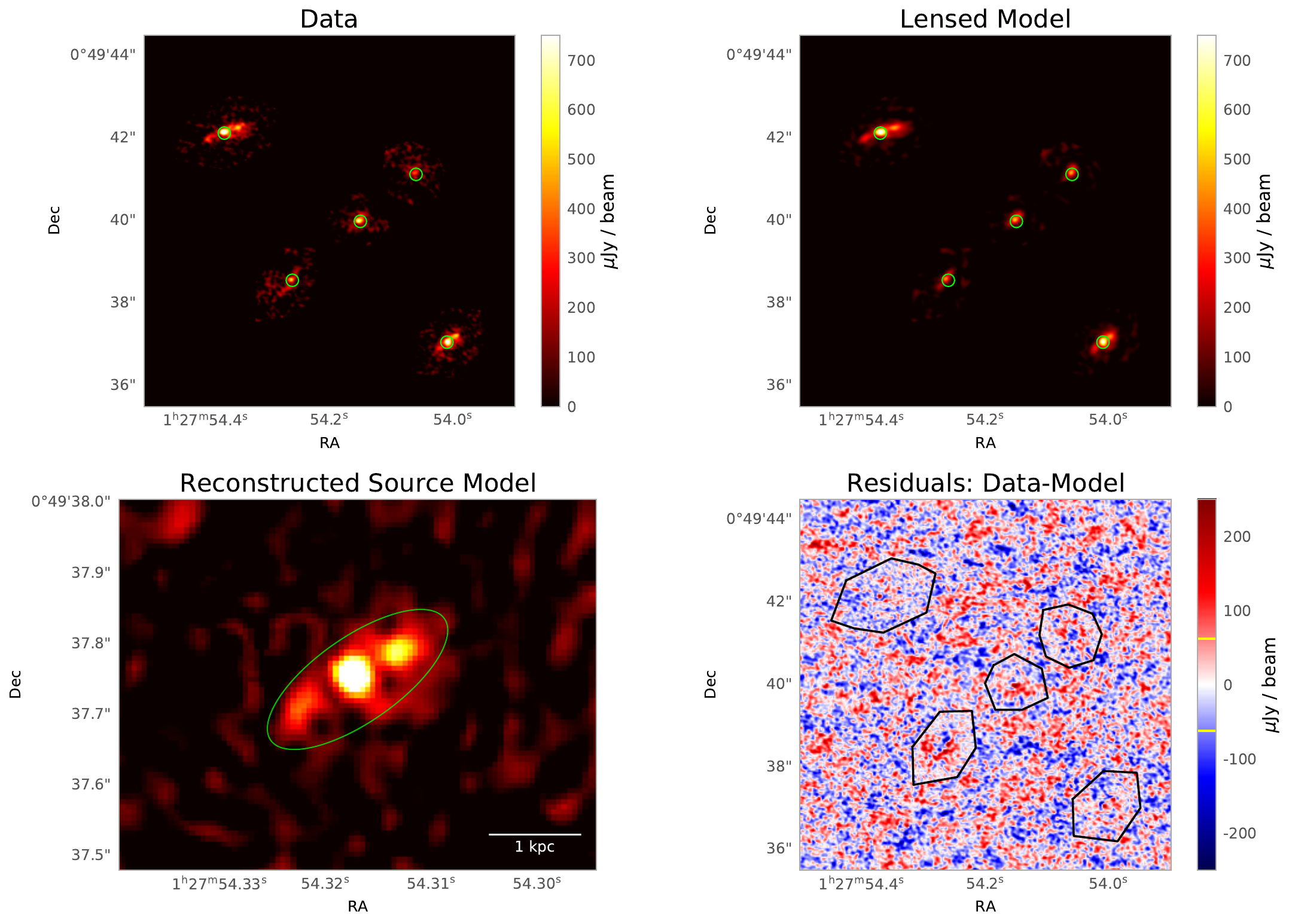}
    \caption{
    Results of pixel-based source reconstruction of HerS-3 using the lens model with a cored halo from Fig.~\ref{fig:model-core}.
    \textit{Top left:} The ALMA 292~GHz continuum image, masked around the five lensed images. To guide the eye, green circles mark the locations of the peak brightnesses in the observed map.
    \textit{Top right:} Model image, lensed and convolved with the beam. Green circles are the same as in the data image.
    \textit{Bottom left:} Reconstructed source-plane image. The scale bar marks $1$~kpc in the source plane. The green ellipse has been visually estimated to represent the extent of the brightness distribution; it has a semi-major axis of $0\farcs15$ or $1.2$~kpc in the source plane.
    \textit{Bottom right:} Data minus model residuals. Black polygons indicate the masks used for the {\em pixsrc} analysis. Yellow lines in the colorbar represent the noise level of $\rm 63 \, \mu Jy/beam$ measured outside of the masked regions.
    }
    \label{fig:pixsrc-core}
\end{figure*}

\section{Gravitational Lensing Model}\label{sec:Lensing-Model}
In order to recover the source-plane morphology and the intrinsic properties of HerS-3 and to explore the lensing configuration, we derived a lens model using the high-angular resolution ALMA continuum data, with initial lens parameters informed by the {\it HST}, Subaru/HSC, and SDSS data. We performed the lens modeling and source reconstruction using the \textit{lensmodel} package \citep{Keeton2001} with the pixel-based source reconstruction code \textit{pixsrc} \citep[][see Appendix~\ref{sec:Models-with-four-galaxies}]{Tagore&Keeton2014, Tagore&Jackson2016}. 

For an efficient and thorough analysis of the lensing configuration of the foreground group of galaxies, we conducted the lens modeling in a two-stage approach. First, we treated the five images of the Einstein cross as point sources, consistent with their compact sizes and the fact that they are well separated, to assess the constraints imposed by the positions of the lens images on possible solutions and identify the most promising lens models. We took into account the properties of the group of four lensing galaxies, such as their redshifts and shape priors (see Sect.~\ref{sec:Foreground-Group-Galaxies} and Appendix~\ref{sec:lens}), assuming that the galaxies G1, G2, G3, and G4 represent the four mass components of the lensing configuration. Secondly, after identifying the most promising model, we reconstructed the full extended source structure and compared the predicted size, orientation, and brightness of each individual image to what is observed in the dust continuum emission with ALMA. Details of the lens modeling setup, including lists of parameters, priors, and constraints, are provided in Appendix~\ref{sec:Lens-model-setup}.

As described in Appendix~\ref{sec:Models-with-four-galaxies}, the lens models that only include galaxies G1, G2, G3, and G4 are unable to reproduce the lensing properties observed for HerS-3. With mild priors on the shapes of the model galaxies, the model can match the lensed image positions only by making the model galaxy G3 much more elongated than observed (Fig.~\ref{fig:model-nohalo-kappa}a). Even then, the model cannot match the orientation of the extended structure in the SW image (Fig.~\ref{fig:pixsrc-nohalo-ellshr}). If we impose stronger priors on the galaxy shapes, the model cannot reproduce the observed image positions, with offsets up to $0\farcs2$ (Figs.~\ref{fig:model-nohalo-kappa}b and \ref{fig:pixsrc-nohalo-tight}). While we show results assuming that all four galaxies lie at the same redshift ($z_{\rm phot} \approx 1$), the conclusions hold for models where the foreground galaxies have different redshifts.

\begin{figure*}[ht!]
    \centering
    \includegraphics[width=0.6\linewidth]{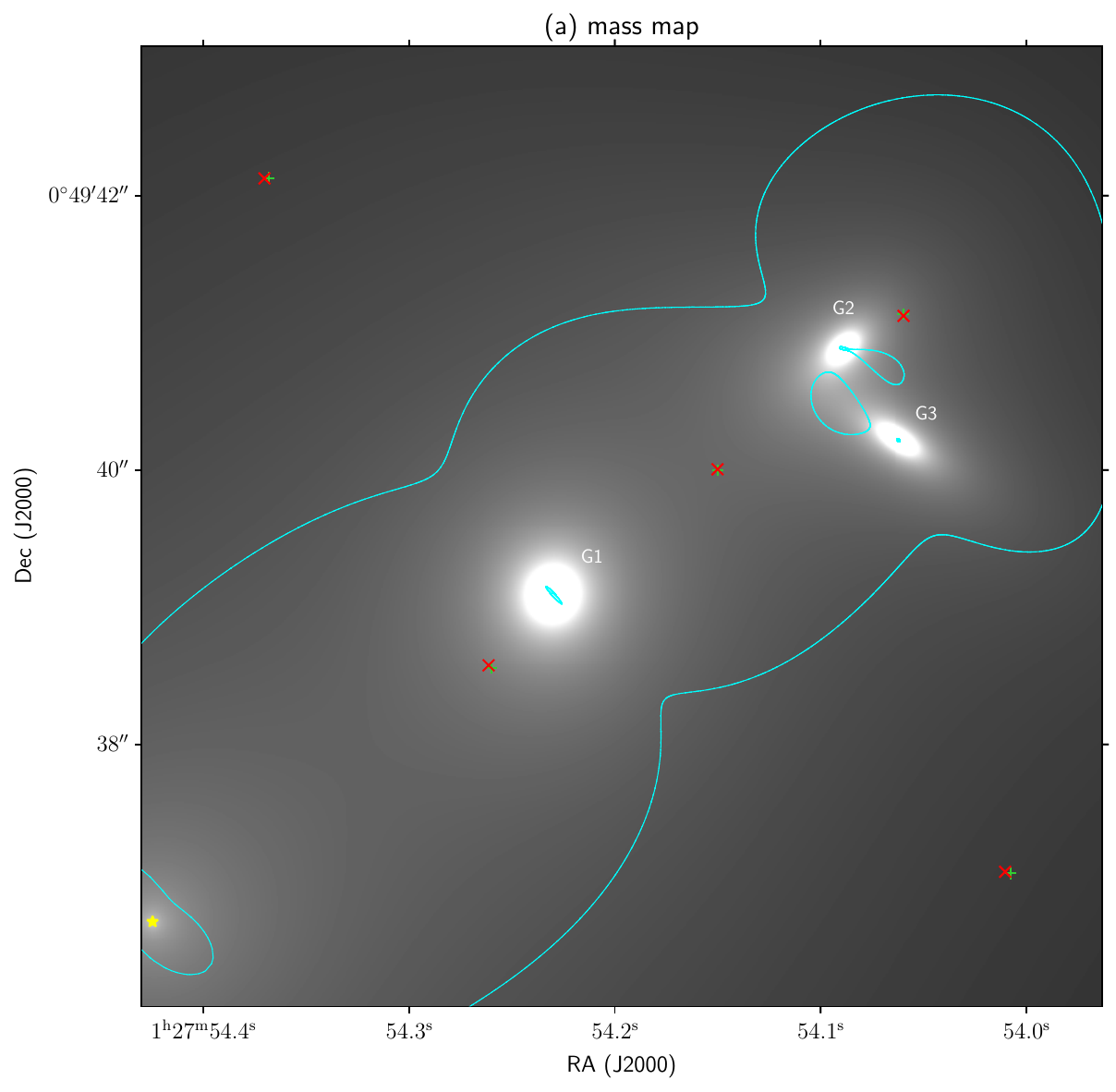}
    \vspace{0.5cm}
    \includegraphics[width=\linewidth]{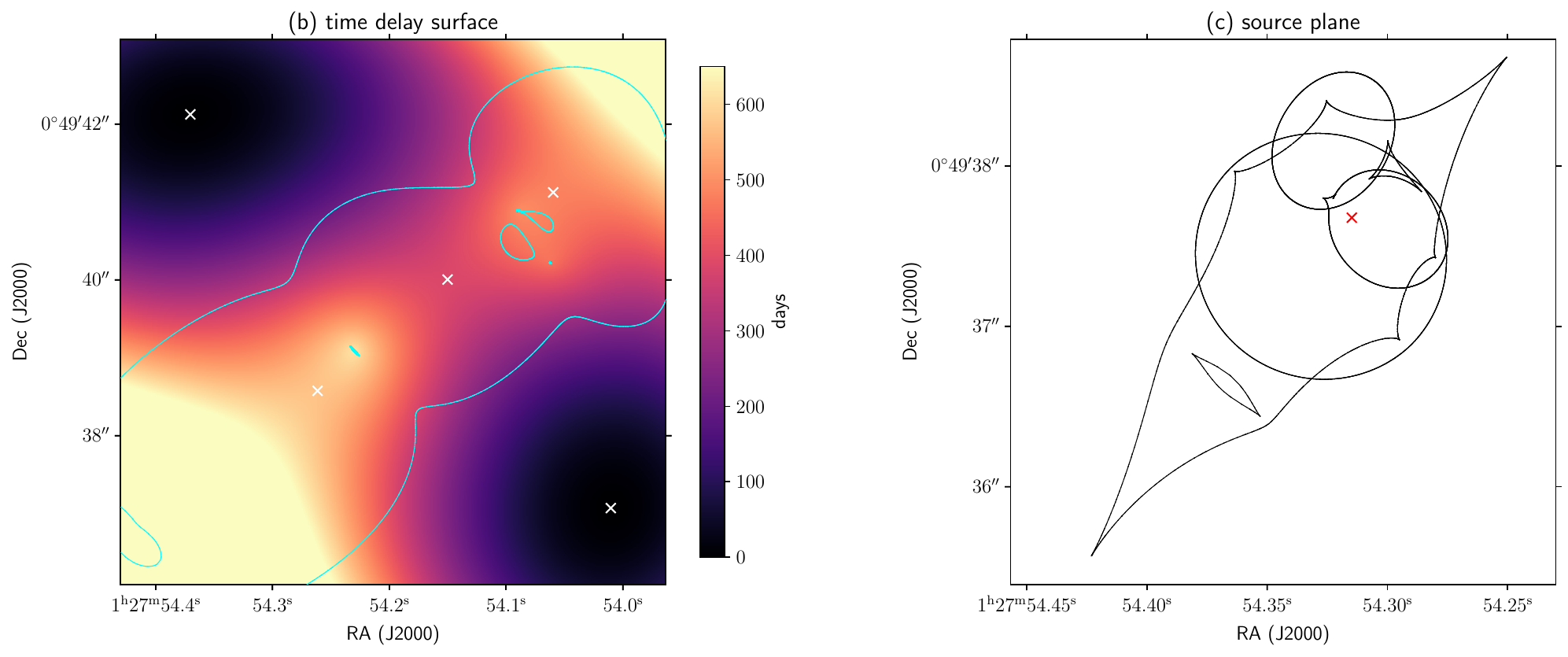}
\vspace{-0.5cm}
    \caption{
    Similar to Fig.~\ref{fig:model-core}, but for the lens model with an NFW halo for the additional mass component.
    }
    \label{fig:model-nfw}
\end{figure*}

\begin{figure*}[ht!]
    \centering
    \includegraphics[width=1\linewidth]{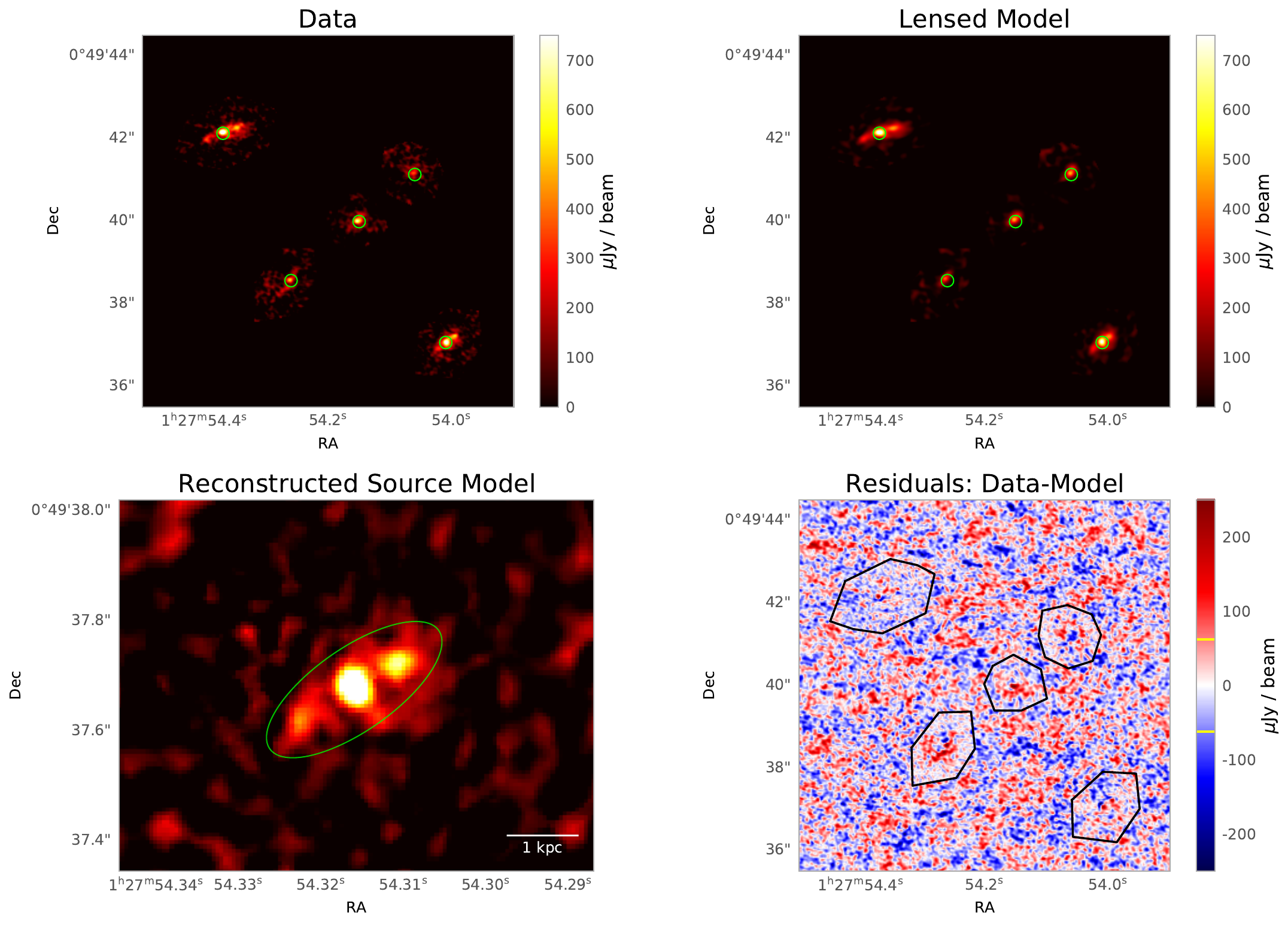}
    \caption{Similar to Fig.~\ref{fig:pixsrc-core}, but for the lens model with the NFW halo from Fig.~\ref{fig:model-nfw}. Here the ellipse in the source panel has a semi-major axis of $0\farcs19$ or $1.5$~kpc.}
    \label{fig:pixsrc-nfw}
\end{figure*}

The failure of those lens models, and the fact that there is no other close-by galaxy at the same redshift in the field, led us to consider the possibility of an additional fifth mass component that could represent a group dark matter halo. We try two models for this additional mass component. The first is a cored isothermal sphere with density profile $\rho(r) = \sigma^2/(2\pi G (s^2+r^2))$ where $s$ is the core radius, which corresponds to a scaled surface mass density for lensing of $\kappa(R) = \Sigma(R)/\Sigma_{\rm crit} = b/(2\sqrt{s^2+R^2})$ (see Appendix \ref{sec:Lens-model-setup} for more details about the parameters). The second model is a spherical Navarro-Frenk-White \citep[NFW;][]{Navarro1996} halo with density profile $\rho(r) = \rho_s r_s^3/(r(r_s+r)^2)$ where $r_s$ is a scale radius and $\rho_s$ is a scale density; the lensing properties of this model are given by \citet{Bartelmann1996}. The additional mass component is assumed to lie at the same redshift as the four visible galaxies. Details of this new lens model are given in Appendix~\ref{sec:Model-with-halo}.

With the additional mass component, the lens model and observed configuration are in excellent agreement, with a rms offset between the observed and model brightness peaks of better than $0\farcs02$. Figure \ref{fig:model-core} depicts the model with a cored halo by showing the surface mass density map, the lensing time delay surface, and the lensing critical curves and caustics. (The model uncertainties are listed in Table~\ref{tab:model_parameters} in Appendix~\ref{sec:Model-with-halo}.) The location of the additional component is constrained to lie 3--5$''$ to the southeast of galaxy G1. The NE and SW images both lie outside the lensing critical curve and at minima of the time-delay surface, implying that they have positive parity. The three components E, C and W all lie inside the critical curve and at saddle-points of the time-delay surface, so they are expected to have negative parity. These results have implications on the kinematics of the images as discussed in Sect.~\ref{sec:Source-Properties}. 

Strictly speaking, the lens model predicts two ``central'' images corresponding to local maxima of the time delay surface, in addition to the five observed images. However, these ``central'' images are extremely demagnified: they could easily be $10^4$ time fainter than the NE image, with specific values depending on the assumed central densities of the foreground lensing galaxies.

The full pixel-based source reconstruction shown in Fig.~\ref{fig:pixsrc-core} provides further support for this lens model. The model image configuration matches the observations in terms of the peak emission, brightness distribution, shape, and orientation for each of the five lensed images.

Figures \ref{fig:model-nfw} and \ref{fig:pixsrc-nfw} show the corresponding results for a lens model with an NFW halo for the additional mass components. Here the rms offset between the observed and model brightness peaks is again better than $0\farcs02$. Many of the features are qualitatively similar to the cored halo model: the halo is centered 3--7$''$ to the southeast of galaxy G1; the lensing critical curves extend from the south-east to the north-west; and the NE and SW images are again local minima of the time delay surface while E, C, and W are all saddlepoints. The pixel-based source reconstruction again matches all the properties observed in the ALMA 292 GHz data. Broadly speaking, then, we conclude that the lensing configuration indicates the presence of a mass component to the south-east, but it cannot precisely constrain the internal structure of this structure.

We briefly explored two ways to generalize the cored halo models. First, we allowed the additional dark matter halo to be elliptical rather than spherical. There is a mild preference ($\sim$1.5$\sigma$) for this halo to be slightly elongated in the North-South direction; the other parameters, including those we use to draw conclusions, did not change substantially. Second, we treated galaxies G1, G2, and G3 with power law mass profiles that were allowed to deviate from an isothermal profile. There is a moderate preference ($\sim$2$\sigma$) for G2 to have a density profile that is steeper than isothermal; again the main lens modeling conclusions were largely unchanged. We defer a more thorough examination of complex lens models to future work.

\begin{table}
\vspace{0.3cm}
    \caption{Predicted lensing masses}
    \label{tab:lensingmass}
    \begin{center}
    \begin{tabular}{ccccc}
\hline\hline
Galaxy 
& \multicolumn{2}{c}{Cored halo model}
& \multicolumn{2}{c}{NFW halo model}
\\
& best & MCMC & best & MCMC
\\
& \multicolumn{4}{c}{log($M_{\rm lens})$ [$\rm M_{\odot}$]}
\\
\hline
G1
& $11.85$ & $11.99_{-0.15}^{+0.08}$
& $12.01$ & $12.02_{-0.06}^{+0.06}$
\\
G2
& $11.53$ & $11.65_{-0.13}^{+0.11}$
& $11.67$ & $11.68_{-0.08}^{+0.08}$
\\
G3
& $11.57$ & $11.65_{-0.12}^{+0.09}$
& $11.68$ & $11.67_{-0.08}^{+0.08}$
\\
G4
& $11.86$ & $11.98_{-0.15}^{+0.13}$
& $12.02$ & $12.05_{-0.11}^{+0.12}$
\\
\hline
    \end{tabular}
    \end{center}
    \vspace{-0.3cm}
\tablenotetext{}{{\bf Notes.}
For each type of halo model, we report the projected lensing mass (including both baryon and dark matter components) within $3''$, assuming a lens redshift $z_l = 1$. Uncertainties indicate the 68\% confidence interval from MCMC sampling of the posterior. For comparison, the stellar masses derived from the optical and near-infrared data are given in Table \ref{tab:SED_fit_results}.
}
\end{table}

    \begin{table*}
        \caption{Predicted lensing magnifications}       
        \label{tab:magnifications}
        \begin{center}
        \begin{tabular}{l|rrr|rrr} 
            \hline\hline
            & \multicolumn{3}{|c}{Cored halo model} & \multicolumn{3}{|c}{NFW halo model} \\
            Image & \multicolumn{2}{|c}{Eigenvalues} & Magnification & \multicolumn{2}{|c}{Eigenvalues} & Magnification \\
\hline
SW
& $ 2.74_{-0.38}^{+0.91}$ & $ 2.16_{-0.31}^{+0.67}$ & $ 5.91_{-1.51}^{+4.47}$  
& $ 2.63_{-0.32}^{+0.39}$ & $ 2.08_{-0.23}^{+0.28}$ & $ 5.46_{-1.20}^{+1.66}$  
\\
NE
& $ 3.81_{-0.57}^{+1.23}$ & $ 1.99_{-0.31}^{+0.68}$ & $ 7.64_{-2.21}^{+5.67}$  
& $ 3.68_{-0.42}^{+0.57}$ & $ 1.90_{-0.25}^{+0.27}$ & $ 7.00_{-1.64}^{+2.16}$  
\\
E
& $-0.88_{-0.32}^{+0.15}$ & $ 1.77_{-0.42}^{+0.75}$ & $-1.55_{-1.39}^{+0.52}$  
& $-0.86_{-0.19}^{+0.15}$ & $ 1.43_{-0.20}^{+0.24}$ & $-1.24_{-0.44}^{+0.30}$  
\\
C
& $-1.36_{-0.47}^{+0.26}$ & $ 1.46_{-0.24}^{+0.53}$ & $-1.97_{-1.68}^{+0.61}$  
& $-1.29_{-0.25}^{+0.21}$ & $ 1.34_{-0.15}^{+0.19}$ & $-1.74_{-0.53}^{+0.44}$  
\\
W
& $-1.17_{-0.44}^{+0.31}$ & $ 1.57_{-0.18}^{+0.43}$ & $-1.85_{-1.32}^{+0.63}$ 
& $-1.07_{-0.24}^{+0.21}$ & $ 1.55_{-0.17}^{+0.22}$ & $-1.66_{-0.53}^{+0.43}$  
\\
total
&&& $19.00_{-5.44}^{+14.66}$
&&& $17.06_{-4.01}^{+5.21}$
\\
     \hline      
        \end{tabular}
        \end{center}
\vspace{-0.3cm}
\tablenotetext{}{{\bf Notes.}
For each type of halo model, we report the two eigenvalues of the lensing magnification tensor (which indicate the linear magnification factors in the principal directions; a negative value indicates parity reversal) and the overall magnification factor (the product of the eigenvalues). Uncertainties indicate 68\% confidence intervals from MCMC sampling of lens models.
}
    \end{table*}         

    \begin{table}
        \caption{Predicted lensing time delays }       
        \label{tab:timedelays}
        \begin{center}
        \begin{tabular}{lcc} 
            \hline\hline
         Image   & Cored halo model & NFW halo model \\
         & \multicolumn{2}{c}{$\Delta t$ (days)} \\
\hline
SW
& $ \hspace{0.29cm} 7.9_{-11.3}^{+12.2}$
& $ \hspace{0.29cm} 2.6_{-15.4}^{+14.6}$
\\
E
& $391.1_{-99.9}^{+76.2}$
& $403.8_{-54.0}^{+65.7}$
\\
C
& $273.6_{-71.2}^{+53.3}$
& $275.9_{-38.2}^{+50.3}$
\\
W
& $344.5_{-85.6}^{+68.4}$
& $349.7_{-44.3}^{+57.2}$
\\
     \hline      
        \end{tabular}
        \end{center}
\vspace{-0.3cm}
\tablenotetext{}{{\bf Notes.}
Predicted time delays relative to the NE image, assuming a lens redshift $z_l = 1$. In most models, NE is the leading image, but, in some cases, the SW image arrives slightly ahead of NE and those correspond to negative time delays.
}
    \end{table}

We can use the lens models to predict various properties of the system. First, we can compute the projected lensing masses of the model galaxies, including both the baryon and dark matter components, within an aperture $R$ to be $M_{\rm lens} = \pi b R \Sigma_{\rm crit}$, where $b$ is the Einstein radius and $\Sigma_{\rm crit}$ is the critical surface density for lensing. We use $R = 3''$ to match the aperture used to compute stellar masses. Table \ref{tab:lensingmass} shows that both the cored and NFW halo lens models predict lensing masses around $10^{12}\,M_\odot$ for galaxies G1 and G4, and slightly lower values for G2 and G3, all of which are reasonable in comparison with the stellar masses.

\begin{figure}[t]
\centering
\includegraphics[width=\columnwidth]{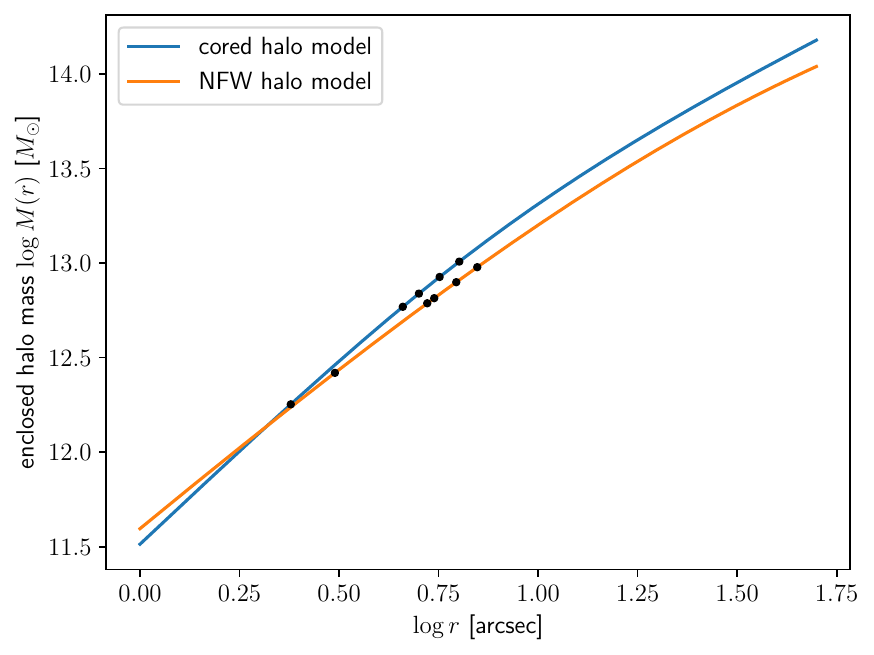}
\vspace{-0.3cm}
\caption{The curves show the projected enclosed dark matter halo mass, $M(r)$, as a function of $r$, for the two lens models using a cored isothermal sphere or a spherical NFW halo (blue and orange lines, respectively). Points mark the locations of the lensed images (relative to the halo center), to indicate the range of radii probed by the lensing data. Between the two models, the points have different $r$ values because the halos have different central positions.
}
\label{fig:model-halomass}
\end{figure}

\begin{figure}[t]
\centering
\includegraphics[width=\columnwidth]{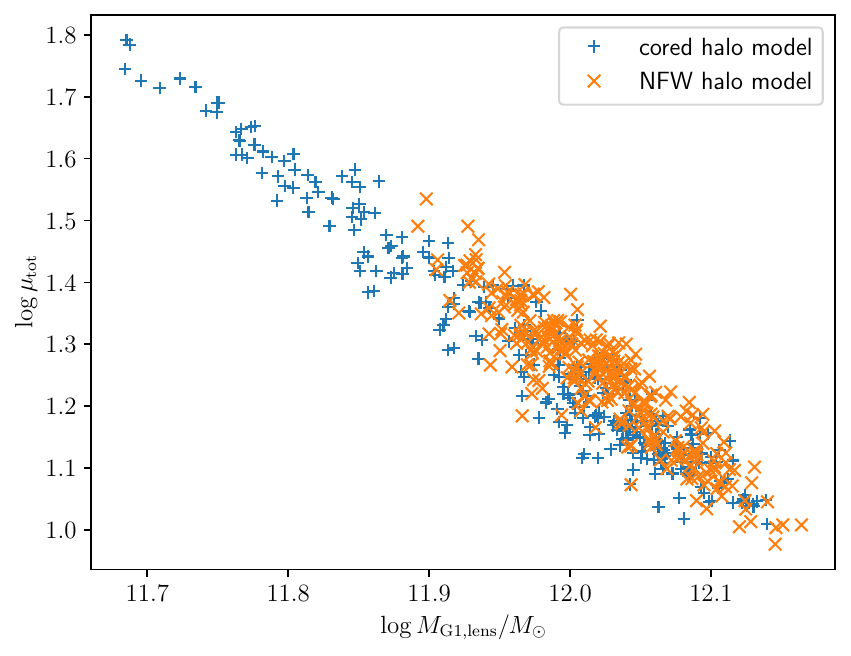}
\vspace{-0.3cm}
\caption{Relationship between the projected lensing mass of galaxy G1 (horizontal axis) and the total lensing magnification (vertical axis) for a selection of lens models drawn from the posterior parameter distributions for the cored halo models and NFW halo models. 
}\label{fig:model-mag-vs-mass}
\end{figure}

We can similarly compute the projected lensing mass of the large dark matter halo in each model. Figure \ref{fig:model-halomass} shows the enclosed mass curve, $M(r)$ as a function of $r$, for both the cored halo and NFW halo models. Over the range of radii spanned by the lensed images, the enclosed dark matter mass ranges from $\log M_{\rm DM}/{\rm M_{\odot}} = 12.2$ up to $13.0$. In both models the enclosed mass curve continues to grow; while these larger radii are not well constrained by the lensing data, it is interesting to see that the cored and NFW halo lens model yield similar results. Because the cored halo model has an isothermal profile, it is straightforward to convert the Einstein radius parameter into an effective velocity dispersion of $\sigma = 633_{-129}^{+185}$ km/s (see Appendix~\ref{sec:Lens-model-setup} for the relation). However, that value may not be especially enlightening as it depends on radii larger than the range spanned by the lensed images. There is no analogously simple quantity for an NFW halo.

We can also compute the lensing magnification factors, which are given in Table \ref{tab:magnifications}. The total magnification of all images is around 19 for the cored halo model and 17 for the NFW halo model, with some non-negligible statistical uncertainties. There is a degeneracy such that lens models can make the halo more massive and the galaxies less massive, or vice versa, over some reasonable range. When the halo is more massive, it contributes a higher surface mass density at the positions of the lensed images, which leads to higher magnification factors. Figure \ref{fig:model-mag-vs-mass} illustrates this degeneracy by showing the relation between the total lensing magnification and the lensing mass of galaxy G1.

Finally, we can predict the lensing time delays, which are listed, in days, for both the cored halo and the NFW halo models in Table~\ref{tab:timedelays}. In most models, the leading image is the NE one, with the SW image arriving a few days later, in between $7.9^{+12.2}_{-11.3}$ and $2.6^{+14.6}_{-15.4}$ days for the cored halo and NFW halo models, respectively. It should be noted that in some cases, the SW image arrives ahead of the NE, resulting in negative time delays. For the three images E, C, and W, which are located at saddle-points of the time-delay surface (Figs.~\ref{fig:model-core} \& \ref{fig:model-nfw}), the time delays relative to the NE image are predicted to be much longer, amounting to about one year with the C, W, and E images in order of arrival time.

\begin{figure*}[ht!]
\centering
\includegraphics[height=5.5cm]{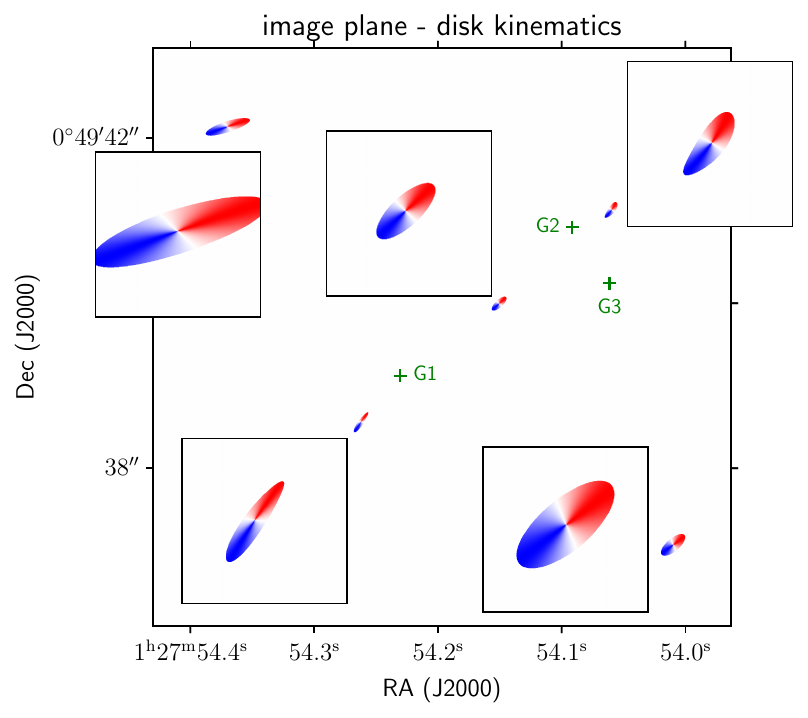}
\includegraphics[height=5.5cm]{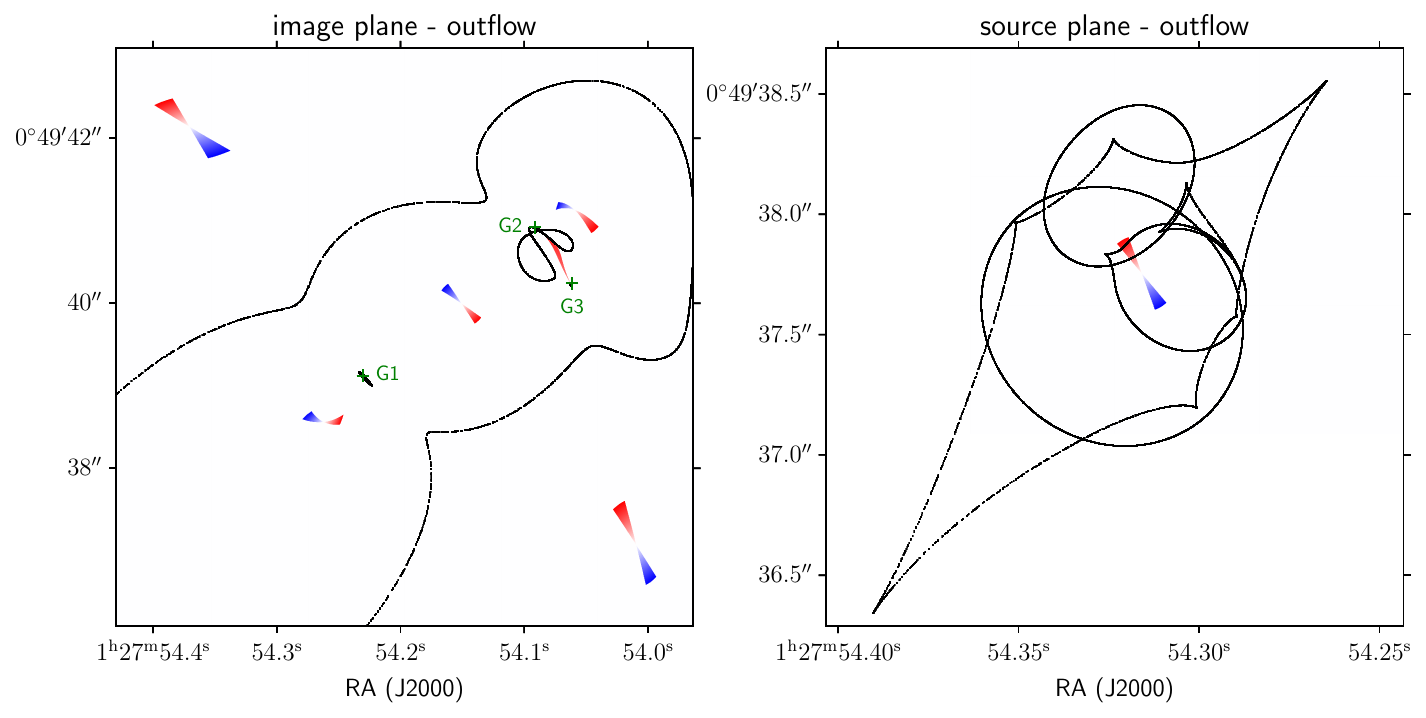}
\caption{
(Left) Model lensed velocity maps assuming a simple rotating disk galaxy with a flat rotation curve and a position angle of 120$^\circ$ based on our source reconstruction. The insets, displaying each image in detail, are 0.5$''$ across. The observed changes in the $\rm ^{12}CO$(9-8) velocity distributions from image to image (as presented in Fig.~\ref{fig:Velocity-Position}) are compatible with the model predictions. (Middle) Velocity map for a model of outflowing gas assumed to be perpendicular to the disk, showing that the axis along which the parity is flipped runs from upper left to lower right. The lensing critical curves are shown in black. (Right) In this model, the outflowing gas is predicted to cross two lensing caustics, creating an additional redshifted filament between G2 and G3 in the velocity map. We assume here that the redshifted outflow lies to the north and the blueshifted outflow lies to the south, but the situation could be reversed depending on the galaxy's inclination. This figure shows results for the cored halo lens model, but results for the NFW halo model are very similar.
}
\label{fig:Velocity-Maps-Parity-Reversal} 
\end{figure*}

\vspace{1cm}
\section{Discussion}\label{sec:Discussion}
In this section, we first present the general properties of the dusty star-forming galaxy HerS-3 (Sect.~\ref{sec:Source-Properties}) derived from all the currently available sub-millimeter to radio data and based on the results of the lensing model described in Sect.~\ref{sec:Lensing-Model}. In Sect.~\ref{sec:dark-matter-halo}, we review the characteristics of the lensing galaxy group placing the findings of this study in a broader context, in particular, by outlining what future observations are needed to further constrain the properties of the massive dark matter halo, and in Sect.~\ref{sec:cosmological-probe}, we describe the potential use of HerS-3 to measure the Hubble constant.  

\subsection{The Starburst HerS-3}\label{sec:Source-Properties}
In the source plane, HerS-3 appears as a nearly edge-on galaxy (Fig.~\ref{fig:pixsrc-core} \& \ref{fig:pixsrc-nfw}) with a bright central region surrounded by two weaker emission peaks. The size of the galaxy (major and minor axis), estimated from the 1~mm dust continuum emission, is $0\farcs30 \times 0\farcs12$ ($\rm \sim 2.3 \times 0.9~kpc^2$) or $0\farcs38 \times 0\farcs15$ ($\rm \sim 2.9 \times 1.2~kpc^2$) for the models with the cored halo and the NFW profile, respectively, with a position angle $\rm PA \sim 120^{\circ}$. Based on the results of the source plane reconstruction, we can derive the inclination angle from the minor to major axis ratio $b/a$ of the disk as $i = {\rm cos}^{-1} (b/a)$, yielding $i \sim 66^\circ$ for both the cored halo and NFW profile models. The rotation in the galaxy’s disk, traced in the $^{12}$CO(9$-$8) emission line (Fig.~\ref{fig:Velocity-Position}), indicates a gradient from red to blue-shifted gas along the west-east axis, which is best seen in the NE and SW images that have the highest S/N data. This is consistent with the predictions of the lensing model as illustrated in the left panel of Fig.~\ref{fig:Velocity-Maps-Parity-Reversal}. Higher angular resolution observations are required to further explore the kinematics of this starburst and reveal the details of the velocity field, in particular in the E, C, and W images. In addition, the blue-shifted absorption lines of OH$^+$, which are detected in each of the five images, indicate a strong outflow activity with gas expelled from the starburst with velocities of at least $\sim 350 \, \rm km \, s^{-1}$. 

HerS-3 appears therefore to be a nuclear starburst with a morphology that is comparable to the nearby galaxy NGC~253, known to possess a starburst-driven wind \citep{Bolatto2013, Perez-Beaupuits2018}. As in NGC~253, the outflow axis in HerS-3 is most likely perpendicular to the active nucleus, specifically along the NE-SW axis where the parity reversal is occurring (middle panel of Fig.~\ref{fig:Velocity-Maps-Parity-Reversal}). Observing this outflow at higher-angular resolution, using molecular absorption lines (such as OH$^+$ or OH) or atomic emission lines (e.g, $\rm [C{\small II}]$ 158~$\rm \mu m$), will allow us to explore in greater detail the morphology of the wind activity in HerS-3, akin to what has been achieved in other high-$z$ active star-forming galaxies \cite[e.g.,][]{Butler2021, Spilker2020a, Spilker2020b, Herrera-Camus2021}, and to trace the parity reversal expected in each of the three saddle-point images, providing a basis to further refine the lensing model.

\subsubsection{Molecular Gas Excitation and Properties}\label{Molecular-Gas-Excitation-and-Properties}
To investigate the CO line excitation and estimate the physical conditions of the molecular gas in HerS-3, we modeled the CO line fluxes using a Large Velocity Gradient (LVG) treatment of statistical equilibrium and radiative transfer that is described in Appendix~\ref{sec:LVG-model}. In total, we have measurements of four $^{12}$CO transitions for HerS-3, namely: $J$=9$-$8 ($11.3\pm0.95 \, \rm Jy \, km s^{-1}$; Table~\ref{tab:derived-global}), $J$=5$-$4 and 3$-$2 \citep[$8.27\pm0.68$ and $5.25\pm0.36 \, \rm Jy \, km s^{-1}$, respectively;][]{Cox2023}, and $J$=1$-$0 ($0.71\pm0.08 \, \rm Jy \, km s^{-1}$; Prajapati et al. in prep). The velocity-integrated line fluxes here used are from the global spectrum of HerS-3 and have not been corrected for gravitational magnification.  As the CO spectral line energy distribution (SLED) of HerS-3 displays a strong $\rm ^{12}CO$(9$-$8) emission line, we used a two-excitation component model, assigning different parameters to each of the components. The  two-component best-fitting solution of the CO SLED of HerS-3 is shown in Fig.~\ref{fig:CO-SLED}. In Appendix~\ref{sec:LVG-model}, the posterior probability distributions of the molecular gas density $n_{\rm H_2}$, the gas temperature $T_{\rm kin}$, and the CO column density per velocity $N_{\rm CO}/{\rm d}v$ of the source for the cold and warm components are presented in Fig.~\ref{fig:CO-SLED-Posterior-Probablity-Distribution}. The fit reproduces the overall SLED very well, explaining the observed $^{12}$CO(3$-$2), (5$-$4) and (9$-$8) relative line fluxes, which a single component model could not have achieved. The physical properties derived from the model are listed in Table~\ref{tab:molecular-gas-properties}.

\begin{figure}[!ht]
    \centering
         \includegraphics[width=1.0\linewidth]{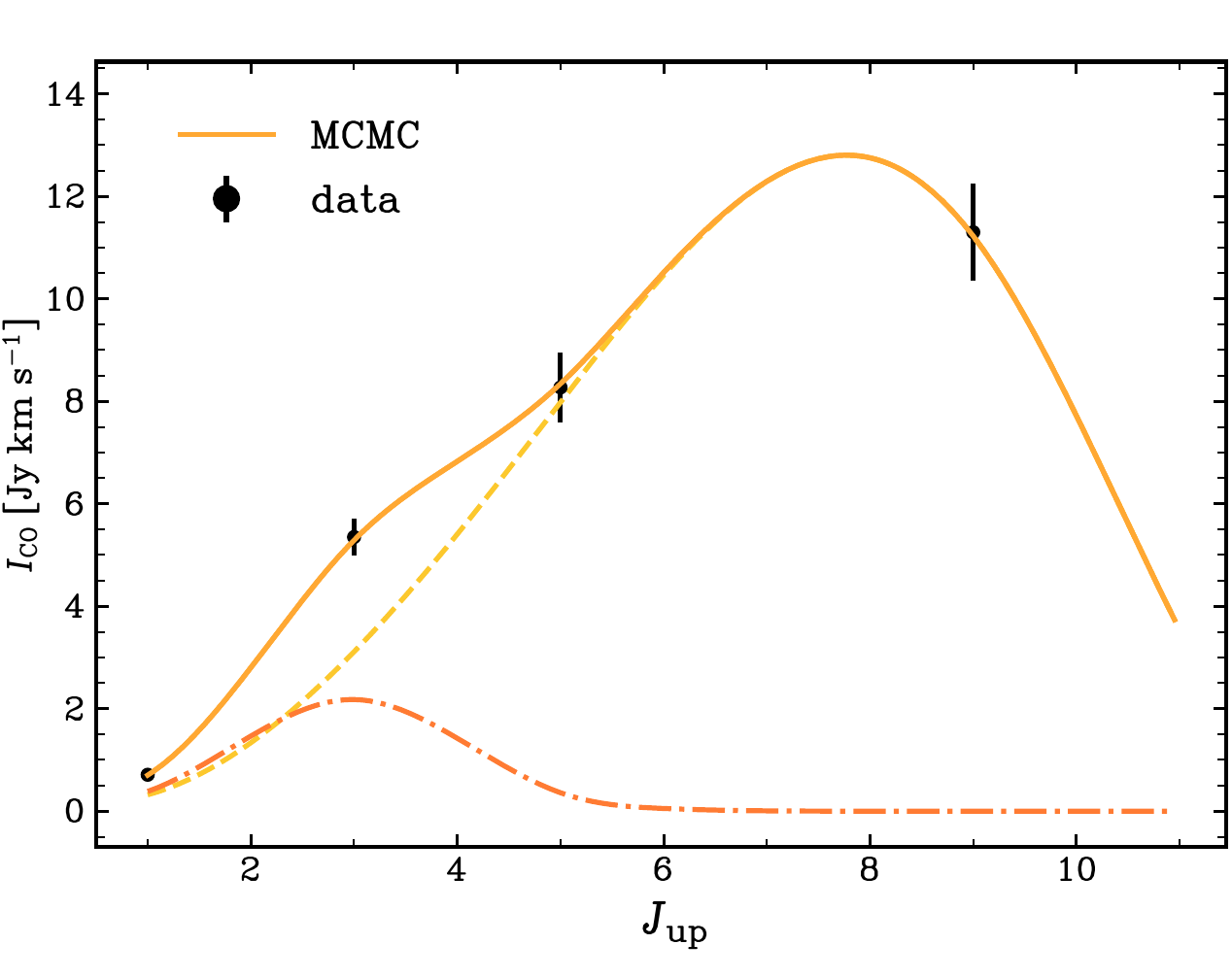}\\
  \caption{Large Velocity Gradient (LVG) two-excitation component model fit to the $^{12}$CO SLED of Hers-3. The velocity-integrated line fluxes (not corrected for gravitational magnification) and their errors are shown as black dots and the solid orange curve shows the best fit from the two-component (MCMC) model corresponding to the maximum posterior possibility. The orange dashed and red dash-dotted lines show the warmer and cooler component fits, respectively. The derived physical properties of the molecular gas of HerS-3 are listed in Table~\ref{tab:molecular-gas-properties}.}
\label{fig:CO-SLED} 
\end{figure}

The molecular gas in HerS-3 is significantly excited with conditions similar to those of the Cloverleaf \citep{Weiss2003, Bradford2009, Riechers2011}, HFLS-3 \citep{Riechers2013} or NGC~6240 \citep{Meijerink2013}. Excitation through the hard radiation (including X-rays) associated with a luminous active galactic nucleus (AGN) is likely not at play as there is no indication for an AGN in HerS-3 (see Sect.~\ref{sec:SED}). It is therefore more plausible that the excitation of the dense molecular gas in HerS-3 is dominated by the infrared photons and collisions (shocks) originating in the massive, intense starburst and the outflowing gas traced in the $\rm OH^+$ absorption lines. This is in line with the CO luminosity  to infrared luminosity ratio of $L_{\rm CO}/L_{\rm IR} \sim 1.6 \times 10^{-4}$ in HerS-3, where $L_{\rm CO}$ has been extrapolated, up to $J$=11-10, from the CO SLED (Fig.~\ref{fig:CO-SLED}). This ratio is higher than for Mrk~231 \cite[$7 \times 10^{-5}$;][]{vanderWerf2010}, where the excitation is dominated by X-rays from the AGN, and the maximum ratio of $L_{\rm CO}/L_{\rm IR} \sim 10^{-4}$ predicted for models of photon and X-ray dominated regions \citep[][and references therein]{Meijerink2013}. As in NGC~6240, where the ratio is $7 \times 10^{-4}$ \citep{Meijerink2013}, the excitation of the interstellar medium in HerS-3 is therefore most likely dominated by shocks, which heat the gas and almost not the dust. Measuring additional, intermediate $J$ level CO transitions would be useful to further constrain the nature of the excitation of the molecular gas in HerS-3.

\begin{table}      
\caption{Physical properties of the molecular gas of HerS-3 derived from the $^{12}$CO SLED two-component LVG model.}
        \label{tab:molecular-gas-properties}              
        \begin{center}
       \vspace{-0.4cm}
        \begin{tabular}{lccc} 
            \hline\hline       
           Component & log($n_{\rm H_2}$) & log($T_{\rm kin}$) & ${\rm log}_{10}(N_{\rm CO/dv})$ \\
                     &   ($\rm cm^{-3}$)  & (K) & $({\rm cm^{-2} \, km \, s^{-1}})$ \\
            \hline
            Low-excitation  & $3.08^{+0.92}_{-1.08}$ & $1.21^{+0.16}_{-0.09}$  & $17.22^{+0.95}_{-0.73}$ \\
            High-excitation & $3.22^{+0.89}_{-0.74}$ & $2.33^{+0.25}_{-0.26}$ & $18.24^{+0.78}_{-0.69}$ \\
          \hline      
        \end{tabular}
        \end{center}
        \vspace{-0.3cm}
            \tablenotetext{}{{\bf Notes.} The values listed are the gas density ($n_{\rm H_2}$), the kinetic temperature ($T_{\rm kin}$), and the CO column density ($N_{\rm CO/dv}$), where the errors are $3\sigma$, as derived from the two-component LVG model.}  
    \end{table}

The $\rm ^{12}$CO(1$-$0) emission line flux enables us to estimate the line luminosity, $L^{\prime}_{\rm CO(1-0)}$, and the molecular gas mass, $M_{\rm H_2}$, associated with HerS-3, without the uncertainties related to the conversion when using higher-$J$ CO lines. We derive the values of  $L^{\prime}_{\rm CO(1-0)}$ and $M_{\rm H_2}$ using the following equations \citep[e.g.,][]{Carilli-Walter2013}:
\begin{eqnarray}
    L^{\prime}_{\rm CO} &=& 3.25 \times 10^7 \, \frac{S_{\rm CO} \Delta v_{\rm CO} \, D^2_{\rm L}}{\nu^2_{\rm CO,rest} \, (1+z)} \, \, \rm{K \, km \, s^{-1} \, pc^{-2}}
\\
    M_{\rm H_2} &=& \alpha_{\rm CO} \times L^{\prime}_{\rm CO(1-0)} \, \, \rm M_{\odot}
\end{eqnarray}
where $S_{\rm CO} \Delta v_{\rm CO}$ is the measured flux density of the line in $\rm Jy \, km \, s^{-1}$, $D_{\rm L}$ the luminosity distance in Mpc, $\nu^2_{\rm CO,rest}$ the rest frequency of the line in GHz, and $\alpha_{\rm CO}$ is the $\rm ^{12}CO(1-$0)-to-$\rm H_2$ conversion factor. We here adopt a value of $\alpha_{\rm CO} = 3 \, \rm M_{\odot} \, (K \, km \, s^{-1} \, pc^2)^{-1}$ as derived by \cite{Dunne2021}, and we multiply it by the usual factor of 1.36 to account for the helium contribution. The $^{12}$CO(1-0) line luminosity is $\mu L^{\prime}_{\rm CO} = 3.07 \pm 0.35 \times 10^{11} \, \rm K \, km \, s^{-1} \, pc^{-2}$, and the molecular gas mass $\mu M_{\rm H_2} = 1.25\pm0.14 \times 10^{12} \, \rm M_{\odot}$. The resulting molecular gas to dust mass ratio is $\delta_{\rm GDR} \sim 168$ for the optically-thin MBB dust mass (and 246, in the case of the general MBB), which remains consistent with the mean ratio of $128^{+54}_{-35}$ derived by \cite{Dunne2021} for a sample of {\it Herschel}-selected starburst galaxies \citep[see also][]{Dunne2022, Stanley2023, Berta2023}.     

From these properties, we can estimate the star formation efficiency (SFE) of HerS-3, which is a measure of the star formation rate (SFR) per unit gas, and the gas depletion time, $t_{\rm depl}$, which is the inverse of the SFE, indicating the time it would take to deplete the gas reservoir with a constant star formation rate and no other on-going processes, such as gas accretion or outflows. The star formation efficiency can be expressed as ${\rm SFE = SFR}/ M_{\rm H_2}$ $\rm yr^{-1}$, where $\rm SFR (M_{\odot} \, yr^{-1}) = 1.09 \times 10^{-10} \, {\it L}_{\rm IR} \, (L_{\odot})$ and $\rm L_{\rm IR}$ is the rest-frame (8 to 1000~$\rm \mu m$) infrared luminosity \citep[][corrected for a \cite{Chabrier2003} initial mass function]{Kennicutt1989}. For HerS-3, using the infrared luminosity derived in Sect.~\ref{sec:SED}, we estimate $\rm \mu SFR = 8.2 \times 10^3 \, M_{\odot} \, yr^{-1}$ and a star formation efficiency $\rm SFE = 6.5 \times 10^{-9} \, yr^{-1}$, yielding a depletion time of $t_{\rm depl} = 152 \, \rm Myr$. This depletion time is significantly shorter than typical Main Sequence galaxies and is instead consistent with infrared luminous starburst galaxies at high-$z$ \citep[see, e.g.,][and references therein]{Stanley2023, Berta2023}.  

\vspace{1cm}
\subsubsection{Intrinsic Properties of HerS-3}
\label{sec:Intrinsic-Properties}
Depending on the type of dark matter halo profile adopted in the lensing model, the overall gravitational magnification factor changes from $\rm \mu = 19^{+14}_{-5}$ for the cored halo model to $\rm \mu = 17^{+5}_{-4}$ for the NFW halo model (Table~\ref{tab:magnifications}).  As explained in Sect.~\ref{sec:Lensing-Model}, this difference is due to the degeneracy between the halo mass and the galaxy masses and deeper observations at higher angular resolution of both the lensing group and HerS-3 will be needed to further constrain the lensing properties and, hence, reduce the significant errors on the magnification factor.  Adopting a value of $\mu \sim 18$, based on the current estimates, we derive the following intrinsic properties of HerS-3: $L_{\rm IR} \sim 4.2 \times 10^{12} \, \rm L_{\odot}$, for the total infrared luminosity;  $\rm SFR \sim 455 \, \rm M_{\odot} yr^{-1}$, for the star formation rate; $M_{\rm H_2} \sim 6.9 \times 10^{10} \, \rm M_{\odot}$, for the molecular gas mass; and $M_{\rm dust} \sim 4.1 \times 10^{8} \, \rm M_{\odot}$ (or $2.8 \times 10^{8} \, \rm M_{\odot}$, in the case of the general MBB), for the dust mass.

\subsection{Lensing Group and Dark Matter Halo}
\label{sec:dark-matter-halo}
The observed properties of the five images of the HerS-3 Einstein cross can be precisely matched by a lensing model that includes, in addition to the group of four visible galaxies at $z_{\rm phot} \sim 1$, a fifth massive component at the same redshift, located $\sim 3''$ to $7''$ (i.e., $\sim$25 to 60~kpc) south-east of the brightest galaxy G1 (Sect.~\ref{sec:Lensing-Model}). The fact that no galaxy is seen at that position in the {\it HST} F110W image points to the presence of a massive dark matter halo associated with the lensing group. Heuristically, the location to the south-east of this massive component is indicated by the curvature of the arc defined by the NE, C, and SW images. As noted above, the need for an additional halo seems robust, but conclusions about its properties depend on assumptions, such as the radial profile (which we have shown explicitly by comparing cored and NFW halo models) and our simplifying approach that the halo is spherical. Even so, it is worthwhile to examine what we can infer about the system from current data and lens models.

Since we assumed in the lensing model that the dark matter halo is at the same redshift as the visible galaxies, we also tested the effect of placing the halo at different redshifts. Taking the best models of each class (cored and NFW halo), we explored the redshift range $z=0.6$ to 1.2, and re-optimized all of the model parameters. For both models, only minor changes resulted at the different redshifts in the core density and the mass of the dark matter halo.   

As reported in Appendix~\ref{sec:zphot}, the four visible galaxies G1, G2, G3, and G4 lensing HerS-3 are part of a larger group containing ten more detected members at $z_{\rm phot} \sim 1$ within a 1~arcmin (500~kpc) radius from the brightest galaxy G1. The stellar mass of the four galaxies located at the center of the group is estimated to be $M_* \simeq 3.6 \times 10^{11} \,\rm  M_{\odot}$ (Table~\ref{tab:SED_fit_results}), accounting for 60\% of total stellar mass of the group (Appendix~\ref{sec:zphot}). Compared to their total (baryonic and dark matter) predicted by the lensing model of $M_{\rm tot} = 3 \times 10^{12} \, \rm M_{\odot}$ (Table~\ref{tab:lensingmass}), the ratio of the stellar to total mass for the four galaxies G1--G4 is $M_*/M_{\rm tot} \sim 12\%$.

Galaxy groups are small aggregates of galaxies, comprising several to $\sim 50$ galaxies within $\rm \sim 1 \, Mpc$, that are embedded within their dark matter halos. They are less massive than galaxy clusters, which have masses up to a few $\rm 10^{13} \, M_{\odot}$, and have thus smaller lensing cross sections than clusters. Being more numerous than clusters, groups of galaxies are known to be potential powerful systems to probe the properties and structure of dark matter halos \citep[e.g.,][]{Fox2001}. The total mass of the group, containing the galaxies lensing HerS-3, is in line with the median mass of known galaxy groups, $M_{\rm group} \sim 3 \times 10^{12} \, \rm M_{\odot}$ \citep[e.g.,][]{Einasto2024}. The lensing group is also a relatively compact group with a radius of $\rm \approx 500 \, kpc$, where the four most massive galaxies (G1, G2, G3, and G4) are quenched systems with low star-formation rates ($0.02 \leq \rm SFR/{\rm (M_{\odot} \, yr^{-1})} \leq 1.5$; see Sect.~\ref{sec:Foreground-Group-Galaxies}).  

It is interesting to note that the position of the massive dark matter halo component, needed to explain the observed properties of the HerS-3 Einstein cross, is located in between the galaxies G1--G3 and G4, at a position approximately corresponding to the center of mass of the entire group. From the estimated lensing mass of the dark matter halo, $\log(M_{\rm DM}/{\rm M_{\odot}}) \sim 13.0$, using the mass enclosed within the largest sampled radius (Fig.~\ref{fig:model-halomass}), the $z\sim1.0$ galaxy group lensing HerS-3 follows the stellar-to-halo mass relation derived in the $\log(M_{\rm DM}/M_*)-\log(M_{\rm DM})$ plane for $z \sim 1$ galaxies with $\log(M_*/M_{\rm DM})\le-1.2$ and $\log(M_{\rm DM}/{\rm M_{\odot}}) \sim 13$ \citep[see Fig.~6 in][]{Girelli2020}.

Based on the current available observational data, the lensing model, which successfully reproduces the properties of the five images of the HerS-3 Einstein cross, reveals a massive dark matter halo at the center of a galaxy group, with no counterpart seen in the {\it HST} near-infrared image. The four brightest lensing galaxies G1--G4 are distributed around the dark matter halo at distances ranging from $\rm \sim 25 \, to \, 80~kpc$. All other members of the group are located at much larger distances (see Appendix~\ref{sec:zphot}). This configuration is reminiscent of a non-relaxed galaxy group, suggesting that the group at $z_{\rm phot} =1$ lensing HerS-3 is not virialized. However, additional observations will be needed to further constrain the lensing model and, henceforth, the properties of both the massive dark matter halo and the galaxy group, so as to better understand its evolutionary state.

Finally, as an example of further refinements in the analysis of the lensing configuration, our lens models indicate that the center of the background source, HerS-3, lies 0.6--0.8 kpc outside a lensing caustic (see panel c in each of Figs.~\ref{fig:model-core} and \ref{fig:model-nfw}). If the source exhibits an outflow to the north-east that is $\gtrsim$1~kpc in extent, as expected based on the outflow activity traced by the $\rm OH^+$ absorption (Sect.~\ref{sec:Source-Properties}), that emission would cross two lensing caustics, creating an additional lensed arc-like image in between the galaxies G2 and G3 (see Fig.~\ref{fig:Velocity-Maps-Parity-Reversal}). This is a clear prediction of current lens models that could directly be tested with future high-angular resolution observations of molecular or atomic tracers of the outflow activity in HerS-3, either with the James Webb Space Telescope ({\it JWST}) via atomic emission lines to trace the extended ionized gas of the outflow or via $\rm H_2$ ro-vibrational lines to probe the hot turbulent gas, or using ALMA to detect the high-velocity emission wings of the [C{\small II}] emission line. Observing this arc-like image and deriving its characteristics will provide the basis to refine the lens model and constrain the properties of the dark matter halo by distinguishing between the core-halo or NFW profile.  

\subsection{HerS-3 as a cosmological probe}
\label{sec:cosmological-probe}
HerS-3 provides a great cosmological probe for measuring gravitational time delays and derive the value of the Hubble constant \cite[see, e.g.,][]{Schmidt2023, Kelly2023}. The active star-forming activity of this galaxy makes the event of a supernova explosion possible or could result in intrinsic starburst variability that could be followed to measure the time delay of the brightness variations between the five observed images in the plane of the lens. Both lensing models predict lensing time delays between the five images of the Einstein cross in HerS-3 that are comparable, in particular for the E, C, and W images, and within a factor of 3 for the SW image relative to the NE one. Whereas the delays are in the range of a few days for the NE and SW images (with arrival times that can swap between them), the E, C, and W images have delays that amount to about a year, with the central image predicted to appear first, with a time delay of about 275 days relative to the NE image. Long-term observations and monitoring will therefore be needed to follow the changes and the corresponding time delays in the east-west images, once an event or indications of variability have been detected in the north-west or south-east image.        

\vspace{1cm}
\section{Conclusions}
\label{sec:Conclusions}
We have presented results based on NOEMA, ALMA, and VLA observations of the dusty star-forming galaxy HerS-3 at $z=3.0607$, which was found to be gravitationally amplified in an Einstein cross with a fifth bright image at the center. The exceptional lensing configuration of HerS-3 has never been seen before and represents the first detection of an Einstein cross at sub-millimeter wavelengths. HerS-3 is lensed by four galaxies located at the center of a larger galaxy group at $z_{\rm phot} \sim 1.0$, seen in the \textit{HST} F110W image, and a massive dark matter halo, which is inferred from the lensing model in order to reproduce all the properties of the five images of the Einstein cross. Our main findings are the following:  
\begin{enumerate}
  \item All five images of the Einstein cross of HerS-3 have been detected with NOEMA in various molecular lines, including the $^{12}$CO(9$-$8) and $\rm H_2O$(2$_{02}$-1$_{11}$) emission lines, which have similar line profiles and central velocities, and the $\rm OH^+(1_1$-$0_1)$ and $\rm OH^+(1_2$-$0_1)$ lines seen in absorption, unequivocally indicating that they share similar redshifts and are therefore the lensed images of HerS-3. The images have wide separations, extending to $7\farcs5$ for the north-east and south-west images, which are unusually large compared to other Einstein crosses. The high angular resolution ($0\farcs1$) 1~mm dust continuum measured with ALMA reveals that each image is extended in the east-west direction with position angles and brightnesses changing from one image to the other. The NOEMA and ALMA data are complemented with VLA imaging data tracing the radio continuum at 6~GHz.  
  \item At a redshift of $z_{\rm phot} \sim$ 1.0, the lensing group is composed of three, close-by massive ($\rm M_* \sim 10^{11} \, M_{\odot}$) quenched systems (G1, G2, and G3), with a fourth galaxy (G4) $13\farcs5$ to the south-east of G1 with a stellar mass comparable to G1. The Einstein cross is centered in between the three close-by galaxies of the group. These lensing galaxies belong to a larger group with ten more members within a $\sim$500~kpc radius around G1.  
  \item Taking only into account the four visible massive galaxies G1--G4 that are close to HerS-3 and located at the center of the galaxy group, the lensing models were unable to reproduce the properties of the five images of the Einstein cross observed in the high-angular resolution ALMA continuum image. The fact that there is no other visible galaxy at the same redshift in the nearby field around the lensing group (the other galaxies of the group, which are far away and much less massive, are unlikely to contribute significantly to the lensing of HerS-3), led us to consider the possibility of a massive additional component that could be the group dark matter halo. Only by adding this massive component, which is constrained to lie $\sim 3-7''$ (or 25 to 60~kpc) to the south-east of G1, does the source reconstruction match the properties of the five images, namely their position, brightness, spatial extent, and orientation.    
  \item Lensing models adopting the cored isothermal sphere and the spherical NFW halo for the dark matter halo yield equally good results for the source reconstruction. However, the internal structure of the dark matter halo cannot be constrained with the current data. In particular, the distribution of the surface density map is different for each of the dark matter halo profiles. As a result, the halo mass may have any value from $\rm log(\it M_{\rm DM}/\rm M_{\odot})=12.2$ to 13.0, and the total lensing magnification is in between $\mu \sim 17$ and 19 (with significant uncertainties) for the NFW and cored halo profiles, respectively.
  \item In the source plane, HerS-3 appears as a starburst galaxy with a highly inclined rotating disk traced in $^{12}$CO(9$-$8). Blue-shifted absorption lines of OH$^+$ are detected in each image, revealing a strong outflow activity with gas expelled at velocities of at least $\rm 350 \, km \, s^{-1}$ from the nuclear region. The molecular gas reservoir of HerS-3, with an estimated mass of $M_{\rm H_2} \sim 6.9 \times 10^{10} \, \rm M_{\odot}$, has an average density of $n_{\rm H_2} \sim 3000 \, \rm cm^{-3}$ and high excitation conditions with a kinetic temperature $T_{\rm kin} \sim 270 \, \rm K$, likely dominated by shocks. With a total infrared luminosity of $L_{\rm IR} \sim 4.2 \times 10^{12} \, \rm L_{\odot}$ and a star formation efficiency of $\rm SFE = 6.5 \times 10^{-9} \, yr^{-1}$, the depletion timescale is $t_{\rm depl} = 152 \, \rm Myr$, placing HerS-3 in the regime of infrared luminous starbursts.  
\end{enumerate}
The remarkable lensing configuration of HerS-3 and the results that have been obtained to date provide a foundation to study in greater depth the properties of this system. Future higher angular resolution and deeper observations, with facilities such as ALMA in the sub-millimeter and the {\it JWST} in the near and mid-infrared, will enable us to further explore the morphology and the kinematics of HerS-3 down to spatial scales of $\rm \leq 100 \, pc$ in the source plane, and to obtain up to three orders of magnitude better near-infrared images than the {\it HST} of the lensing galaxy group and the environment of the dark matter halo, which could reveal weaker components of the galaxy group and additional lensing effects. The prospect of such high-quality data opens the way for further improvements in the lensing model, including testing observational predictions related to the parity reversal in the east, central and west images of the Einstein cross and the effect expected if the powerful wind of HerS-3 does cross two lensing caustics. Henceforth, it will be possible to better constrain the mass distribution of the galaxy group lensing HerS-3 and understand its evolutionary state, derive with greater precision the properties of the dark matter halo, such as its profile and associated mass, place stringent limits on any associated stellar emission, and, eventually, show evidence of sub-structures. 

The HerS-3 system with its exceptional Einstein cross with a fifth central image has been revealed by the present study to be a unique astrophysical laboratory to explore, at small spatial scales, a nearly edge-on dusty starburst galaxy at the peak of cosmic evolution and, importantly, to study the characteristics of the $z_{\rm phot} \sim 1.0$ galaxy group lensing HerS-3 and the properties of the associated massive dark matter halo.

\begin{acknowledgements}
This work is based on observations carried out under project number W21DF with the IRAM NOEMA Interferometer. IRAM is supported by INSU/CNRS (France), MPG (Germany) and IGN (Spain). This paper makes use of the following ALMA data: ADS/JAO.ALMA~2022.1.00145.S. ALMA is a partnership of ESO (representing its member states), NSF (USA) and NINS (Japan), together with NRC (Canada), MOST and ASIAA (Taiwan), and KASI (Republic of Korea), in cooperation with the Republic of Chile. The Joint ALMA Observatory is operated by ESO, AUI/NRAO and NAOJ. This work is also based on observations carried out under program VLA/22A-211 using the National Radio Astronomy Observatory’s (NRAO) Karl G. Jansky Very Large Array (VLA). The National Radio Astronomy Observatory is a facility of the National Science Foundation operated under cooperative agreement by Associated Universities, Inc. The anonymous referee is thanked for providing useful comments that helped to improve the contents of this paper. R. Eliot and D. Herbey are kindly thanked for their initial help with the IRAM data reduction. We would also like to thank L.-D. Roger for useful discussions about some aspects of this study. This work benefited from the support of the project Z-GAL ANR-AAPG2019 of the French National Research Agency (ANR). E.B. and E.M.C. are supported by Padua University grants Dotazione Ordinaria Ricerca (DOR) 2022–2024 and by the Istituto Nazionale di Astrofisica (INAF) grant Progetto di Ricerca di Interesse Nazionale (PRIN) 2022 2022383WFT “SUNRISE" (CUP C53D23000850006). T.J.L.C.B. gratefully acknowledges the financial support from the Knut and Alice Wallenberg foundation through grant no. KAW 2020.0081. A.J.B. acknowledges support from the Radcliffe Institute for Advanced Study at Harvard University. D.R. gratefully acknowledges support from the Collaborative Research Center 1601 (SFB~1601 sub-projects C1, C2, C3, and C6) funded by the Deutsche Forschungsgemeinschaft (DFG)–500700252. LM acknowledges financial support from the Inter-University Institute for Data Intensive Astronomy (IDIA), a partnership of the University of Cape Town, the University of Pretoria and the University of the Western Cape, and from the South African Department of Science and Innovation’s National Research Foundation under the ISARP RADIOMAP Joint Research Scheme (DSI-NRF Grant Number 150551) and the CPRR Projects (DSI-NRF Grant Number SRUG2204254729).
\end{acknowledgements}


\bibliography{references.bib}
\bibliographystyle{aasjournal}



\begin{appendix}
    
\section{Photometric redshift of the foreground group of lensing galaxies}\label{sec:zphot}
\begin{figure*}[ht!]
\centering
\includegraphics[width=0.8\textwidth]{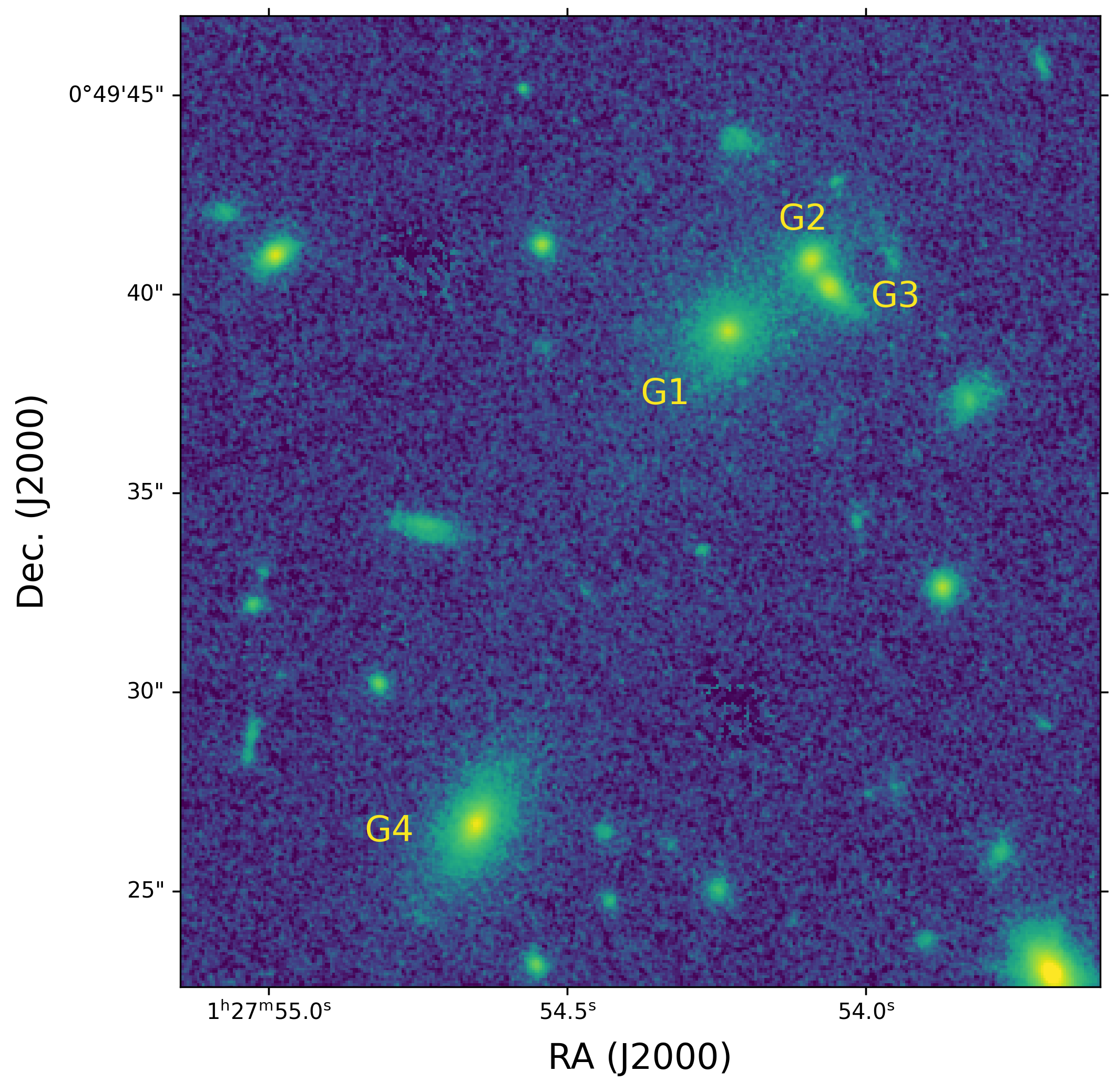}
\caption{The {\it HST} \textit{F110W} image showing the group of the four massive lensing galaxies (G1, G2, G3 and G4) at $z_{\rm phot}\sim$1.
}
\label{fig:HST-all-sources}. 
\end{figure*}

In this section, we derive the characteristics of the foreground group of lensing galaxies from the near-infrared {\it HST} F110W image (Fig.~\ref{fig:HST-all-sources}) and the available SDSS and Subaru/HSC photometry. Extracting the lens's photometry is necessary to characterize the properties of each member of the galaxy group, such as their fluxes, surface brightness distributions, SEDs, and photometric redshifts. However, this extraction is complicated when sources are heavily blended, making simple aperture photometry techniques not straightforward (e.g., galaxies G2 and G3 in Fig.~\ref{fig:HST-all-sources}). To address this issue, we applied surface brightness modeling techniques to fully disentangle the 2D emission of each galaxy. 

Some key ingredients are necessary to model the surface brightness distribution of a galaxy; these are the characteristic PSF of the observations, a noise map, and a source mask covering all sources but those that will be modeled. Additionally, given the large number of sources present in an {\it HST}-like FOV, it is useful to have a segmentation image where each source detected above a given threshold is labelled with a non-zero integer. This segmentation image will not be used to compute photometry but to estimate the blending between sources and the sources' centroids to be able to automatically perform the fitting. We detected sources up to $\sim 60^{\prime\prime}$ ($\sim 500$ kpc) from the main lensing group of G1, G2, and G3. For the \textit{HST}, the PSF model and noise map are available from \citet{Borsato2024}, and for the Subaru data these were both downloaded from the archive. Whereas for the SDSS u band, we only downloaded the available noise map. In order to compute the missing input ingredients, we first produced the segmentation image. We stacked the Subaru/HSC data computing a mean Subaru image and its associated noise map. We convolved the stacked images and noise map with a small Gaussian kernel to reduce the pixel-to-pixel variations ensuring well defined detection segments. We used the \texttt{detect\_sources} \texttt{photutils} task to identify all sources having at least 10 connected pixels above $SNR=2$. We decided to base the detection on the Subaru data and not on the \textit{HST} data in order (i) to ensure a reasonable detection threshold across most of the available bands and (ii) to avoid including in the segmentation map all those {\it HST} sources that are too faint to have a Subaru/SDSS counterpart and are thus not usable for the SED fitting. This kind of detection approach severely suffers from source blending, i.e., when two sources are close enough to be connected above the threshold level and are identified as one source. To partially address this issue, we applied the \texttt{deblend\_sources} \texttt{photutils} task that uses multi-thresholding and watershed segmentation to separate blended source segments. The resulting separation is sharp and does not include contamination between the deblended sources. In order to make full use of the \textit{HST} image quality, we first projected the Subaru-based segmentation image to the \textit{HST} and then performed the deblending on it. To produce the source mask, we first enlarged each deblended segment and then computed their intersections. When two or more sources intersected, we defined a group mask as the merged enlarged intersecting segments. With the groups defined, we computed the source centers in the \textit{HST} through \texttt{photutils} \texttt{source\_catalog} given the \textit{HST}-deblended segmentation image. Lastly, we made cutouts of each group around their group mask, masking all not belonging sources that were partially included in the cutout. We then projected each cutout and mask from the \textit{HST} to the other bands. With the source mask computed, we measured and subtracted the background using the \texttt{photutils} class \texttt{background2D}. This class estimates the background map using (sigma-clipped) statistics in grid boxes covering the input image. The background is estimated as the median value computed in the masked and sigma-clipped box region. We adopted $30\times30$ pixel boxes. By combining the boxes, it is possible to have a 2D low-resolution map of the image background. Then, the final background map is calculated by interpolating the low-resolution background map.

\begin{figure}
    \centering
    \includegraphics[width=0.9\linewidth]{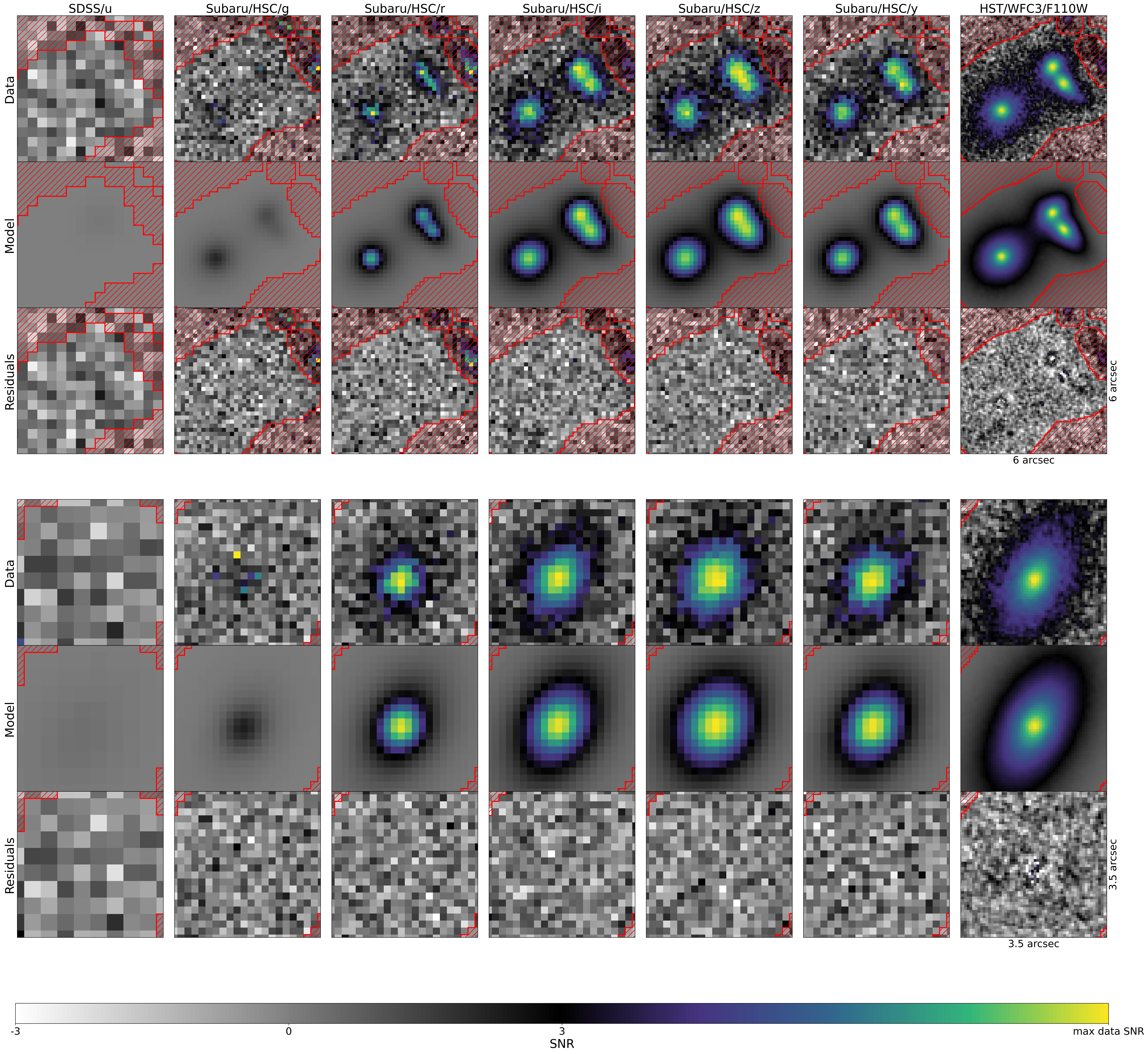}
    \caption{Results of the surface brightness modeling for the lensing galaxies G1, G2, G3 (upper panel), and G4 (lower panel) in the SDSS/u, Subaru/HSC/g, r, i, z, Y, and \textit{HST}/WFC3/F110W. In each row, we show the same $5\farcs6\times5\farcs6$ cutouts of the observed data, best-fitting models, and residuals. The data, models and residuals are all normalized to the corresponding noise map. The color map is defined to be in linear greyscale where the signal-to-noise ratio is between $-3$, and 3, and in logarithmic scale from a signal-to-noise ratio of 3 to the maximum observed in the cutout otherwise. The red dashed regions mark the portions of the data that have been masked during the fit and that were therefore not used in the minimization process.}
    \label{fig:B_subtraction_results}
\end{figure}

\begin{figure*}[t!]
\centering 
\includegraphics[width=0.9\linewidth]{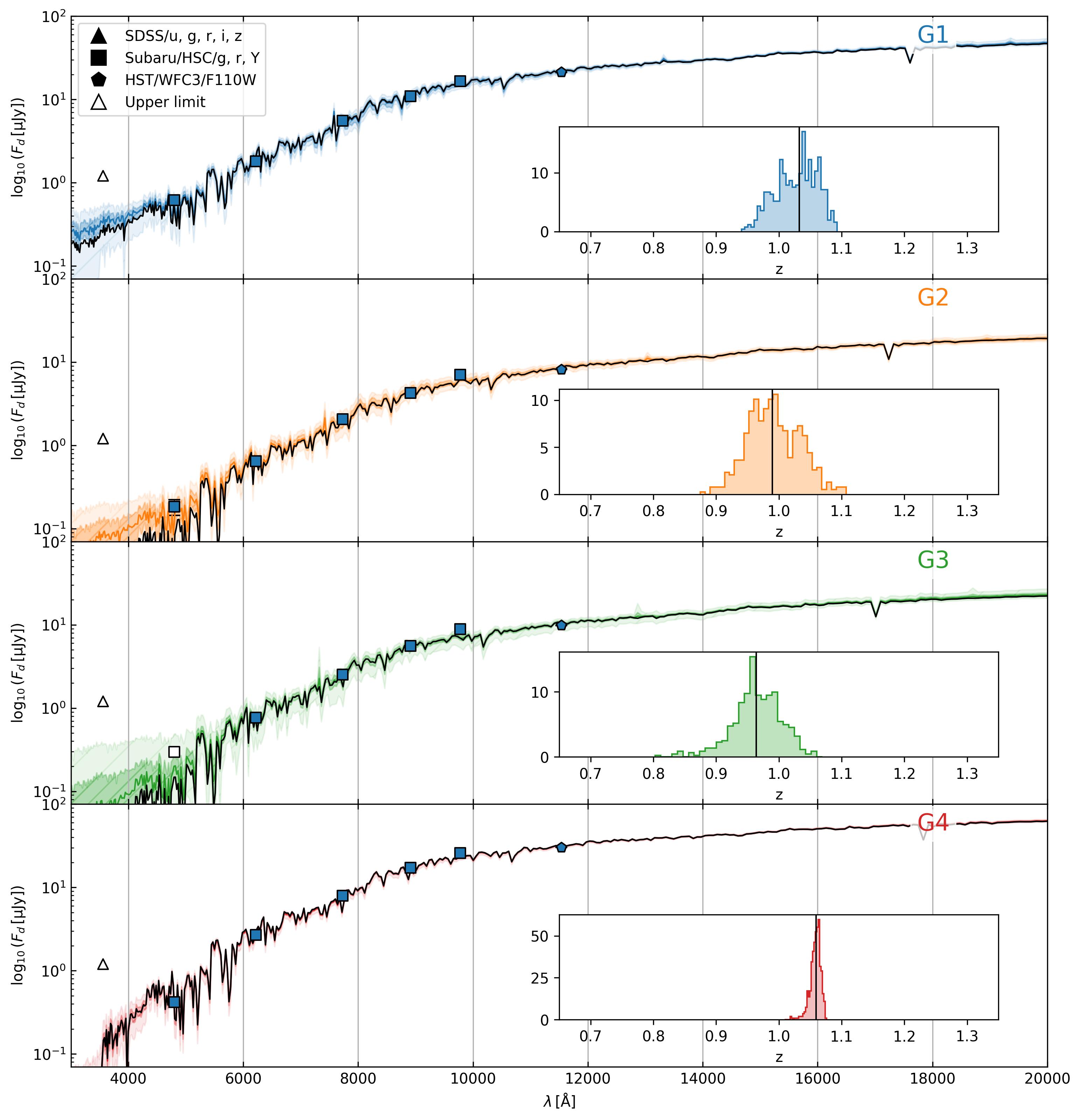} 
\caption{Best-fitting SEDs for the lensing galaxies G1, G2, G3, and G4 in blue, orange, green, and red, respectively (from top to bottom). In each panel, the SED is shown in black, while the $1\sigma$ and $3\sigma$ uncertainty intervals are highlighted with dashed and more transparent bands. The observed flux densities are displayed with filled black symbols, when there is a detection, and hollow symbols, when an upper limit is used. Each dataset is represented by a different symbol (shown in the upper left box) with the triangles tracing the SDSS data, the squares the Subaru/HSC data and the pentagons the \textit{HST}/WFC3 data. The posterior distributions of the photometric redshifts of the lensing galaxies are shown in separate insets in each of the corresponding panels.}
\label{fig:Photo-z}
\end{figure*}

We performed the surface brightness modeling using PyAutoGalaxy \citep{Nightingale2023}. For each group, the procedure adopted for the fitting is as follows: we first modeled the \textit{HST} data; when the group was made up of multiple galaxies, we performed a first preparatory fit, where we masked all but one of the blended sources fitting it with a S\'ersic profile iteratively. We used the results of these fits to initialize the simultaneous fit of all group galaxies by taking Gaussian priors centered on the maximum likelihood values of the common parameters and with a width corresponding to that of the posterior distribution scaled to allow an adequate sampling of the parameter space \citep[see Sec.~6.1 in][for details]{Nightingale2018}. 

With the \textit{HST} fit performed, we moved to fitting all the other bands. To simplify the fitting procedure, we now assumed that the galaxies' relative positions, flattenings, and position angles are the same in all bands and equal to the maximum likelihood \textit{HST} results. However, the effective surface brightness, effective radius, and S\'ersic index were left free but given narrow priors as before. We introduced two new parameters to model possible residual astrometric offsets between the \textit{HST} and each other given band. This approach seemed to work well for most cases, while significantly reducing the computing time and complexity of the model. As a concluding step, we inspected all the surface brightness fits, editing the model and refitting the data where significant residuals were found similarly to what was done in \citet{Borsato2024}. We found that it was enough to add unresolved components in the \textit{HST} band for the more complex sources. The substructures that required the addition of complexity in the \textit{HST} were in most cases washed away by the lower resolution and sensitivity of both the Subaru and SDSS data. 

Notably, some galaxies were undetected in the SDSS/u and Subaru/HSC/g bands. We randomly placed $3''$ apertures in the SDSS FOV to compute upper limits in these bands. The images were masked to remove the source's emission. Then, we computed the flux distribution through these apertures and estimated the standard deviation, weighting each flux by the number of unmasked pixels in the aperture. To measure the fluxes of the lenses, we integrated the surface brightness profiles up to a circularized radius of $3^{\prime\prime}$. This was done for two reasons: first, to provide a straightforward comparison to the upper limits in the SDSS/u and SDSS/g bands and, second, to limit the flux from the tails of the surface brightness model of the SE lens in the \textit{HST} data. The $3^{\prime\prime}$ aperture was large enough to include all significant model contributions given the constraints possible with the data available for a galaxy at $z\sim1$. If the flux estimate is compatible with 0, i.e., $F_d - 3*\sigma_{F_d} \leq 0$, we adopt as the upper limit the $\rm 95^{th}$ percentile of the flux distribution. In Fig.~\ref{fig:B_subtraction_results}, we show the results of the surface brightness modeling for the four lensing galaxies G1, G2, G3, and G4, done in the SDSS/u filters, the Subaru/HSC (g, r, i, z, Y) filters and the $HST$ WFC3/F110W filter. In Table~\ref{tab:lensing-galaxies-flux-densities}, we report the flux densities of the G1, G2, G3, and G4 lensing galaxies derived from the SDSS, Subaru/HSC and \textit{HST} observations.

\begin{figure*}[t!]
\centering
\includegraphics[width=0.9\linewidth]{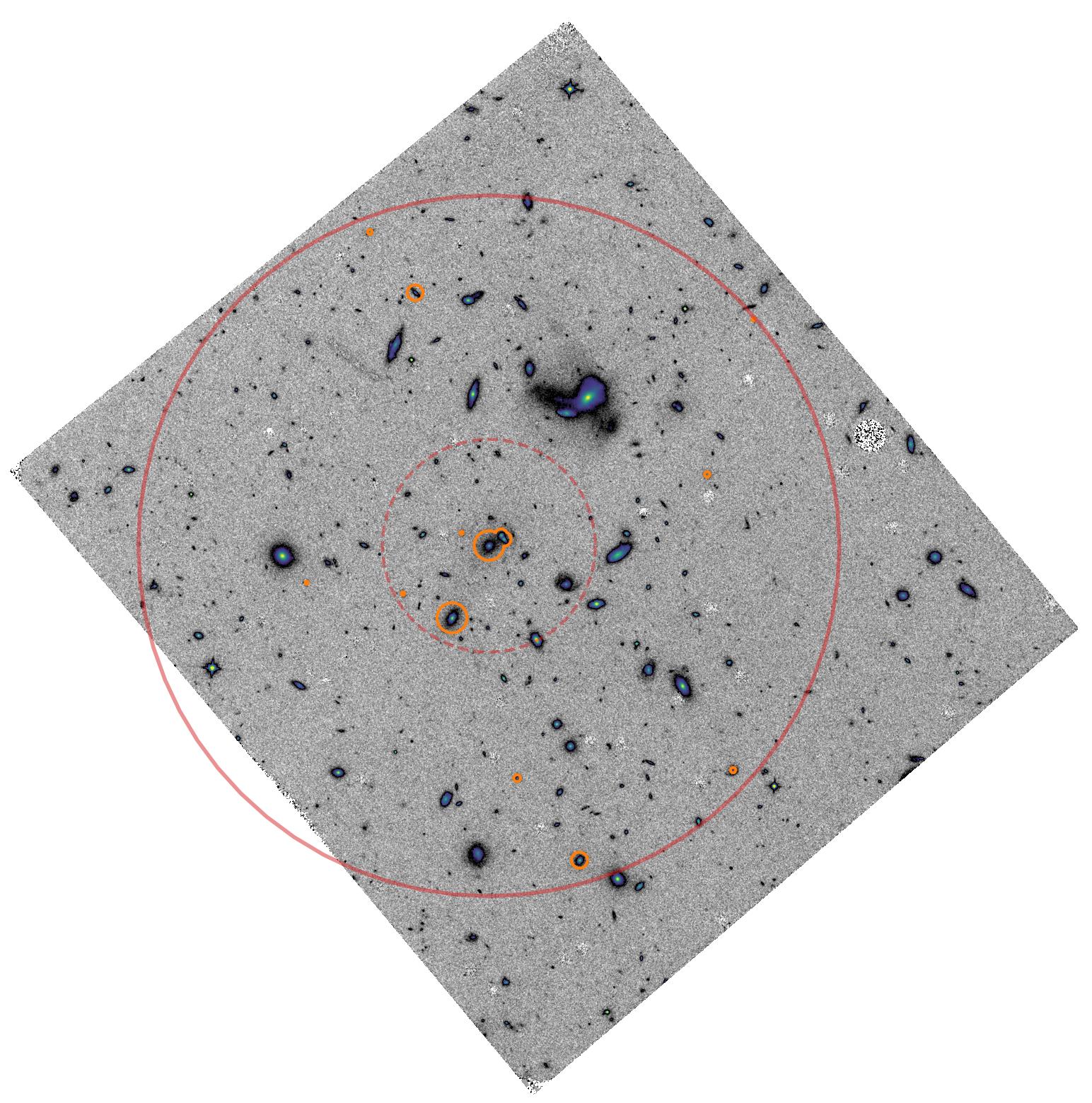} 
\caption{Galaxies that are candidate group members in the field around G1, and have photometric redshifts in the range $0.9<z_{\rm phot}<1.1$, are shown in orange and overplotted on the \textit{HST} image. The sizes of each candidate member galaxy are normalized to their best-fitting stellar mass. The continuous red circle corresponds to a radius of $60^{\prime\prime}$ ($\sim500$~kpc), while the dashed circle corresponds to a radius of $20^{\prime\prime}$ ($\sim150$~kpc), with both circles centered on the position of the galaxy G1. The image is shown in greyscale for pixels having $-3<SNR<5$ and in viridis in the range $5\leq SNR<100$ in order to highlight the sources in the \textit{HST} image.}
\label{fig:members}
\end{figure*}

We fit the lenses' spectral energy distribution (SED) using the \texttt{bagpipes} code\footnote{Available at \url{https://github.com/ACCarnall/bagpipes.git}} \citep{Carnall2018}. We parametrized the star formation history of each lens with an exponential burst, and we included a variable dust absorption following the Calzetti absorption law \citep{Calzetti2001}, as well as a variable stellar metallicity and redshift. The wavelength coverage of our observations and their broad-band nature do not allow us to constrain the main group lenses' (G1, G2, G3, and G4) stellar metallicity and dust absorption well. This effect produced broad posterior distribution reaching very high and unlikely metallicities ($\gtrsim 2\, Z_{\odot}$) and dust absorptions ($\gtrsim 1\,{\rm mag}$). These fits generally produced photometric redshift posteriors $\sim 1$, low star formation rates, $0.01 \lesssim  \rm SFR\, / ({\rm M_{\odot}\, yr^{-1}}) \lesssim 1$, and stellar masses of $\rm \sim 10^{11} M_{\sun}$, confirming that these galaxies are massive (at least in their stellar component) quenched systems. Due to degenerations between the SED parameters in the fit, a better constraint of the stellar metallicities and dust absorptions is necessary to avoid affecting the photometric redshift solution. Hence, as prior for the metallicities and dust absorptions, we adopted the values measured by \cite{Carnall2022} for a sample of 114 $z\sim1$ passive and massive galaxies with $\log_{10}(M_*/{\rm M_\sun)} > 10.8$ that matched reasonably well our lensing galaxies. These priors force the metallicity ($Z$) and dust extinction ($A_V$) of our lenses to span reasonable values. Under these assumptions, the median photometric redshifts we found are $z_{\rm G1} = 1.03_{-0.03}^{+0.03}$, $z_{\rm G2} = 0.99_{-0.05}^{+0.04}$, $z_{\rm G3} = 0.97_{-0.03}^{+0.04}$, and $z_{\rm G4} = 1.06_{-0.01}^{+0.01}$. The uncertainty intervals are computed from the 16$^{\rm th}$ and 84$^{\rm th}$ percentiles of the full posterior distribution. All galaxies have redshift distributions that are contained within $z=0.9$ and $z=1.1$. Given the differences between group members and possible residual systematics related to the 2D deblending and SED modeling, we assume the group to be at a redshift of $z_{phot} \sim 1$. The best-fitting SEDs are shown in Fig.~\ref{fig:Photo-z}, and the properties of the galaxies are listed in Table~\ref{tab:SED_fit_results}. 

\begin{table}[]
    \begin{center}
    \caption{Flux densities of the G1, G2, G3, and G4 lensing galaxies derived from the SDSS, Subaru/HSC and \textit{HST} observations.}
        \label{tab:lensing-galaxies-flux-densities}
    {\scriptsize
    \begin{tabular}{c c c c c c c c}
\hline
     Lens &  SDSS/u  & Subaru/g & Subaru/r & Subaru/i & Subaru/z & Subaru/Y & HST/F110W \\
\hline
      G1  & $<1.21$ & $0.62\pm0.07$ & $1.81\pm0.11$ & $5.53\pm0.15$ & $11.96\pm0.27$ & $16.65\pm0.45$ & $21.21\pm0.06$ \\
      G2  & $<1.21$ & $0.18\pm0.04$ & $0.64\pm0.07$ & $2.06\pm0.11$ & $4.28\pm0.24$ & $7.08\pm0.35$ & $8.17\pm0.03$ \\
      G3  & $<1.21$ & $<0.30$ & $0.77\pm0.08$ & $2.51\pm0.14$ & $5.56\pm0.28$ & $8.87\pm0.43$ & $9.85\pm0.03$ \\
      G4  & $<1.21$ & $0.42\pm0.05$ & $2.69\pm0.09$ & $7.99\pm0.13$ & $17.23\pm0.25$ & $25.86\pm0.46$ & $30.37\pm0.10$ \\
\hline
    \end{tabular}
    }
        \end{center}
    \tablenotetext{}{{\bf Notes.}  All the flux densities and upper limits are in units of $\mu {\rm Jy}$. The upper limits are computed at $3\sigma.$}
\end{table}

For the remaining galaxies within a radius of $\sim 60^{\prime\prime}$ ($\sim 500$~kpc) from the main lensing group, given that many of the neighbouring sources were fainter, we performed three additional SED fits for each galaxy, where we reduced the model complexity by fixing the metallicity to sub-solar, solar and super-solar, i.e., $Z/Z_{\odot} = 0.5$, $Z/Z_{\odot} = 1$ and $Z/Z_{\odot} = 1.5$. Again, if all the best-fitting SEDs predicted a photometric redshift $z_{\rm phot} \sim 1$ and low specific star formation rates, we used the \cite{Carnall2022} templates as above. To decide which best-fitting SED model to use, we prioritized the \cite{Carnall2022} SED when available, otherwise we used the all-parameters-free models, unless either the metallicity or the dust extinction reached extreme values. Then, we selected as candidate group members all the galaxies that had the chosen best-fitting median photometric redshift between 0.9 and 1.1. This resulted in a total of 14 candidate members, including G1, G2, G3, and G4, with a total stellar mass of $M_* \sim 6 \times 10^{11} \, M_{\odot}$. A summary plot of the galaxy group with all the candidate members is shown in Fig.~\ref{fig:members}. It is important to note that most of the stellar mass of this galaxy group at $z_{\rm phot} \sim 1$ is accounted for by the four galaxies G1, G2, G3, and G4, which contain $\sim 60\%$ of the total mass within $500$~kpc and $\sim 90\%$ within $250$~kpc (where only two additional candidate group members were identified).  These findings suggest that the lensing galaxies G1--G4 belong to a more extended and complex galaxy group. This result is in line with the lensing modeling analysis, which indicates the presence of a massive dark matter halo, which is located in between G1 and G4 and is close to the mass center of the group, in order to reproduce the observed properties of the HerS-3 Einstein cross (see Sect.~\ref{sec:Lensing-Model}).

\section{Details of the lens model}\label{sec:lens}
We conducted the lens modeling in two stages. First, because the ALMA images are well separated and fairly compact, we treated them as point images for initial parameter space exploration. This approach is computationally efficient and facilitates a thorough exploration. Once we identified promising models, we reconstructed the full extended source structure using the \texttt{pixsrc} software by \citet{Tagore&Keeton2014} and \citet{Tagore&Jackson2016}.

\subsection{Lens model setup: Parameters and constraints}
\label{sec:Lens-model-setup} 
For the point image analysis, we used the positions in Table~\ref{tab:cont-individual-images} and computed offsets relative to galaxy G1 (see coordinates in Table~\ref{tab:SED_fit_results}). We assumed position uncertainties of $0\farcs03$. We used the ALMA 292 GHz flux densities (and their uncertainties) as constraints on the relative brightnesses.

We treat the galaxies G1, G2, G3, and G4 as singular isothermal ellipsoids. These models implicitly assume that the galaxies have dark matter halos that extend beyond the baryonic components, and the baryons and dark matter combine to have a total density profile of the form $\rho(r) = \sigma^2/(2\pi G r^2)$, which corresponds to a scaled surface mass density for lensing of
\begin{equation}
  \kappa(R) = \frac{\Sigma(R)}{\Sigma_{\rm crit}}
  = \frac{b}{2R}
  \qquad\mbox{where}\qquad
  b = 4\pi \left(\frac{\sigma}{c}\right)^2 \frac{D_{ls}}{D_{s}}
\end{equation}
Here $\sigma$ is the velocity dispersion, $D_s$ and $D_{ls}$ are angular diameter distances from the observer or lens to the source, respectively, and $\Sigma_{\rm crit}$ is the critical surface mass density for lensing. We can generalize to an elliptical model by replacing $R$ with an elliptical radius.

Each isothermal mass component is described by five parameters\footnote{We treated G4 as spherical for simplicity, so it has three parameters rather than five.}: two position coordinates, the Einstein radius parameter $b$, and an ellipticity and position angle.
We constrain the galaxy positions using the offsets given in Table \ref{tab:SED_fit_results} with uncertainties of $0\farcs03$. For a fixed aperture $R$, the lensing mass of an isothermal ellipsoid is $M_{\rm lens} = \pi b R \Sigma_{\rm crit}$, where $\Sigma_{\rm crit}$ is the critical density for lensing. Therefore we place lognormal priors on the relative Einstein radii using the relative stellar masses (from Table \ref{tab:SED_fit_results}), with uncertainties of $0.1$ dex. We fit the ellipticity and position angle using the quasi-Cartesian components $e_c = e \cos 2\theta_e$ and $e_s = e \sin 2\theta_e$. In initial modeling, we found that the model galaxies sometimes became highly elongated. In an attempt to limit such behavior, we imposed mild Gaussian priors on $e_c$ and $e_s$ using the observed ellipticities and position angles, with uncertainties of $0.20$ in each component (see Table \ref{tab:galaxy-shape-priors}). Last but not least, we included external shear characterized by the quasi-Cartesian components $\gamma_c$ and $\gamma_s$ to represent tidal effects from other mass in the vicinity of the lens galaxies or along the line of sight. We imposed Gaussian priors on the shear of $\gamma_{c,s} = 0.00 \pm 0.05$ for mild priors and $0.00 \pm 0.02$ for strong priors.

Since the precise lens redshifts ($z_l$) are unknown, we initially assume that all four galaxies are at the same redshift. In that case the modeling can be done in angular units without assuming a specific value for the lens redshift, but we do need to assume $z_l$ in order to convert from angular units to physical units. We explored the possibility that the galaxies have different redshifts; while some details changed, the overall conclusions remained the same, so we defer further consideration of multi-plane lens models until more information is known about the redshifts.

To summarize: minimal models have a total of 20 mass model parameters (18 for the four galaxies and two for external shear) and three source parameters (position and brightness). There are a total of 34 constraints or priors on the models: 15 constraints from the image positions and brightnesses, eight additional constraints from the galaxy positions, three priors on the relative Einstein radii, and eight shape priors.

\begin{figure*}[ht!]
    \centering
    \includegraphics[width=1\linewidth]{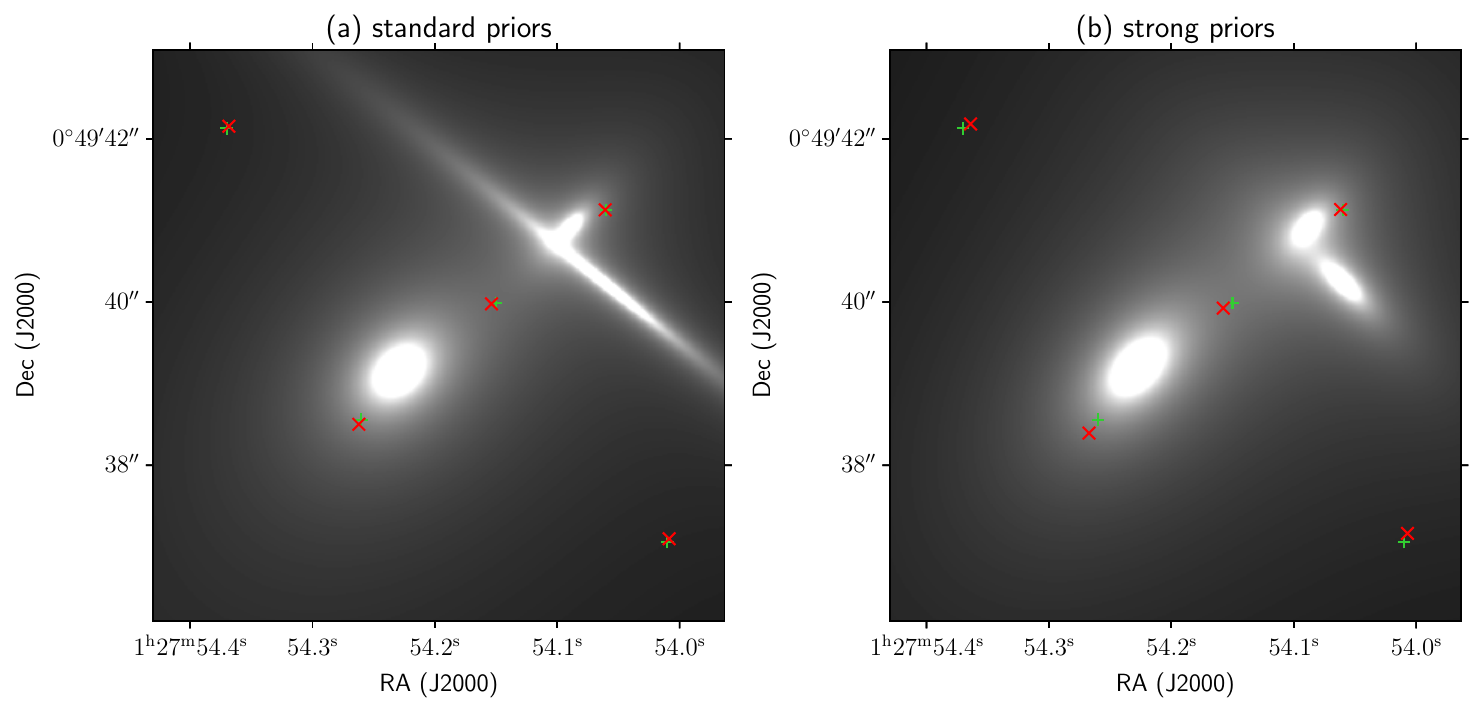}
    \caption{The gray scale shows surface mass density maps for lens models that only include galaxies G1, G2, G3, and G4, all assumed to lie at the same redshift ($z_{\rm phot} \sim 1$). The green $+$ symbols show the observed image positions, while the red $\times$ symbols show the positions of the images predicted by the lens model.
    Panel (a) shows results with mild shape priors; the model G3 is highly elongated, making this model implausible.
    Panel (b) shows results with strong shape priors; here the galaxy shapes are reasonable, but this model cannot reproduce the image positions well.
    In both panels, galaxy G4 is outside of the frame to the left.}
    \label{fig:model-nohalo-kappa}
\end{figure*}

\begin{table}
    \caption{Galaxy shape priors} 
    \label{tab:galaxy-shape-priors}  
    \begin{center}
    \begin{tabular}{crrcc} 
        \hline\hline
        Galaxy & \multicolumn{1}{c}{$e_c$} & \multicolumn{1}{c}{$e_s$} & Standard priors & Strong priors \\
        \hline
        G1 & $-0.04$ & $-0.17$ & $\pm0.20$ & $\pm$0.05 \\
        G2 & $ 0.09$ & $-0.38$ & $\pm0.20$ & $\pm$0.05 \\
        G3 & $ 0.05$ & $ 0.57$ & $\pm0.20$ & $\pm$0.05 \\
    \hline      
    \end{tabular} 
    \end{center}
\vspace{-0.2cm}
\tablenotetext{}{{\bf Notes.} The quasi-Cartesian shape parameters are $e_c = e \cos 2\theta_e$ and $e_s = e \sin 2\theta_e$, where $e$ is the ellipticity and $\theta_e$ is the position angle measured East of North.
Note that G4 is assumed to be spherical.
Columns 4 and 5 give the standard deviation of Gaussian priors on $e_c$ and $e_s$ in the standard and strong cases, respectively.
We also include Gaussian priors on external shear of the form $\gamma_{c,s} = 0.00 \pm 0.05$ for standard priors, and $0.00 \pm 0.02$ for strong priors.}
\end{table}

\subsection{Models with just four galaxies}
\label{sec:Models-with-four-galaxies}

\begin{figure*}[ht!]
    \centering
    \includegraphics[width=1.0\linewidth]{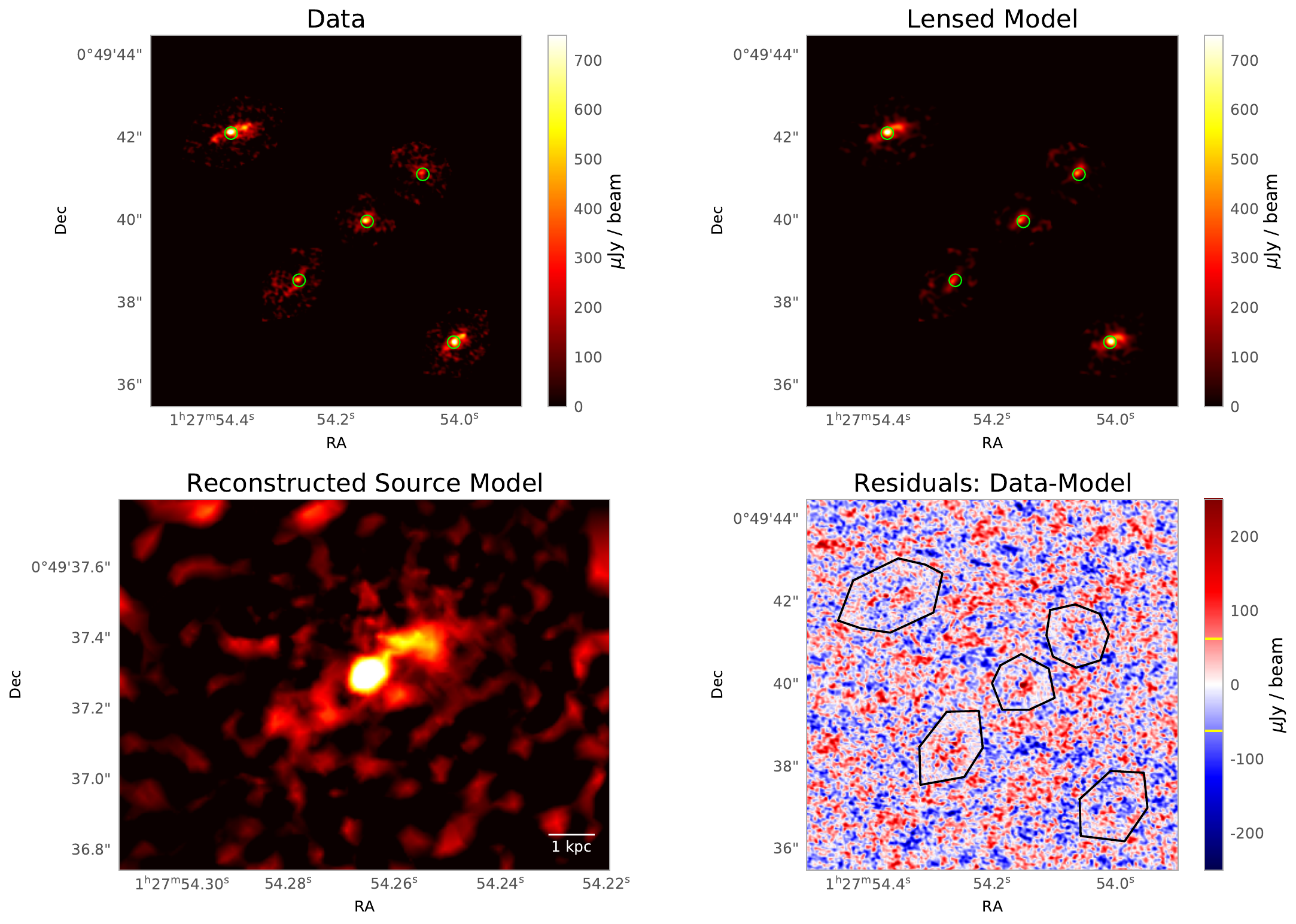}
    \caption{
    Results of pixel-based source reconstruction of HerS-3 using the lens model with just the four main galaxies (G1, G2, G3 and G4) and standard priors (as shown in Fig.~\ref{fig:model-nohalo-kappa}a).
    \textit{Top left:} The ALMA 292~GHz continuum image, masked around the five lensed images. To guide the eye, circles mark the image coordinates given in Table \ref{tab:cont-individual-images}.
    \textit{Top right:} Model image, lensed and convolved with the beam. Green circles represent the same image peak brightness positions as in the data image.
    \textit{Bottom left:} Reconstructed source-plane image. The scale bar marks $1$~kpc in the source plane.
    \textit{Bottom right:} Data minus model residuals. Black polygons indicate the masks used for the {\em pixsrc} analysis. Yellow lines in the colorbar represent the noise level of $\rm 63 \, \mu Jy/beam$ measured outside of the masked regions. 
    }
    \label{fig:pixsrc-nohalo-ellshr}
\end{figure*}

\begin{figure*}[ht!]
    \centering
    \includegraphics[width=1.0\linewidth]{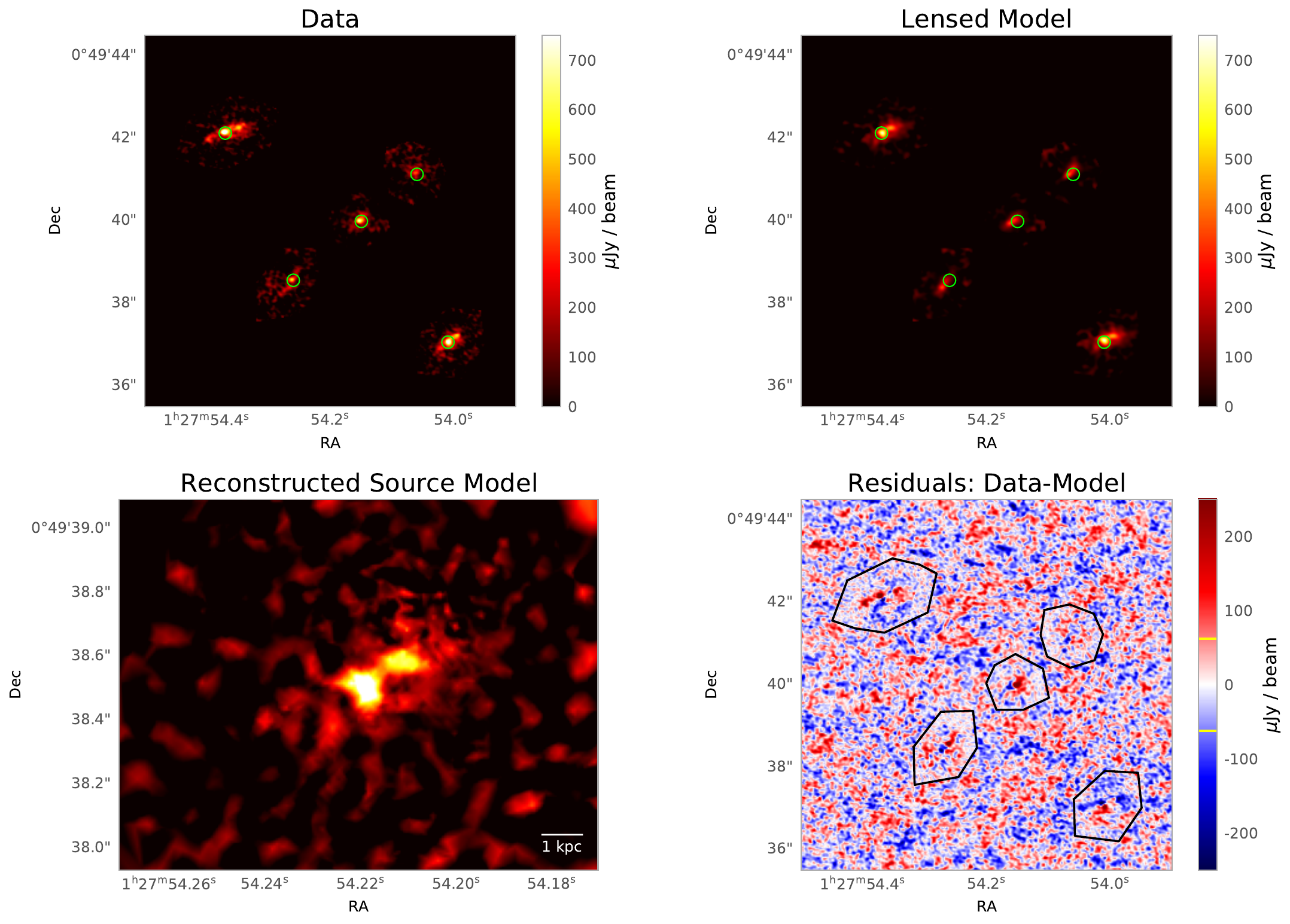}
    \caption{
    Similar to Fig.~\ref{fig:pixsrc-nohalo-ellshr}, but for the lens model only including the four main galaxies with strong priors (as shown in Fig.~\ref{fig:model-nohalo-kappa}b).
    }
    \label{fig:pixsrc-nohalo-tight}
\end{figure*}

We first consider lens models that include four mass components representing the four observed galaxies G1, G2, G3, and G4 (plus external shear). Figure~\ref{fig:model-nohalo-kappa} depicts the results by showing the surface mass density map as well as the observed and predicted image positions. The only way for this type of model to match the images reasonably well is to make one of the model galaxies highly elongated (panel a). This model has a total $\chi^2 = 104$, with contributions of $12$ from the image positions (corresponding to an rms offset in the image positions of $0\farcs05$), $40$ from the image brightnesses, and $52$ from the priors (compared to 11 degrees of freedom). Because the model G3 is unreasonably elongated, we consider strengthening the shape priors by reducing the uncertainties on the $e_c$ and $e_s$ shape components to 0.05. The resulting model has more plausible galaxy shapes but it cannot reproduce the image positions very well (panel b): this model has $\chi^2 = 300$, with $90$ from the image positions (an rms offset of $0\farcs13$), $52$ from the image brightnesses, and $158$ from the priors.

Even though these models have clear shortcomings, it is still worthwhile to see how they handle the full ALMA 292 GHz data. We use the pixel-based source reconstruction code \textit{pixsrc} from \cite{Tagore&Keeton2014}. This code reconstructs the source surface brightness on a grid in the source plane in a way that minimizes a penalty function with two terms. The first term is a goodness of fit comparing the model image (lensed and convolved with the beam) to the ALMA data. The second ``regularization'' term penalizes deviations from a smooth surface brightness distribution to prevent the model from overfitting noise. Note that we apply \textit{pixsrc} to the real space data (not the $uv$ plane) and adopt the noise scaling technique of \citet{Riechers2008} to account for spatially correlated noise \citep[see also][]{Sharon2019}.

We construct a data mask around each of the five observed images of the background galaxy, including a small margin of background in each region as required by \textit{pixsrc}. Outside of the masked regions, we measure a noise level of $\rm 63 \, \mu Jy/beam$. With the noise scaling approach, we find that increasing the noise level in \textit{pixsrc} to $\rm 80 \, \mu Jy/beam$ provides the best balance between goodness of fit and regularization. For the beam, we use an elliptical Gaussian with FWHM $0\farcs12$ and $0\farcs10$ along the major and minor axes, respectively, and a position angle of $74^\circ$.

Figures \ref{fig:pixsrc-nohalo-ellshr} and \ref{fig:pixsrc-nohalo-tight} show results of pixelated source reconstructions for the models with mild and tight shape priors, respectively. The inability to match the observed positions is apparent. Moreover, the models fail to match the orientation of the extended structure in the SW image, as well as general features of the E, C, and W images.

\subsection{Models with an additional mass component}
\label{sec:Model-with-halo}
In many lens systems with multiple galaxies in close proximity, the galaxies are part of a group or cluster \citep[see, e.g.,][and Sect.~\ref{sec:zphot}]{Natarajan2024}. We therefore consider a new set of models with a fifth mass component that could represent a common dark matter halo. We treat the additional mass component as a cored isothermal sphere or a spherical NFW halo. In both cases the halo is described by four parameters (position, Einstein radius or density parameter, and scale radius), bringing the total number of mass model parameters to 27.

\begin{figure*}
    \centering
    \includegraphics[width=1.0\linewidth]{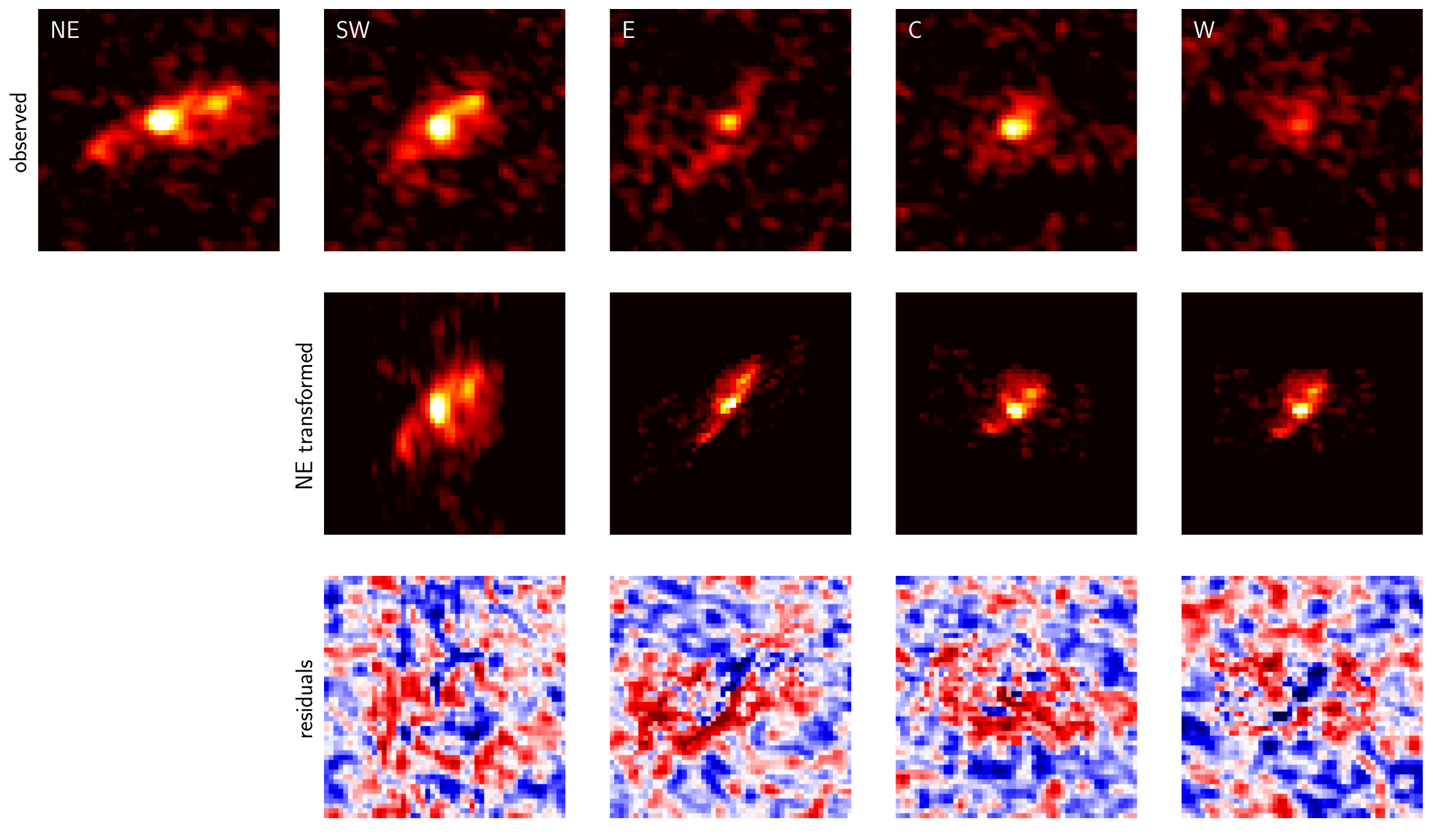}
    \caption{
    Illustration of the lensed image shape analysis using linear transformations.
    The top row shows cutouts around the five lensed images in the ALMA 292 GHz map.
    The middle row shows transformed versions of the NE image computed using the best lens model with a cored halo.
    The bottom row shows residuals; such residuals are used to compute an image shape penalty when exploring the model parameter space.
    Note that this analysis does not properly account for beam effects, but it is still useful for assessing how well different models can match the image shapes.
    }
    \label{fig:model-core-xform}
\end{figure*}

We introduce an approximate method to impose constraints from the extended images shapes without requiring a full source reconstruction. In the absence of beam effects, lensing conserves surface brightness, so images shapes are determined by spatial distortions. In the vicinity of an image $i$, distortions are governed by the local lensing magnification tensor $\boldsymbol{\mu}_i$ such that small displacements in the image plane, $\Delta\boldsymbol{x}_i$, are related to small displacements in the source plane, $\Delta\boldsymbol{u}$, by $\Delta\boldsymbol{x}_i = \boldsymbol{\mu}_i\,\Delta\boldsymbol{u}$. Repeating the same analysis for a different image $j$, we see that we can describe the transformation from image $i$ to image $j$ as $\Delta\boldsymbol{x}_j = \left(\boldsymbol{\mu}_j \, \boldsymbol{\mu}_i^{-1}\right) \Delta\boldsymbol{x}_i$. We can apply the idea to lens modeling as shown in Fig.~\ref{fig:model-core-xform}. Specifically, given a model we can compute the tensor $\boldsymbol{\mu}_j \, \boldsymbol{\mu}_{\rm NE}^{-1}$ that would transform the NE image to each of the other image locations (middle row). We can then compare the transformed images to the observed images (top row) and use the residuals (bottom row) to compute a shape penalty for the lens models. This method is less precise than a full pixelated source reconstruction, in particular because it does not account for beam smearing, but it is much faster so it can be easily incorporated into the parameter space exploration.

Models with the additional mass component are quite successful, and are discussed fully in Section \ref{sec:Lensing-Model}. For completeness, we list all of the parameter values here in Table \ref{tab:model_parameters}.

\begin{table*}[]
    \caption{Parameters for lens models with four galaxies and an additional dark matter halo}
        \label{tab:model_parameters}
    \begin{center}
    \begin{tabular}{crrrr}
    \hline\hline
Parameter & \multicolumn{2}{c}{Cored halo model} & \multicolumn{2}{c}{NFW halo model} \\
& \multicolumn{1}{c}{best} & \multicolumn{1}{c}{MCMC} & \multicolumn{1}{c}{best} & \multicolumn{1}{c}{MCMC} \\
\hline
$\log b_1$ & $-0.27$ & $-0.13_{-0.15}^{+0.08}$ & $-0.11$ & $-0.09_{-0.06}^{+0.06}$ \\
$x_1$ & $0.00$ & $0.00_{-0.03}^{+0.03}$ & $0.00$ & $0.01_{-0.03}^{+0.03}$ \\
$y_1$ & $0.02$ & $0.02_{-0.03}^{+0.03}$ & $0.00$ & $0.01_{-0.03}^{+0.03}$ \\
$e_{c1}$ & $0.03$ & $0.06_{-0.10}^{+0.10}$ & $0.06$ & $0.07_{-0.10}^{+0.11}$ \\
$e_{s1}$ & $-0.23$ & $-0.19_{-0.13}^{+0.15}$ & $-0.03$ & $-0.03_{-0.17}^{+0.16}$ \\
\hline
$\log b_2$ & $-0.58$ & $-0.46_{-0.13}^{+0.11}$ & $-0.44$ & $-0.43_{-0.08}^{+0.08}$ \\
$x_2$ & $2.13$ & $2.14_{-0.03}^{+0.03}$ & $2.12$ & $2.13_{-0.03}^{+0.03}$ \\
$y_2$ & $1.79$ & $1.79_{-0.03}^{+0.03}$ & $1.80$ & $1.79_{-0.03}^{+0.02}$ \\
$e_{c2}$ & $0.03$ & $0.02_{-0.14}^{+0.13}$ & $-0.01$ & $-0.01_{-0.12}^{+0.13}$ \\
$e_{s2}$ & $-0.36$ & $-0.37_{-0.13}^{+0.15}$ & $-0.37$ & $-0.38_{-0.13}^{+0.16}$ \\
\hline
$\log b_3$ & $-0.55$ & $-0.47_{-0.12}^{+0.09}$ & $-0.43$ & $-0.44_{-0.08}^{+0.08}$ \\
$x_3$ & $2.52$ & $2.51_{-0.03}^{+0.03}$ & $2.52$ & $2.51_{-0.03}^{+0.03}$ \\
$y_3$ & $1.13$ & $1.13_{-0.03}^{+0.03}$ & $1.13$ & $1.14_{-0.03}^{+0.03}$ \\
$e_{c3}$ & $-0.14$ & $-0.20_{-0.13}^{+0.16}$ & $-0.24$ & $-0.23_{-0.12}^{+0.15}$ \\
$e_{s3}$ & $0.60$ & $0.55_{-0.14}^{+0.09}$ & $0.49$ & $0.52_{-0.13}^{+0.10}$ \\
\hline
$\log b_4$ & $-0.25$ & $-0.13_{-0.15}^{+0.13}$ & $-0.09$ & $-0.07_{-0.11}^{+0.12}$ \\
$x_4$ & $-6.33$ & $-6.33_{-0.03}^{+0.03}$ & $-6.33$ & $-6.33_{-0.03}^{+0.03}$ \\
$y_4$ & $-12.33$ & $-12.33_{-0.03}^{+0.03}$ & $-12.34$ & $-12.34_{-0.03}^{+0.03}$ \\
\hline
$\log b_h$ & $0.89$ & $0.75_{-0.20}^{+0.22}$ & $-0.91$ & $-0.87_{-0.09}^{+0.14}$ \\
$x_h$ & $-2.36$ & $-3.05_{-1.06}^{+0.73}$ & $-2.92$ & $-3.42_{-1.87}^{+0.87}$ \\
$y_h$ & $-1.97$ & $-2.56_{-1.00}^{+0.67}$ & $-2.38$ & $-2.79_{-1.73}^{+0.83}$ \\
$\log s_h$ & $0.70$ & $0.57_{-0.23}^{+0.21}$ & $1.52$ & $1.47_{-0.31}^{+0.28}$ \\
\hline
$\gamma_c$ & $0.02$ & $0.02_{-0.01}^{+0.01}$ & $0.03$ & $0.03_{-0.01}^{+0.01}$ \\
$\gamma_s$ & $-0.06$ & $-0.06_{-0.02}^{+0.02}$ & $-0.07$ & $-0.07_{-0.03}^{+0.03}$ \\
\hline
    \end{tabular}
    \end{center}
\tablenotetext{}{{\bf Notes.}
We list the best value along with the median and 68\% confidence interval for each parameter in each of the two types of lens models with a cored isothermal sphere halo and a spherical NFW halo. Note: in a high-dimensional parameter space with an asymmetric posterior distribution, the best value may not fall within the 68\% confidence interval of the marginalized distribution for some parameters.
All positions $x$ and $y$ are given in arcsec relative to the observed position of galaxy G1.
For galaxies G1--G4, $b$ is the Einstein radius parameter in arcsec.
Galaxies G1--G3 are allowed to be elliptical, while G4 is assumed to be spherical.
For the cored halo, $b_h$ is the Einstein radius parameter and $s_h$ is the core radius, both in arcsec.
For the NFW halo, $b_h$ is the NFW lensing strength $\kappa_s = \rho_s r_s / \Sigma_{\rm crit}$ (which is dimensionless), and $s_h$ is the scale radius $r_s$ in arcsec.
$e_c$ and $e_s$ are the quasi-Cartesian components of ellipticity, while $\gamma_c$ and $\gamma_s$ are the analogous parameters for external shear (which are all dimensionless).
}
\end{table*}

\section{Radiation transfer Large Velocity Gradient Model}\label{sec:LVG-model}
To investigate the CO line excitation of HerS-3 and estimate the physical conditions of the molecular gas, we derived a Large Velocity Gradient (LVG) radiative transfer model for the CO spectral line energy distribution (SLED) of HerS-3 using the Markov chain Monte Carlo (MCMC) implementation of RADEX \citep{vanderTak2007}, as presented in \cite{Yang2017}. This one-dimensional non-LTE radiative transfer code assumes spherical symmetry and describes the velocity gradients, ${\rm d}v/{\rm d}r$, in the expanding sphere approximation. It performs a MCMC sampling of the parameter space that includes the molecular hydrogen density ($n_{\rm H_2}$), the gas kinetic temperature ($T_{\rm k}$), the CO column density per unit velocity gradient ($N_{\rm CO}/{\rm d}v$), and the solid angle of the source ($\Omega_{\rm app}$), and it samples the posterior probability distributions functions from the CO line fluxes modeled by RADEX. A Bayesian approach is used to fit the line fluxes from the model, and the code {\em emcee} is adopted to perform the MCMC calculation. 

\begin{figure*}[!ht] 
    \centering
        \includegraphics[width=.48\linewidth]{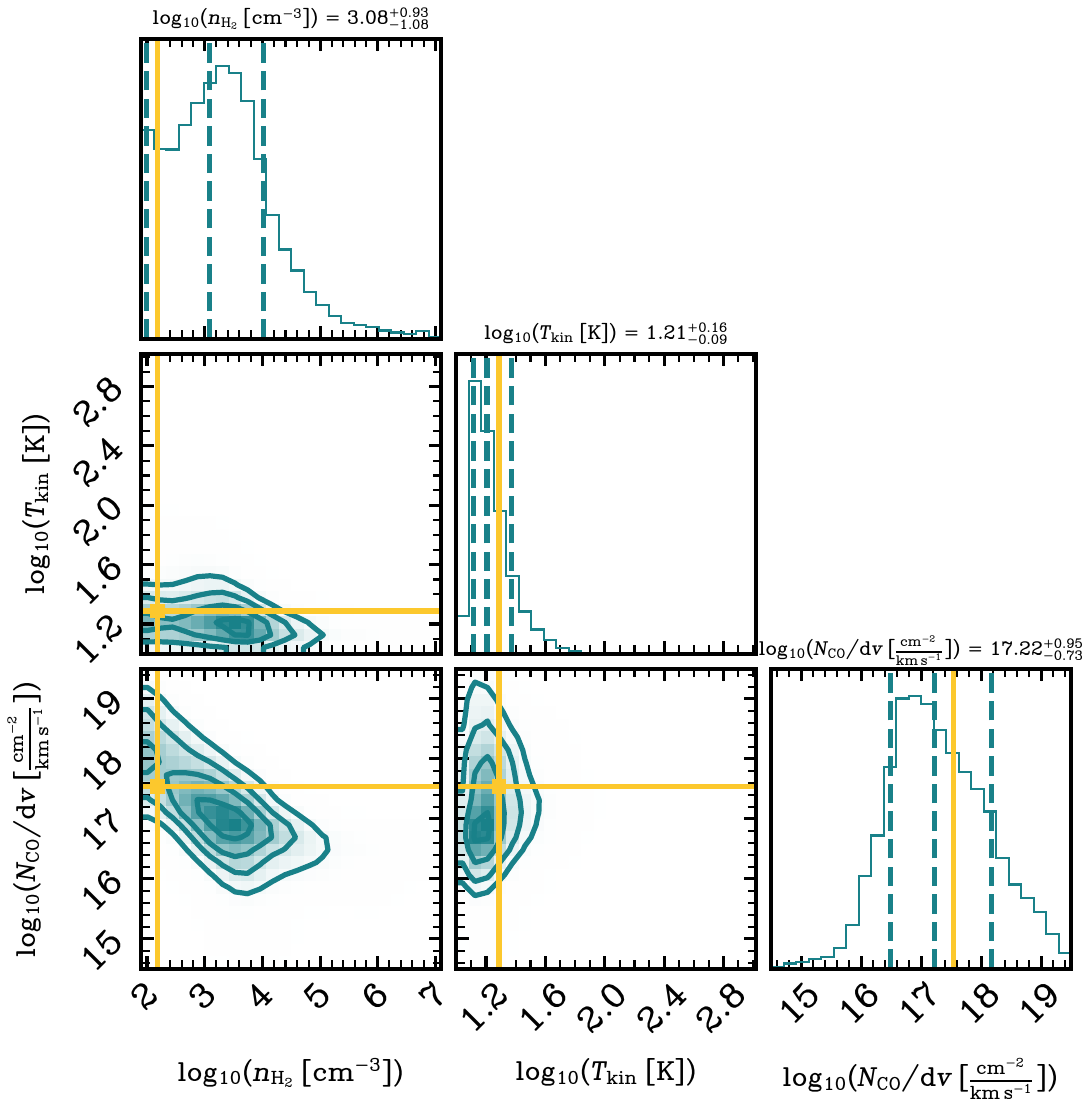}
        \includegraphics[width=.48\linewidth]{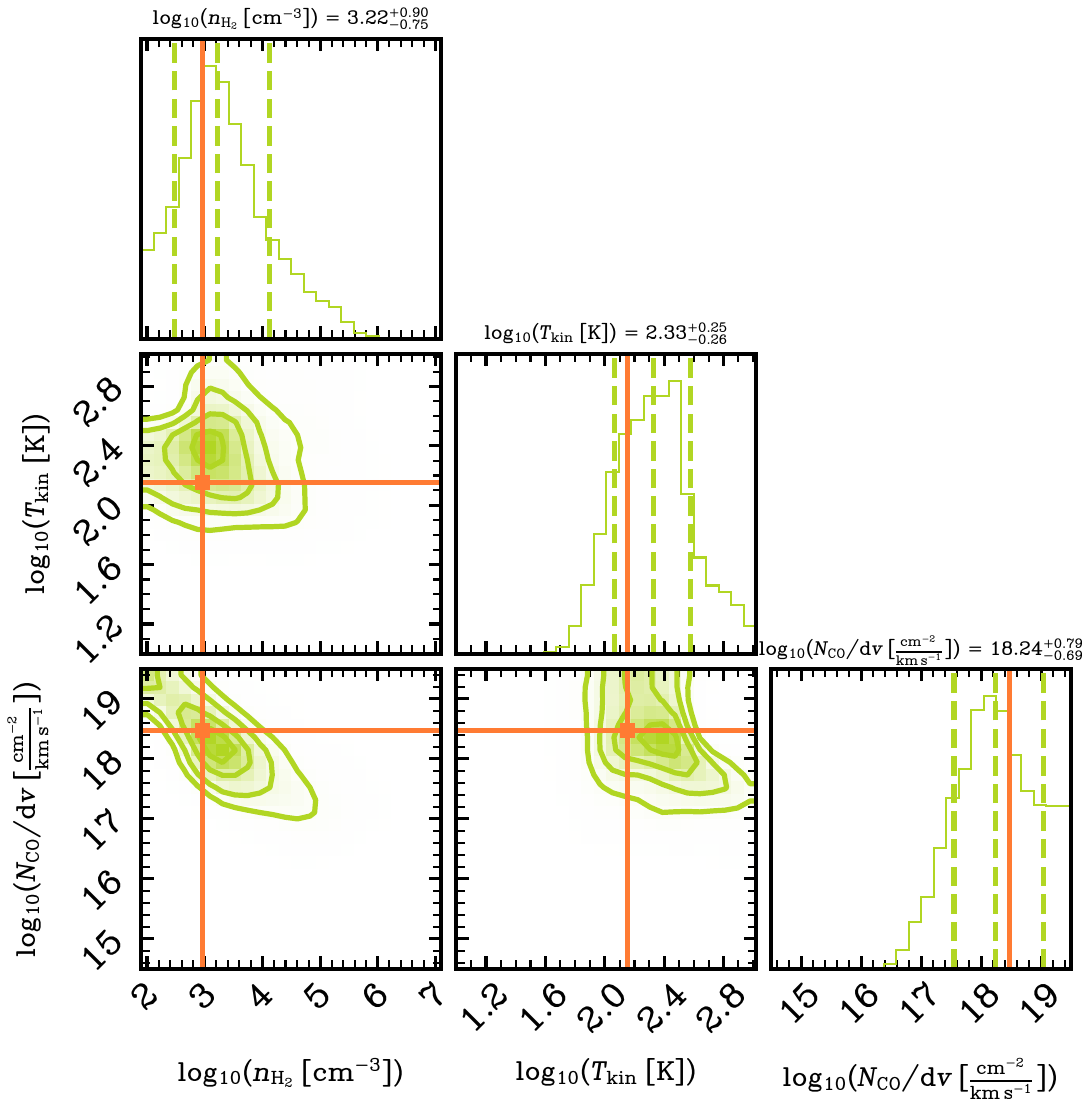}    
  \caption{The posterior probability distributions of the molecular gas density $n_{\rm H_2}$, gas temperature $T_{\rm kin}$ and CO column density per velocity $N_{\rm CO}/{\rm d}v$ of the source for the cold (left) and warm components (right), with the maximum posterior possibility point in the parameter space shown in orange lines and points, while dotted lines mark the $\pm 1\sigma$ range in the distributions. The contours are in steps of $0.5\sigma$ starting from the center. The resulting parameter values are listed above the corresponding histograms.}
\label{fig:CO-SLED-Posterior-Probablity-Distribution} 
\end{figure*}

As the CO SLED of HerS-3 displays a strong $\rm ^{12}CO$(9$-$8) emission line (Fig.~\ref{fig:CO-SLED}), we used a two-excitation component model, assigning different parameters to each of the components and assuming two additional priors on the relative sizes and temperatures within the Bayesian analysis. The results of the two-component model are shown in Fig.\ref{fig:CO-SLED}. Figure~\ref{fig:CO-SLED-Posterior-Probablity-Distribution} displays the posterior distributions of the molecular gas density $n_{\rm H_2}$, the gas temperature $T_{\rm kin}$, and the CO column density per velocity $N_{\rm CO}/{\rm d}v$ of the source for the cold and warm components.

\end{appendix}

\end{document}